\documentclass{aa} 

\usepackage{color, colortbl}    
\usepackage{graphicx}
\usepackage[varg]{txfonts} 
\begin{document}

   \title{CODEX: Role of velocity substructure in the scaling relations of galaxy clusters}

   \subtitle{}

   \author{S. Damsted\inst{1}
          \and
          A. Finoguenov\inst{1}
          \and
          N. Clerc\inst{2}
          \and
          I. Davalgait$\dot{\rm e}$\inst{1}
          \and
          C.~C. Kirkpatrick\inst{1}
          \and
          G. A. Mamon\inst{3}
         \and
         J. Ider Chitham\inst{4}
          \and
          K. Kiiveri\inst{1}
          \and
          J. Comparat\inst{4}
          \and
          C. Collins\inst{5}
          }

   \institute{Department of Physics, University of Helsinki,
              Gustaf Hällströmin katu 2, 00560 Helsinki\\
              \email{sanna.damsted@helsinki.fi}
         \and
    IRAP, Universit{\'e} de Toulouse, CNRS, UPS, CNES, F-31028 Toulouse, France
    \and
    Institut d’Astrophysique de Paris (UMR 7095: CNRS \& Sorbonne Université), F-75014 Paris, France             
    \and
    MPE, Giessenbachstr. 1, Garching 85748, Germany
    \and
    Astrophysics Research Institute, Liverpool John Moores University, IC2, Liverpool Science Park, 146 Brownlow Hill, Liverpool L3 5RF, UK
    }

   \date{Received ; accepted }

 
  \abstract
   {The use of galaxy clusters as cosmological probes relies on a detailed understanding of their properties. They define cluster selection and ranking linked to a cosmologically significant cluster mass function. Previous studies have employed small samples of clusters, concentrating on achieving the first calibrations of cluster properties with mass, while the diversity of cluster properties has been revealed via detailed studies. }
   {The large spectroscopic follow-up on the CODEX cluster sample with SDSS and NOT enables a detailed study of hundreds of clusters, lifting the limitations of previous samples. We aim to update the spectroscopic cluster identification of CODEX by running the spectroscopic group finder on the follow-up spectroscopy results and connecting the dynamical state of clusters to their scaling relations.}
   {We implemented a reproducible spectroscopic membership determination and cleaning procedures, based on the redMaPPer membership, running the spectroscopic group finder on the follow-up spectroscopy results and cleaning the membership for spectroscopic outliers. We applied the Anderson-Darling test for velocity substructure and analysed its influence on the scaling relations. We also tested the effect of the X-ray-to-optical centre offset on the scaling relations.}
   {
    We report on the scaling relations between richness, X-ray luminosity, and velocity dispersion for a complete sample of clusters with at least 15 members. Clusters with velocity substructure exhibit enhanced velocity dispersion for a given richness and are characterized by 2.5 times larger scatter. Clusters that have a strong offset in X-ray-to-optical centres have comparable scaling relations as clusters with substructure. We demonstrate that there is a consistency in the parameters of the scaling relations for the low- and high-richness galaxy clusters. Splitting the clusters by redshift, we note a decrease in scatter with redshift in all scaling relations. We localize the redshift range where a high scatter is observed to $z<0.15$, which is in agreement with the literature results on the scatter. We note that the increase in scatter for both high- and low-luminosity clusters is $z<0.15$, suggesting that both cooling and the resulting active galactic nucleus feedback are at the root of this scatter.}
   {}


   \maketitle
%

\section{Introduction}

Galaxy clusters are important cosmological probes, whose abundance and spatial distribution deliver competitive measurements of late-epoch large-scale structure growth (e.g. \citealt{vikhlinin09},  \citealt{2020IderChitham}, \citealt{finoguenov2020}, \citealt{Kirkpatrick2021}, \citealt{lindholm2021}; see \citealt{Clerc_Finoguenov2022} for a review). Currently, cosmological tests that rely on the scaling relations between X-ray luminosity and cluster richness are limited by systematics (e.g. \citealt{lindholm2021}); a better understanding of these relations is required to improve the robustness of cluster cosmological constraints.

Determinations of scaling relations for massive clusters have been possible thanks to targeted cluster studies carried out by  LoCuSS \citep{Smith16, Mulroy2019, Farahi19}, WtG \citep{mantz2016}), CCCP, CFHTLS, COSMOS \citep{Kettula2015}, CODEX \citep{2019Capasso,Kiiveri2021}, XMM-XXL \citep{Lieu16}, and eFEDS  \citep{Chiu22} teams. While the best-studied samples cover tens of clusters, the new samples, including the current study, explore the trends of hundreds of clusters. The aim of studying larger samples is to improve the current cluster models, and one of the benefits of a large sample is the opportunity to study subsamples for consistency. While scaling relations typically take normalization, slope, and scatter into account, the physical effects that include these parameters are mergers (or, generally speaking, a dependence on recent episodes of mass accretion), sloshing, cooling, and active galactic nucleus feedback \citep[see][for a review]{kravtsov_borgani2012}.   

Scaling relations can also be predicted using numerical simulations \citep[e.g.][]{kravtsov_borgani2012}. While predicting the baryonic properties of groups and clusters is hindered by our poor understanding of feedback \citep[for recent progress, see e.g.][]{Eckert21,pop22}, the link between the total cluster mass and velocity dispersion of galaxies has been well established \citep{Carlberg97,Mamon2013}. \citet{Kirkpatrick2021} used the large SPectroscopic IDentification of ERosita Sources (SPIDERS) database \citep{Clerc2016, blanton17} to place cosmological constraints on the modelling of velocity dispersion and its statistical uncertainty.

Scaling relations are conventionally characterized by the logarithmic slope, normalization, and intrinsic scatter, for which a log-normal distribution is assumed.  
In this paper we extend the spectroscopic database of CODEX and apply established methods for substructure detection to investigate the influence of substructure on scaling relations. Our work extends past efforts to understand cluster richness as a mass proxy, pioneered by \cite{rozo11,rozo14,rozo15}, to its current widely used definition.

An important aspect of our study is the large sample size, which includes nearly a thousand clusters, and access to extensive spectroscopy for the member galaxies. By design, the bright ($>0.4L_*$) red-sequence galaxies of each cluster are sampled at a sufficiently high rate, through multiple pointings using SDSS fibre spectroscopy or multi-object slit spectroscopy from the Nordic Optical Telescope (NOT), extending the completeness of SDSS data from $r_{AB}=$17.7 to 19.5 (SDSS) and 21 (NOT) for cluster red-sequence members. As such, our study prefaces the expectations of the DESI survey \citep{desioverview}, which reaches $r=$20.
This bright-member strategy enables the discovery of population trends, such as the link between cluster bulk properties and brightest central galaxy (BCG) properties \citep{2019Erfanianfar, 2021Sohn} as well as constraints on cosmology \citep{2015Bocquet, 2020IderChitham}. Such an observational strategy prevents us from retrieving fine details on the kinematic structure of individual clusters in the same way that dense spectroscopic observations do. Such detailed diagnostics include the anisotropy of galaxy orbits \citep{2013Biviano,2019Mamon}, cluster mass profiles \citep{2013Rines}, and mass accretion rates \citep{2021Pizzardo}, or the galaxy luminosity functions and velocity dispersion functions \citep{2017Sohn}. These observational strategies are, however, not incompatible, as illustrated by stacking analyses of poorly sampled systems which provide access to averaged, finely detailed kinematics measurements \citep[e.g.][]{2019Capasso, 2019Mamon}. Future large-aperture instrumentation, such as DESI \citep{desi} and 4MOST \citep{dj19}, is likely to deliver large numbers of scarcely sampled systems, and analyses of the kind presented in this paper are crucial to preparing for the flow of spectroscopic data analysis from upcoming surveys. Deeper spectroscopic data can be used to spatially identify substructure \citep{Burgett2004}, which is beyond the reach of our dataset.

This paper is structured as follows: In Sect. \ref{data} we describe the data used and present a catalogue of substructure in CODEX clusters. We study the cluster scaling relations in Sect. \ref{results} and present concluding remarks in Sect. \ref{Discussion}. Throughout this paper we use $H_0=70\,\rm km\,s^{-1}\, Mpc^{-1}$ and a flat $\Lambda$ cold dark matter model with $\Omega_{\rm m}=0.3$. Unless stated otherwise, uncertainties are quoted for the 68\% confidence level. The regression analysis performed in this paper uses a natural log (ln) of quantities. Plots show decimal log (log) values for the sake of convenience.

\section{Data}\label{data}

\subsection{Samples of galaxy clusters and red sequences}

In this work we use the data collected by the CODEX survey \citep{finoguenov2020}. The CODEX cluster catalogue is constructed based on the ROSAT All-Sky Survey (RASS) source detection and redMaPPer identification \citep{2014Rykoff}. redMaPPer provides a catalogue of red-sequence cluster member galaxies, which is publicly released as a targeting catalogue of SDSS\footnote{This catalogue is publicly available on the SDSS website at \url{https://www.sdss.org/dr17/data_access/value-added-catalogs/?vac_id=spiders-target-selection-catalogues}.}.

redMaPPer searches for red sequences in photometric and sky positional data using X-ray positions as initial putative galaxy cluster centres. For each red sequence, the algorithm determines each galaxy's probability of being an actual member, $p_{\rm mem}$,  based on its colours, position, and magnitude. The sum of these probabilities across a given red sequence measures the cluster's optical richness, $\lambda$. A correction factor is applied to $\lambda$ to account for survey masks and local survey depth variations in this process. The redshift value that best matches the colour and luminosity distribution of the red sequence, $z_{\lambda}$, is identified with the photometric redshift of the candidate cluster. The position of the BCG in the red sequence defines the optical centre of the cluster. A typical radius is defined as $R_{\lambda} = 1 h^{-1} {\rm Mpc\ } \left(\lambda/100 \right)^{0.2}$ \citep{2014Rykoff}. The circular region around the optical centre of radius $R_{\lambda}$ is the aperture within which the richness, $\lambda$, is calculated. In general, the positions of the X-ray and the optical centres differ, and this offset will be interpreted later in this paper as an indicator of system dynamical states.

\subsection{Observations}

Targeted spectroscopic follow-up of CODEX clusters was performed as a part of SDSS-III \citep{SDSSIII} and continued as a part of the SPIDERS programme of the SDSS-IV survey \citep{blanton17}. To date, spectroscopic redshifts of 66,274 red-sequence galaxies have been measured. The cluster characterization based on these data has been made public through several SDSS releases.
In assembling the data for this work, we used SDSS DR14 and DR16 SDSS \citep{SDSSDR14, SDSSDR16}. The redshift measurement uncertainty for typical cluster galaxies targeted in SDSS is a fraction of $10^{-4}$, corresponding to a velocity uncertainty of 20\,km\,s$^{-1}$ at $z=0.2$, much below the expected velocity dispersion within a galaxy cluster. We, therefore, neglect this source of uncertainty in what follows.

In addition to SDSS spectroscopy, we include here the results of CODEX observations performed at the NOT under programmes 48-025, 51-035, 52-026, and 53-020 (PI A. Finoguenov) and during the NOT observing school of FINCA, which covers five clusters with a richness greater than 50 and a red-sequence redshift greater than 0.3. Due to crowded fields in this range, slit masks are more successful at obtaining enough redshifts to validate the clusters and to measure a velocity dispersion. The slit masks allowed the collection of 15-20 spectra per cluster. Target galaxies were selected similarly to the SPIDERS observations \citep{Clerc2016}. The slit width used was 1.5~arcsec, resulting in a spectral resolving power of 500. The exposure time was 2700~sec per cluster on average over the 400-700 nm waveband. The average seeing over the programmes is around 1~arcsec. Using observations of 144 galaxies by both NOT and SDSS, we determined the velocity error of the NOT slit mask mode to be $77\pm5$ km s$^{-1}$. This error is negligible for the error budget of our high-z velocity dispersion measurement, given our selection of rich clusters with a median velocity dispersion of $\rm 800\,km\,s^{-1}$.

The limitation in target magnitude introduces a redshift-dependent sampling depth at $z>0.3$, as we detail in the next sections. For this work, we only retained systems that have high sampling rates, with $>14$ confirmed spectroscopic members, and we considered the $z>0.3$ part of the sample separately. Our selection of red-sequence galaxies for the dynamical analysis of the cluster is a bonus from the point of view of systematics associated with the galaxy type, as the orbits of early-type galaxies are more isotropic \citep{biviano04}.

\subsection{Cluster membership and sampling rate}

As a first step in the construction of the membership catalogue for this work, we assigned the spectroscopic redshifts to the red-sequence cluster members and removed duplications.

In the second phase, we performed a rejection of outliers. 
Although we intended to refine their approach in this paper, it is useful at this stage to recall the steps followed by \citet{Kirkpatrick2021} in producing a catalogue of 2740 clusters in the 5350\,deg$^2$ area surveyed by SPIDERS.
In their approach, an automated standard iterative 3$\sigma$ clipping routine flags out galaxies ostensibly deviating from the bulk of the redshift distribution of red-sequence members in a cluster. Fed with this information, a series of collaborative visual inspections adjusted the membership for about one-third of the entire sample, yielding the measurements of the cluster redshift and velocity dispersion published in \cite{Kirkpatrick2021}.

\begin{figure}[ht]
    \centering
    \includegraphics[width=\hsize]{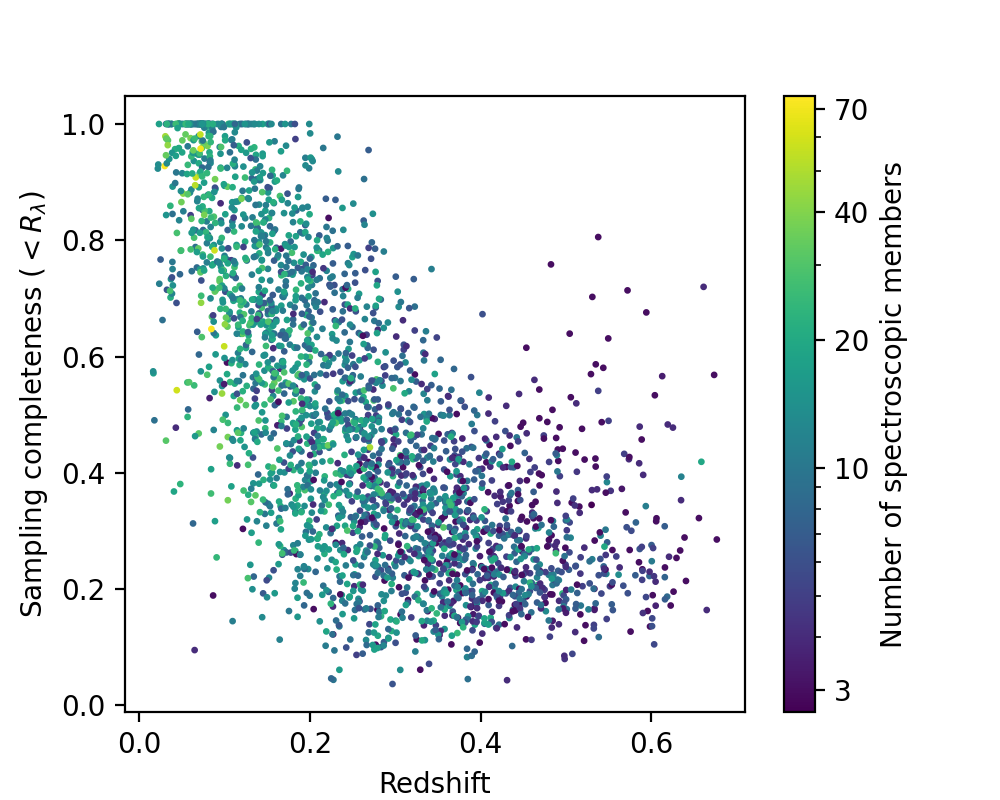}
    \caption{Spectroscopic sampling rate of the 2740 clusters in \citet{Kirkpatrick2021} as a function of cluster redshift, colour-coded by the number of spectroscopic members. The completeness, $S$ (Eq.~\ref{eq:samplingrate}) is estimated relative to photometrically identified red-sequence members weighted by their membership probability. The latter is updated with the membership flag retrieved from spectroscopic observations and described in \citet{Kirkpatrick2021}. These estimates are performed within a distance $R_{\lambda}$ from the X-ray centre of each cluster (see the main text for details).}
    \label{fig:completeness_rate}
\end{figure}

In Fig.~\ref{fig:completeness_rate} we show the targeting efficiency based on the catalogue of 2740 clusters presented in \citet{Kirkpatrick2021}. We use colour to highlight the trends associated with a large number of members. Spectroscopic completeness, $S,$ is estimated within an aperture radius, $R_{\lambda}$, from the X-ray centre of each cluster. This sampling rate was computed relative to the sum of the photometric membership probabilities estimated by redMaPPer ($0.05 \leq p_{\rm mem} \leq 1$), those probabilities being updated to $p_{i}=0$ or 1 with knowledge of actual spectroscopic membership, based on visual inspection. Membership probabilities of red-sequence galaxies without a spectroscopic redshift were left unchanged (i.e.~$p_{i} = p_{\rm mem}$). In other terms, with $N_{{\rm mem},z}$ the number of red-sequence members spectroscopically identified as cluster members, we defined the spectroscopic completeness, $0\leq S \leq 1$, as
\begin{equation}
    \label{eq:samplingrate}
    S = \frac{N_{{\rm mem},z}}{\sum_i p_{i}} = 
    \left(1+ 
    \frac{\sum_{i,p_{i}<1} p_{i}}{N_{{\rm mem},z}}
    \right)^{-1} 
   \ ,
\end{equation}where the second sum sign runs over all red-sequence members not associated with a spectroscopic redshift measurement.

Such a definition disregards situations where the membership assignment algorithm splits a red sequence into several components; or where the algorithm merges multiple red sequences into a single entity.
Due to the limitations on the fibre magnitude of targets, there is a trend of decreasing target sampling rate with redshift.

For the present study, we extended the visual inspection procedure of \citet{Kirkpatrick2021} to the 5032\,deg$^2$ area of CODEX not observed by SPIDERS, thereby producing membership flags, cluster redshifts and velocity dispersions for $\lambda>40$ CODEX clusters over the whole 10382\,deg$^2$ survey area of CODEX survey \citep{finoguenov2020}. The catalogue published along with this paper makes these values available.

The procedure described above conspicuously involves human decisions. We wanted to unify the cluster membership selection for those clusters and introduce fully automated methods for membership determination. For that, we used two reproducible approaches. The first approach, coined SPIDERS, relies on the velocity dispersion and median spectroscopic redshift values obtained after visual inspection. To make the cluster membership reproducible, we modified the {\it Clean} routine of \cite{Mamon2013}, replacing the initial evaluation of velocity dispersion with a SPIDERS measurement. We then performed the membership cleaning starting from a full list of redMaPPer galaxies with spectra and using prescriptions of {\it Clean}, which rely on $2.7\sigma_v(R)$ and $R<R_{200c}$ rejection.\ Here, $\sigma_v(R)$ is the line-of-sight velocity dispersion profile predicted for a cluster following the \cite{NFW} model of the mass expected for the supplied SPIDERS value of $\sigma_{v, \rm gap}$ (gapper estimate of velocity dispersion) and the concentration expected from its mass and from performing an aperture correction. After that we proceed with {\it Clean}'s iterative reevaluating the $\sigma_v$, using the remaining members. We call this process reproducible because it allows one to obtain the member catalogue based on the published velocity dispersion, redshift and red-sequence member catalogues.

The main approach to cluster membership determination adopted in this paper uses an application of the self-calibrated group finder (SCGF\footnote{https://www.galaxygroupfinder.net/}) of \cite{Tinker_2021,Tinker_2022} on the spectroscopic data of CODEX red-sequence members. In the SCGF code, satellite galaxies are attached to the groups using the nearest neighbours algorithm, with the brightest galaxy as the group centre. This group finder implementation has a free parameter, $B_{\rm sat}$, that is used to set the threshold probability for a galaxy to be a satellite. It is set to a default value of $B_{\rm sat} = 10$. The code is run with the following parameters: $z_{\rm min}$, $z_{\rm max}$, and a fraction of the sky. For this run, $z_{\rm min} = 0$, $z_{\rm max} = 6.8$, and the fraction of the sky is 0.25. We ignored the resulting mass estimates of SCGF as they require a volume-limited sample of galaxies. We supplied r-band magnitudes for the selection of the brightest galaxy.

\begin{figure}[ht]
    \centering
    \includegraphics[width=\hsize]{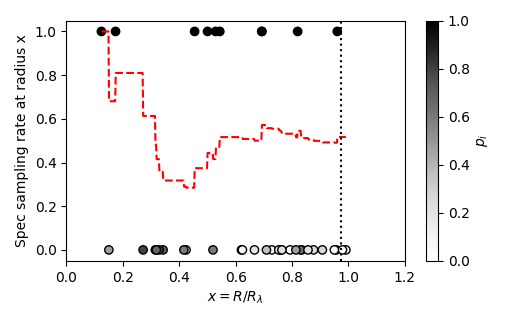}
    \caption{Radial spectroscopic coverage of one confirmed cluster. Each filled circle represents one red-sequence galaxy with $p_{\rm mem} >0.05$. The $x$-axis is their projected distance to the X-ray centre, normalized by $R_{\lambda}$. Spectroscopically identified members are displayed at $y=1$ and coloured in black ($p_i=1$), and interlopers are displayed at $y=0$ and coloured in white ($p_i=0$). Galaxies not assigned a spectroscopic redshift are coloured according to their $p_{\rm mem}$ value ($p_i=p_{\rm mem}$). The dashed red line shows the completeness rate, $S(<x)$ (Eq.~\ref{eq:samplingrate}) within radius $x$.
    The dotted line indicates the virial radius estimated from an X-ray luminosity--mass relation. }
     \label{fig:completeness_rate_id}
\end{figure}

The SCGF code performs a reproducible separation of the projected components, which was manually done for the SPIDERS. It also performs cleaning of spatially isolated galaxies, which are potentially more strongly contaminated by outliers. Application of SCGF to the red galaxies, as opposed to using it on all galaxies, is a standard procedure, known to provide higher sensitivity towards more massive halos \citep[e.g.][]{Tinker_2021}. The result of the SCGF run is a catalogue of 5024  spectroscopic clusters that have at least three members in an area of 10,382 square degrees, with a subset of 3356 clusters having five or more member galaxies, which compares well to our previous release of 2740 clusters in \cite{Kirkpatrick2021} in an area of 5350 square degrees. Using the SCGF member catalogues for clusters with more than four members, we performed the rejection of outliers using {\it Clean} in its standard mode of estimating the first guess of velocity dispersion for each cluster using all its SCGF members. The redshift linking length of the SCGF imposes similar cluster member acquisition criteria as the standard practice adopted in dynamical measurements \cite[e.g.][]{lopes09}.
In Fig. \ref{fig:completeness_rate_id} we show an example illustrating the radial coverage of the SPIDERS programme, with membership derived from the above-described procedure.

\begin{figure}[ht]
    \centering
    \includegraphics[width=\hsize]{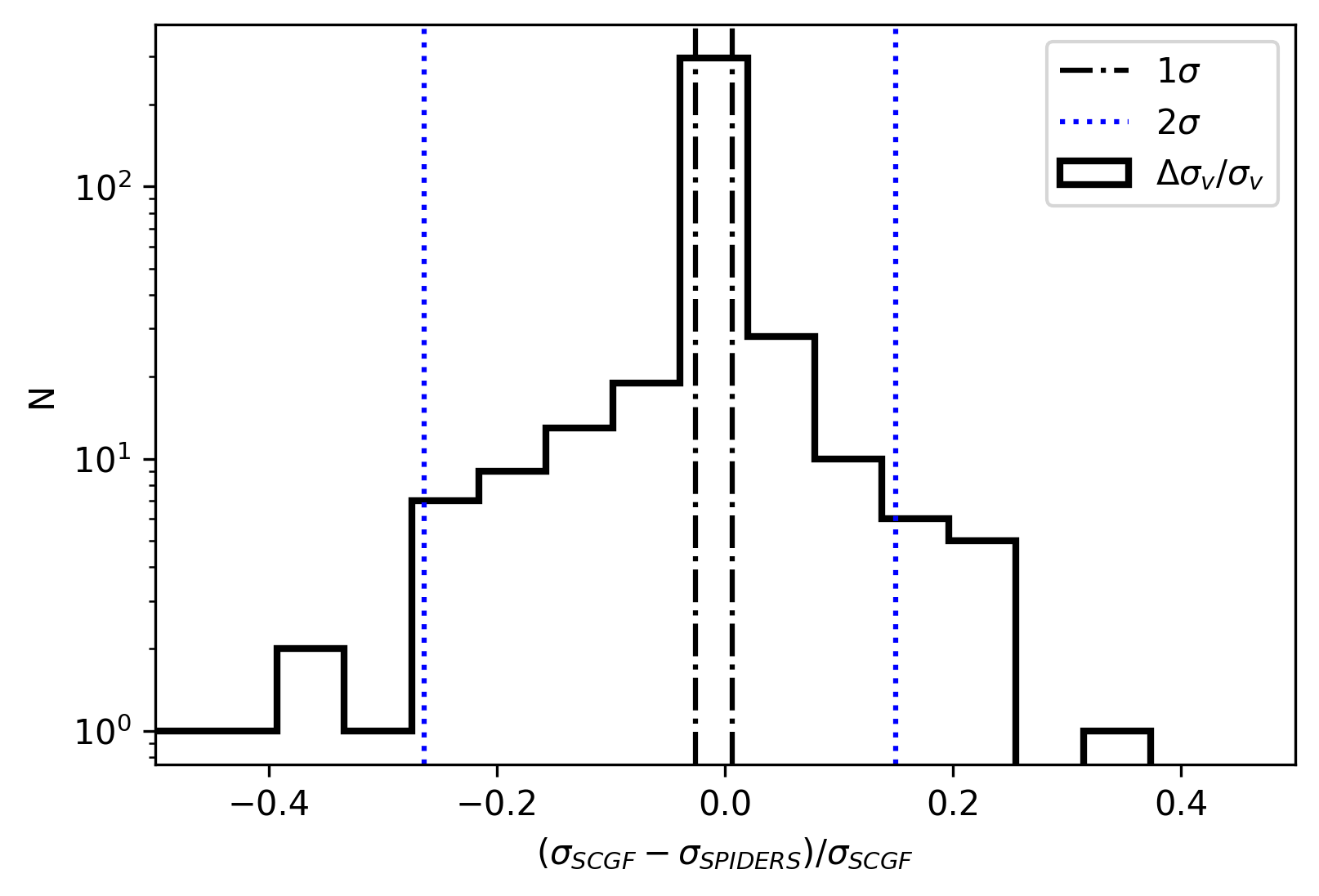}
    \caption{Differences in velocity dispersions for the clusters present in both the SCGF and SPIDERS samples with a minimum of 15 member galaxies. The total number of clusters in bins of velocity dispersion difference is plotted. The 68 (95)\% of the population with the lowest absolute deviation is marked with dash-dotted black  (dotted blue) lines.}
    \label{fig:difference in vdisp histogram} 
\end{figure}

Application of SCGF also enables spectroscopic cluster identification at low richness  (around 20), which was partially left out in the manual screening \cite{Kirkpatrick2021}. After the screening, we retained a catalogue of 2240 clusters with five or more cleaned member galaxies. This catalogue is further used in computing the dynamical properties of the clusters. In addition, SCGF also yields a measurement of the spectroscopic optical cluster centre. Applying the group finder algorithm to clean the cluster membership before performing the dynamical analysis has been previously done by \cite{serra11} and a combination of group finders with and without applying {\it Clean} has been included in the tests of recovering the velocity dispersion of  \cite{Old15}. Comparison to the SPIDERS estimates of velocity dispersion (calculated for this paper) for the clusters in common (Fig. \ref{fig:difference in vdisp histogram}) shows an agreement within 5\% for 90\% of the sample, and the rest having a 20\% scatter.

\begin{table*}
\caption{Fraction and number of Gaussian and non-Gaussian clusters in different redshift bins. Rows 1-2 and 5-6 show the number of Gaussian or non-Gaussian clusters per redshift bin divided by the total number of Gaussian or non-Gaussian clusters in the SPIDERS or SCGF sample. Rows 3-4 and 7-8 show the number of Gaussian or non-Gaussian clusters divided by the total number of clusters in the corresponding redshift bin.}
\label{table: G NG redshift bins}      
\centering          
\begin{tabular}{ l l l l l l}
\hline\hline       
Sample catalogue & Fraction & 
\multicolumn{1}{c}{$z<0.1$}  & 
\multicolumn{1}{c}{$0.1 \leq z<0.2$} &
\multicolumn{1}{c}{$0.2 \leq z<0.3$} &
\multicolumn{1}{c}{$z \geq 0.3$} \\
\hline
SCGF &  $G_{\mathrm {bin}}/G_{\mathrm {total}}$ & 0.41 [176/432]  & 0.34 [146/432] & 0.16 [71/432] & 0.09 [39/432] \\
SCGF &  $NG_{\mathrm {bin}}/NG_{\mathrm {total}}$ & 0.44 [43/98] & 0.35 [34/98] & 0.15 [15/98] & 0.06 [6/98] \\
SCGF &  $G_{\mathrm {bin}}/(G_{\mathrm {bin}}+NG_{\mathrm {bin}})$ & 0.80 [176/219] & 0.81 [146/180] & 0.83 [71/86] & 0.87 [39/45] \\
SCGF &  $NG_{\mathrm {bin}}/(G_{\mathrm {bin}}+NG_{\mathrm {bin}})$ & 0.20 [43/219] & 0.19 [34/180] & 0.17 [15/86] & 0.13 [6/45] \\
SPIDERS & $G_{\mathrm {bin}}/G_{\mathrm {total}}$ & 0.33 [129/391] &0.36 [141/391] & 0.18 [69/391] &  0.13 [52/391] \\
SPIDERS &  $NG_{\mathrm {bin}}/NG_{\mathrm {total}}$ & 0.35 [32/92]  & 0.34 [31/92] & 0.18 [17/92] & 0.13 [12/92] \\
SPIDERS &  $G_{\mathrm {bin}}/(G_{\mathrm {bin}}+NG_{\mathrm {bin}})$ & 0.80 [129/173] & 0.82 [141/160] & 0.80 [105/69] & 0.81 [52/64] \\
SPIDERS &  $NG_{\mathrm {bin}}/(G_{\mathrm {bin}}+NG_{\mathrm {bin}})$ & 0.20 [32/173] & 0.18 [31/160] & 0.20 [17/86] & 0.19 [12/64] \\
\hline                  
\end{tabular}

\end{table*}

\begin{figure}[ht]
    \centering
    \includegraphics[width=\hsize]{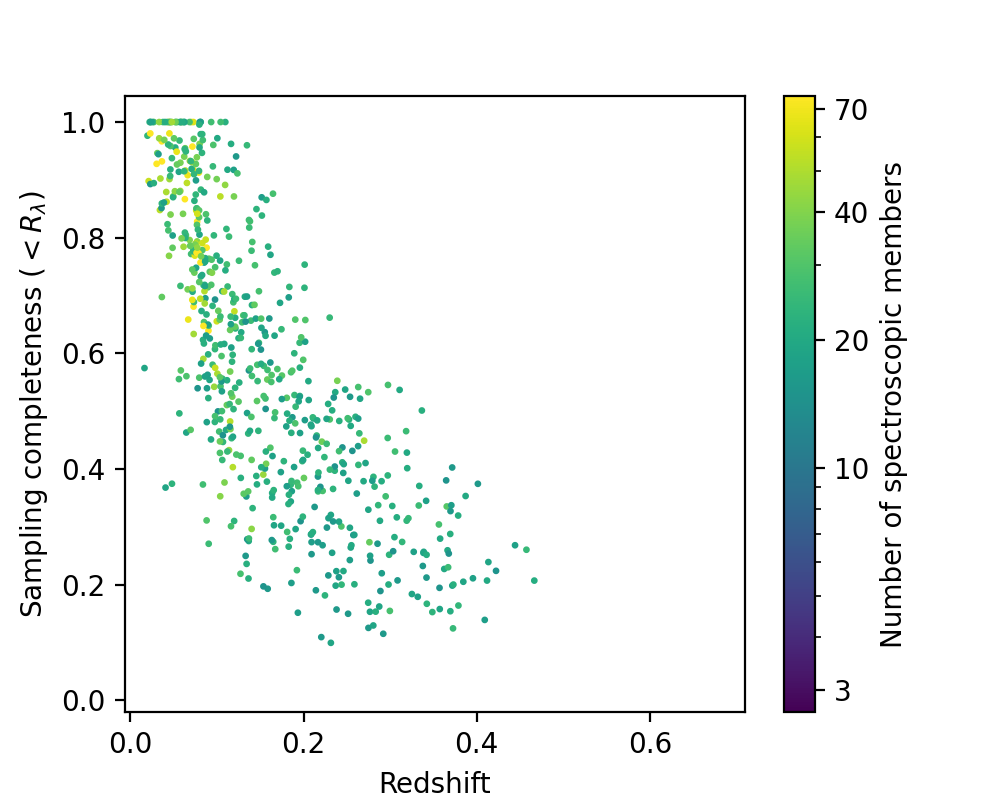}
    \caption{Same as Fig.~\ref{fig:completeness_rate}, but for the main sample of this paper, which consists of 640 clusters and memberships assigned by the SCGF algorithm and the final number of clean members above 14. The calculations presented in this figure were done before applying \emph{Clean}.}
    \label{fig:completeness_rate_fof}
\end{figure}

For the scaling relation part of this study, we further retained only the clusters with at least 15 clean members, where the scatter on $\sigma_v$ becomes less dominant  \citep{Saro13}, and where different velocity dispersion estimators deliver the same result \citep{Kirkpatrick2021}.  As {\it Clean} performs a slight adjustment of the redshift of the cluster, we recomputed the cluster $L_X$; however, in 99\% of cases, the redshifts are within $1.5\times 10^{-3}$ from the values published in \cite{Kirkpatrick2021}. The spectroscopic sampling rate based on SCGF membership assignment (before applying {\it Clean}) is shown in Fig.~\ref{fig:completeness_rate_fof}; the trend behaves similarly as the entire SPIDERS visually inspected sample (Fig.~\ref{fig:completeness_rate}). 
The number of clusters remaining in the SPIDERS and SCGF samples are then 611 and 640, respectively, and 402 clusters are present in both samples. All these clusters have only a single SCGF group component per CODEX cluster remaining after applying the cuts on the number of members. The median number of clean members in the sample is 20 and 95\% of the clusters have fewer than 40 members, without any strong richness trend. This makes this sample fairly uniform in terms of the quality of cluster dynamics data. The spectroscopic sampling improves only for rich clusters at $z<0.1$.
The two catalogues reveal differences in the membership as well as the values of $\sigma_v$.  Given that SCGF is a fully reproducible methods and is easily extendable to other datasets, we selected it as a primary, using the SPIDERS catalogue to verify the robustness of our conclusions.

To study the scaling relations, we further limited the sample by requiring completeness better than 50\%, imposing a selection on richness ($\lambda$) as described in \cite{finoguenov2020}:
\begin{equation}\label{richcut}
    \lambda>33.5\left(\frac{z}{0.15}\right)^{0.8}.
\end{equation}
This introduces a $\sim20$\% reduction in the sample size, leaving 483 clusters in the SPIDERS sample, 530 clusters in the SCGF sample, and 402 clusters in common between the samples. Table~\ref{table: G NG redshift bins} presents the cluster statistics details.

\subsection{Substructure}
To identify the substructure in the cluster catalogues, we applied the Anderson-Darling (AD) test following the procedure of \cite{Hou2009} to the cleaned cluster members. The AD test performs a goodness-of-fit test based on ordered data. We use a cumulative distribution function of a Gaussian distribution as the underlying hypothetical distribution:

\begin{equation}\label{eq:Phi}
    \Phi(x_i)=\frac{1}{2}\left[ 1-\mathrm{erf}\left(\frac{x_i-\mu}{\sqrt{2}\,\sigma_v}\right)\right] ,
\end{equation}
\noindent where $x_i \leq x < x_{N+1-i}$, $\mu$ is the mean and $\sigma_v$ is the velocity dispersion, calculated as a second moment of the velocity distribution \citep{beers90AJ}. We then calculated the test statistic $A^2$ using
\begin{equation}\label{eq:A2}
    A^2 = -N-\frac{1}{N}\sum (2i-1)\, \left\{\ln\Phi\left(x_i\right)+\ln\left[1-\Phi\left(x_{N+1-i}\right)\right] 
    \right\}
\end{equation}
and corrected it for the size of the sample and the requirement to measure both the mean and dispersion of the distribution:
\begin{equation}\label{eq:A2star}
    A^{2*} = A^2 \left(1+\frac{0.75}{N}+\frac{2.25}{N^2}\right) \ .
\end{equation}
\noindent The significance level for conforming to Gaussianity $\alpha_{AD}$, is calculated as
\begin{equation}\label{eq:alpha}
    \alpha_{AD} = a \exp\left(-\frac{{A^{2*}}}{b}\right)\ ,
\end{equation}
\noindent where $a=3.6789468$ and $b=0.1749916$ are fit parameters from \cite{nelson1998}.

\begin{figure}[ht]
    \centering
    \includegraphics[width=\hsize]{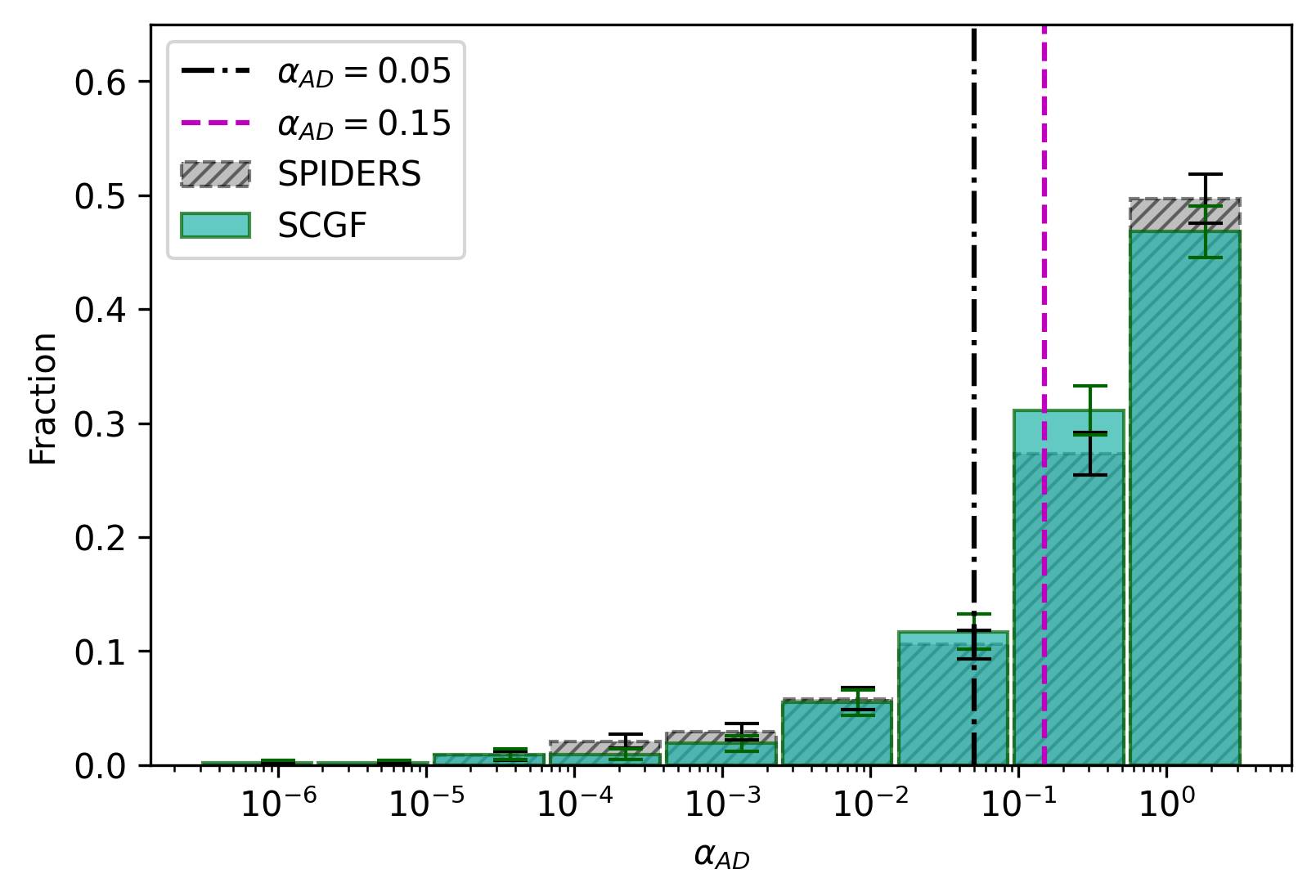}
    \caption{Distribution of clusters over the AD test $\alpha_{AD}$ values for the SPIDERS and SCGF samples. The SPIDERS sample is marked with grey and dashed edges and SCGF with light blue and solid edges. The threshold limit for structure at $\alpha_{AD}=0.05$ is marked with a dash-dotted black line, and the threshold for possible substructure $\alpha_{AD}=0.15$ is marked with a dashed magenta line. The heights of the bars are calculated as the number of clusters per bin divided by the total number of clusters in the sample.}
    \label{fig:hist alpha all}
\end{figure}

We set the limit of substructure detection to $\alpha_{AD} < 0.05$, which implies a $>95$\% confidence on substructure detection. This limit has been previously adopted by \citet{Hou13} on studies of galaxy groups, and we separately test a slightly broader range of substructure susceptibility by checking the clusters with the substructure detection in the $85-95$\% confidence level range. The clusters without substructure are hereinafter called Gaussian clusters, and clusters with substructure  are called non-Gaussian clusters.

\begin{figure}[ht]
    \centering
    \includegraphics[width=\hsize]{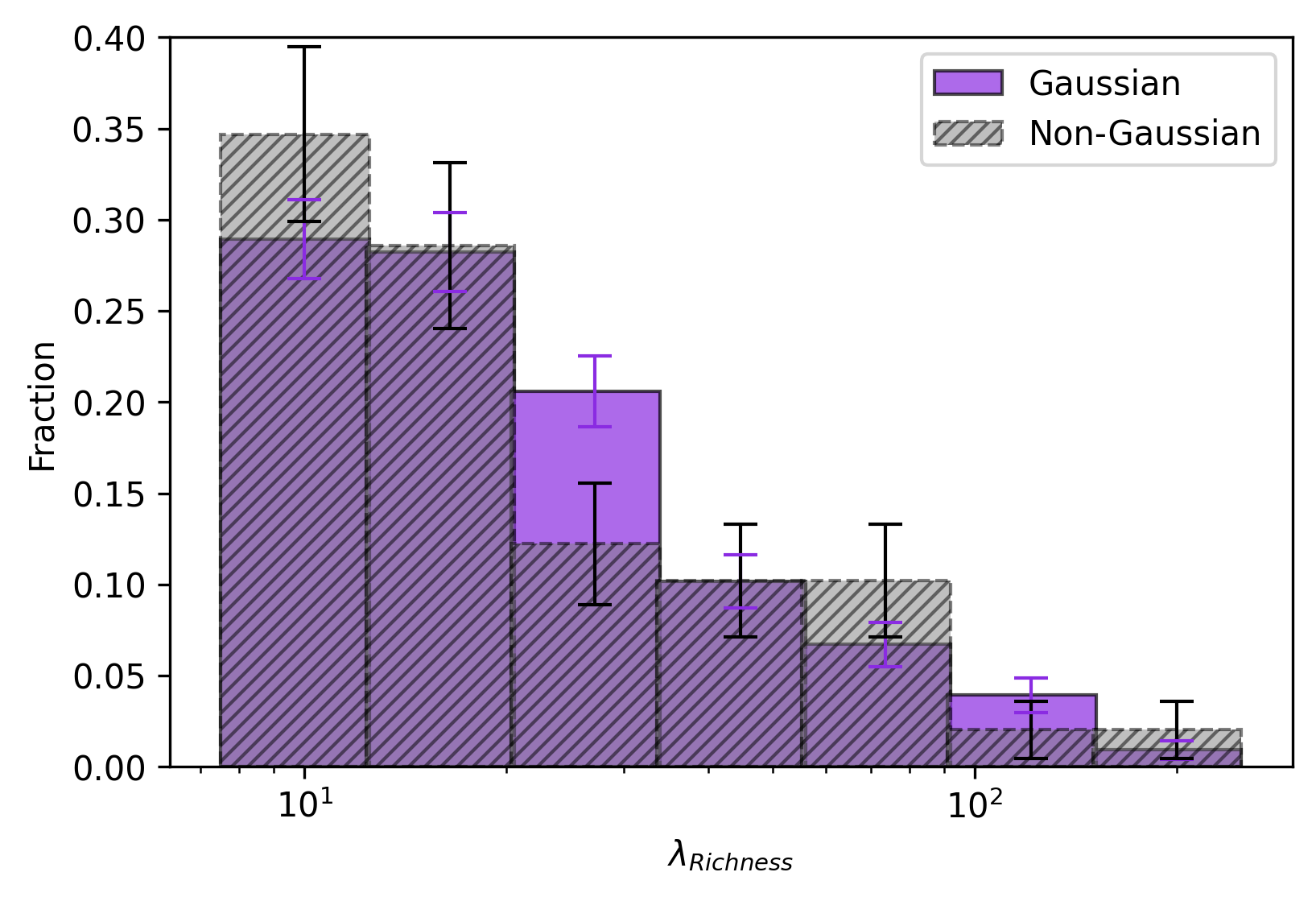}
    \caption{Normalized fractions of Gaussian and non-Gaussian clusters at different richnesses for the SCGF sample. Gaussian clusters are marked with purple and non-Gaussian clusters with hatched grey.}
    \label{fig:hist fraction of G NG richness SCGF}
\end{figure}

\begin{figure}[ht]
    \centering
    \includegraphics[width=\hsize]{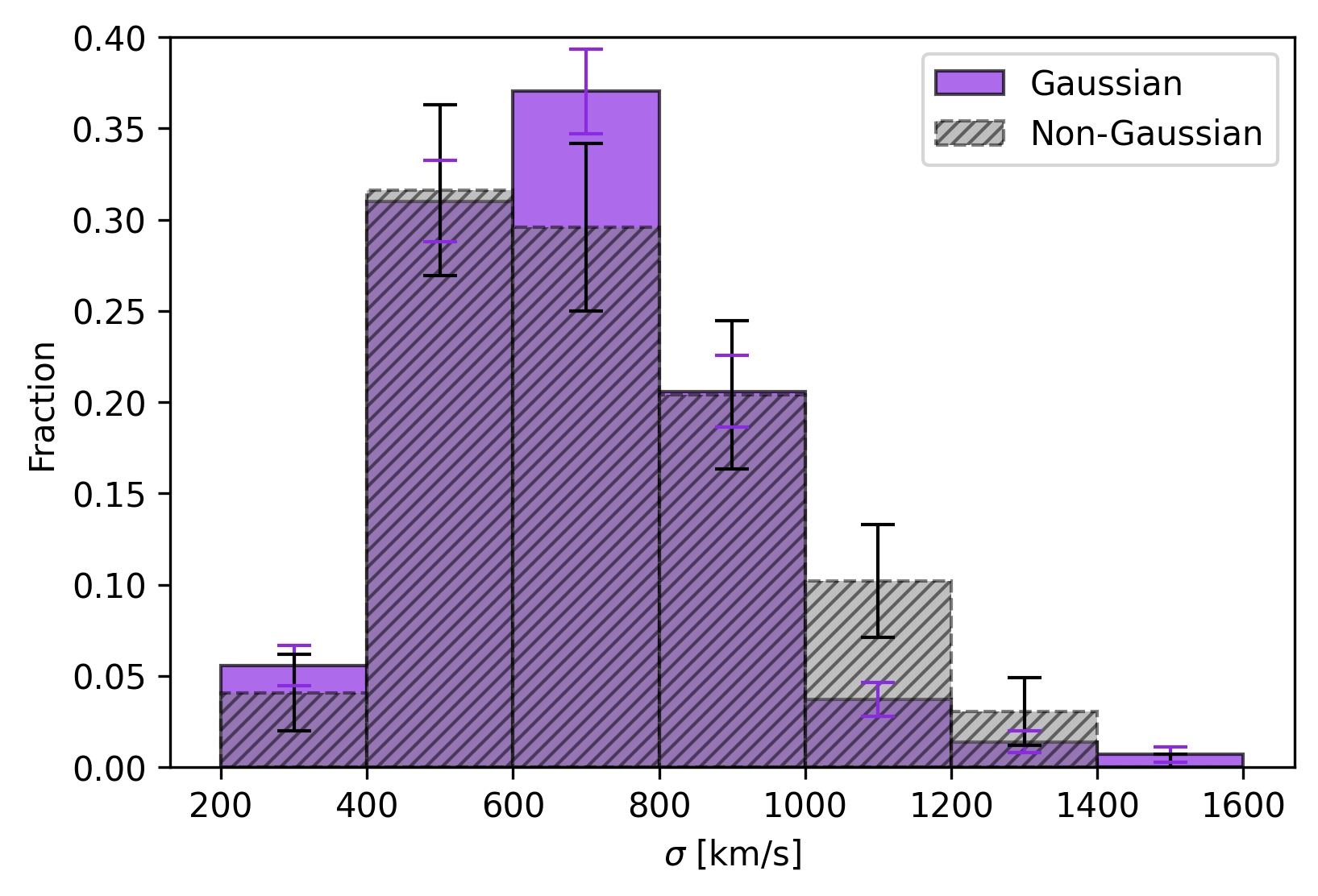}
    \caption{Normalized fractions of Gaussian and non-Gaussian clusters at different velocity dispersions for the SCGF sample. Gaussian clusters are marked with purple and non-Gaussian clusters with hatched grey.}
    \label{fig:hist fraction of G NG vdisp SCGF}
\end{figure}

Of the total 483 clusters in the SPIDERS sample, we found 391 Gaussian and 92 non-Gaussian clusters. For the SCGF sample of 530 clusters, we found 432 Gaussian and 98 non-Gaussian clusters. Of the 402 clusters shared between the two samples, only 24 ($\sim$6\%) were found to have a non-matching Gaussianity flag. This shows that regardless of our choice of membership assignment, the samples agree well on the presence of substructure. Figure \ref{fig:hist alpha all} shows the distribution of clusters in both samples over the values of $\alpha_{AD}$. We show two boundaries used in the subsequent analysis, placed at 95\% and 85\% confidence levels for substructure detection ($1-\alpha_{AD}$). The fractions (heights of bars) and the errors are calculated using Eq. \ref{eq:bimodial error}.

To find a difference between the two samples, we employed two approaches. One is the well-known Kolmogorov-Smirnov (KS) test, and the other one is shuffling. The drawback with the KS test is (1) that it tests for differences in the distributions, which could be location, scale, or other features, while we are only interested in location; and (2) the $p$-values of the KS test are only well known for Gaussian distributions. Shuffling belongs to the broad category of permutation tests, and here the sample sizes require Monte Carlo sampling. The foundation of the method is laid out by \citet{Dwass57}. We use it to test whether a difference in the location (and sometimes in the scale) of the sample can be reproduced by chance. We will be quoting probability values (p-values) of two comparison samples being the same.
No difference in the medians of the $\alpha_{AD}$ distributions between SPIDERS and SCGF samples has been found using a shuffling test (p-value of 0.76) or a KS test (p-value of 0.46).

\begin{figure}[ht]
    \centering
    \includegraphics[width=\hsize]{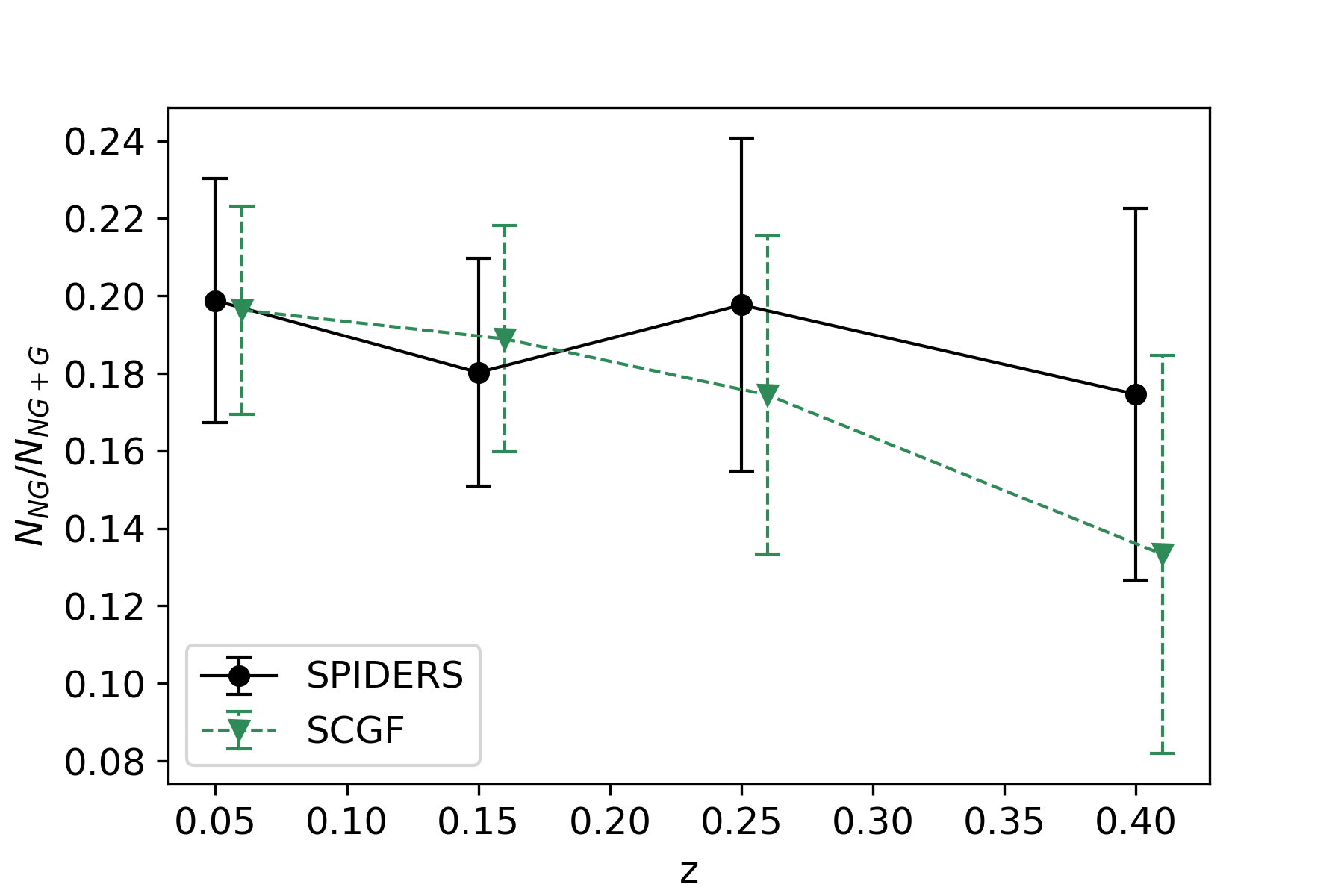}
    \caption{Fraction of non-Gaussian clusters per redshift bin for the SPIDERS and SCGF samples with 68\% binomial confidence intervals. Redshift bins are $z= [0.0, 0.1, 0.2, 0.3, 0.7]$. The SPIDERS sample is plotted using solid black lines and filled circles. The SCGF sample is shown using dashed sea-green lines and filled triangles.
    }
    \label{fig:fractions}

\end{figure}

\begin{figure}[ht]
    \centering
    \includegraphics[width=\hsize]{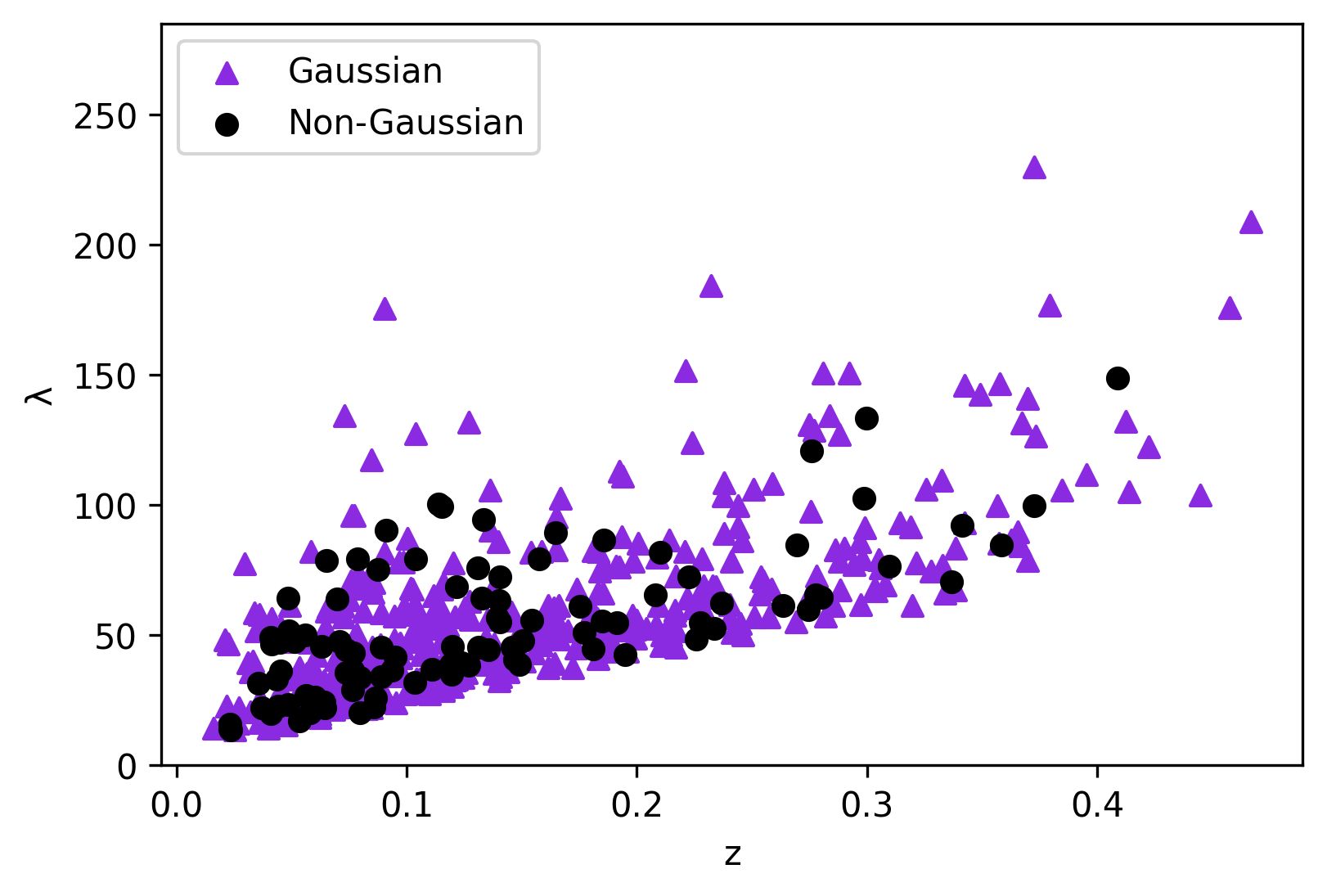}
    \caption{Redshift--richness plane of the SCGF cluster sample, marking the Gaussianity of the clusters, after applying our selection criteria identified by Eq. \ref{richcut}. Gaussian clusters are marked with purple triangles and non-Gaussian clusters with black circles. 
}
    \label{fig:scatter z richness SCGF}
\end{figure}

Figure \ref{fig:hist fraction of G NG richness SCGF} shows the richness distribution for Gaussian and non-Gaussian clusters, revealing no trend in the fraction of non-Gaussian clusters. No difference in the medians of the richness distributions between Gaussian and non-Gaussian samples has been found using a shuffling test (p-value of 0.8307) or a KS test (p-value of 0.91).
Figure \ref{fig:hist fraction of G NG vdisp SCGF} shows the velocity dispersion distributions for Gaussian and non-Gaussian clusters, which shows a marginally higher (97\% confidence) median value as well as a width of distribution for non-Gaussian clusters using a shuffling test. (The KS test p-value between the samples is 0.12. The shuffling test p-values for medians and standard deviations are 0.03 and 0.03.)
Figure \ref{fig:fractions} studies a redshift dependence of the fraction of non-Gaussian clusters, also in comparison to the SPIDERS sample, with no significant trend found, and Fig. \ref{fig:scatter z richness SCGF} shows the redshift-richness plane of the SCGF catalogue. 

In calculating the fractions $p$ (of $x$ successes having a total of $n$ trials) we used the Agresti-Coull method \citep{ac98} to compute the 68\% confidence level interval ($z_{\alpha/2}=1$) using
\begin{equation}\label{eq:bimodial error}
\widetilde{p}\pm z_{\alpha/2}\sqrt{\dfrac{\widetilde{p}(1-\widetilde{p})}{\widetilde{n}}}, \text{where } \widetilde{p}=\dfrac{x+{z_{\alpha/2}^2}/{2}}{n+z_{\alpha/2}^2} \text{and } \widetilde{n}=n+z_{\alpha/2}^2
\end{equation}

The fraction of non-Gaussian clusters decreases marginally with redshift. A detailed tabulation of the split by Gaussianity is given in Table~\ref{table: G NG redshift bins}. 

\begin{figure}[ht]
    \centering
    \includegraphics[width=\hsize]{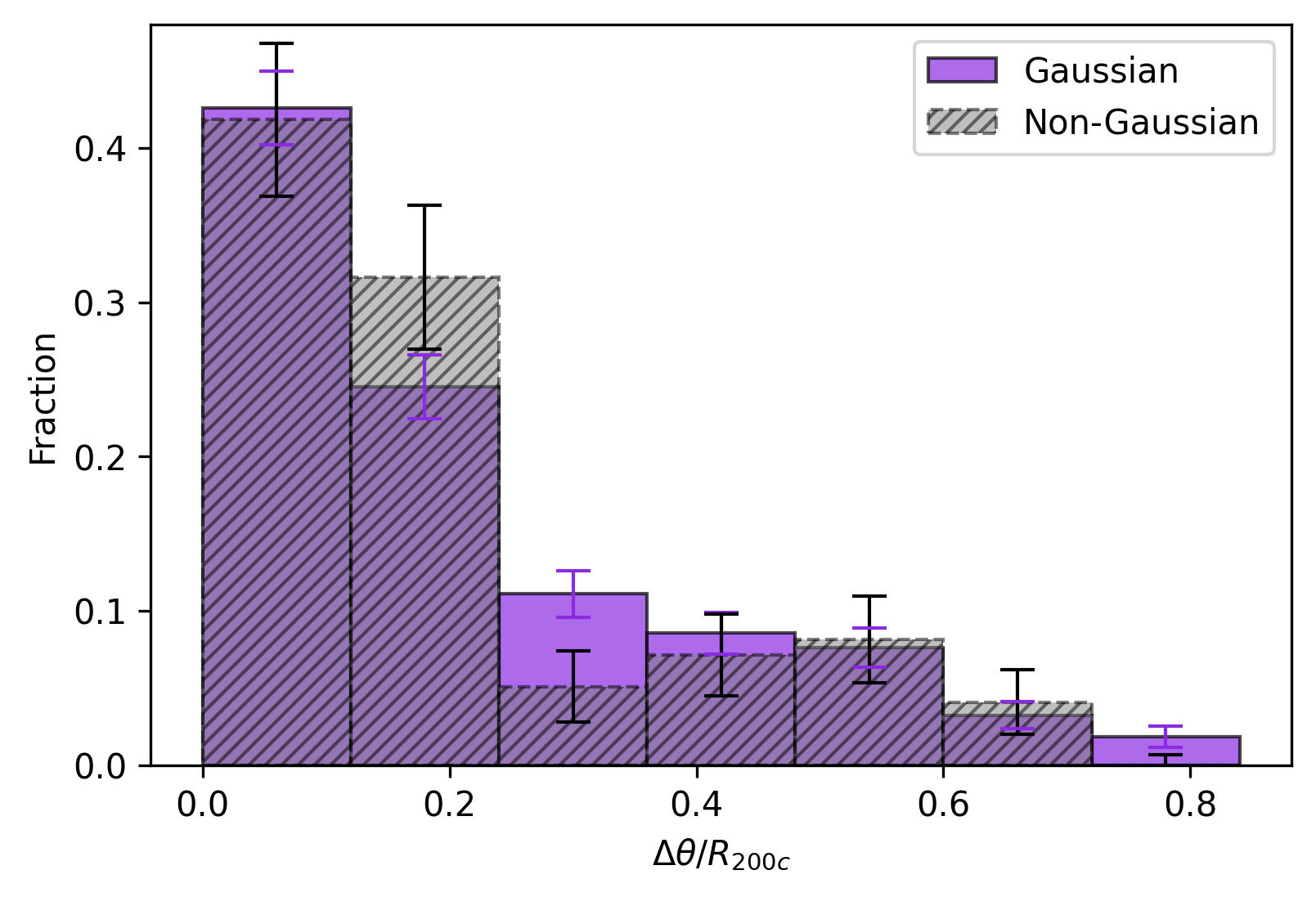}
    \caption{Normalized fractions for the angular separation of SCGF optical and X-ray centres as a fraction of the virial radius of the clusters. Gaussian clusters are marked with purple and non-Gaussian clusters with hatched grey.}
    \label{fig:center separation dist SCGF}
\end{figure}

An additional statistic potentially sensitive to the substructure is an offset between the optical and X-ray centres. SPIDERS optical centre uses the redMaPPer's centre on the brightest red-sequence galaxy, while SCGF reevaluates this based on the available spectroscopy. X-ray centres are taken from the CODEX catalogue of \citet{finoguenov2020}. During spectroscopic campaigns, the 3 brightest red-sequence members received priority targeting, so the BCG spectroscopic validation is highly complete \citep{2019Erfanianfar}, making SCGF centre robust against spectroscopic sampling. In Figs. \ref{fig:center separation dist SPIDERS} and \ref{fig:center separation dist SCGF} we present the distribution of the optical-to-X-ray offset for redMaPPer and SCGF optical cluster centres, illustrating the effect of the substructure. On average, SCGF offsets are smaller, which we attribute to the rejection of outliers. No difference in the offset median value has been found between Gaussian and non-Gaussian samples using a shuffling test (p-value of 0.50) as well as the KS test (p-value of 0.70). The statistical uncertainty on RASS centring is $<0.07R_{200c}$ for 68\% of the sample and $<0.17R_{200c}$ for 99\% of the sample. 
The distribution of the offsets appears bimodal for clusters with substructure, separated at about 0.4$R_{200c}$,  albeit at a low confidence level, based on the bimodality test. The observed separation is due to the difference between the main component identified as being more X-ray luminous versus having a more massive galaxy. The presence of large separations is not limited to non-Gaussian clusters. We consider Gaussian clusters with large offsets as candidates for merging clusters and perform a separate analysis of this class of objects. 

\section{Results}\label{results}

\begin{table*}
\caption{Summary of regression analysis for velocity dispersion  obtained using the gapper method vs the richness scaling relation $\ln(\sigma_v\; km^{-1} s)=\alpha + \beta \ln(\lambda E_z / 47.2)+N(0,\sigma)$ for the SCGF sample.}          
\label{table:linmix parameters vdisp}      
\centering          
\begin{tabular}{ l c c c c}
\hline\hline       
Sample & Intercept $\alpha$ & Slope $\beta$ & Intrinsic scatter $\sigma$ & N clusters\\
\hline
Full & $6.461 \pm{0.016}$ & $0.365\pm{0.029}$ & $0.138\pm{0.017}$ & 530\\
Gaussian & $6.446\pm{0.017}$ & $0.370\pm{0.030}$ & $0.095\pm{0.021}$ & 432\\
Non-Gaussian  & $6.524\pm{0.047}$ & $0.351\pm{0.089}$ & $0.231\pm{0.040}$ & 98 \\
Full $\lambda \geq$ 47.2 & $6.451\pm{0.037}$ & $0.381\pm{0.062}$ & $0.137\pm{0.023}$ & 298\\
Full $\lambda <$ 47.2 & $6.460\pm{0.040}$ & $0.358\pm{0.090}$ & $0.143\pm{0.028}$ & 232\\
Gaussian $\lambda \geq$ 47.2 & $6.438\pm{0.034}$ & $0.378\pm{0.054}$ & $0.081\pm{0.029}$  & 245\\
Gaussian $\lambda <$ 47.2 & $6.460\pm{0.043}$ & $0.405\pm{0.097}$ & $0.118\pm{0.034}$ & 187\\
Non-Gaussian $\lambda \geq$ 47.2 & $6.491\pm{0.125}$ & $0.426\pm{0.220}$ & $0.246\pm{0.058}$ & 53\\
Non-Gaussian $\lambda <$ 47.2 & $6.481\pm{0.097}$ & $0.243\pm{0.210}$ & $0.221\pm{0.065}$& 45\\
$0.05 < \alpha_{AD} \leq 0.15$ & $6.456\pm{0.029}$ & $0.307\pm{0.051}$ & $0.132\pm{0.034}$& 73\\
$\alpha_{AD} > 0.15$ & $6.442\pm{0.012}$ & $0.383\pm{0.022}$ & $0.087\pm{0.017}$& 359\\
$\Delta \theta/R_{200c} \geq 0.3$ & $6.497\pm{0.022}$ & $0.294\pm{0.046}$ & $0.153\pm{0.023}$ & 131\\
$\Delta \theta/R_{200c} < 0.3$ & $6.447\pm{0.012}$ & $0.384\pm{0.022}$ & $0.131\pm{0.014}$& 399\\
$z \geq 0.3$ & $6.560\pm{0.162}$ & $0.238\pm{0.163}$ & $0.077\pm{0.062}$& 45 \\
$0.15 \leq z < 0.3$ & $6.450\pm{0.045}$ & $0.394\pm{0.080}$ & $0.067\pm{0.045}$& 136\\
$z < 0.15$ & $6.460\pm{0.021}$ & $0.372\pm{0.045}$ & $0.164\pm{0.020}$& 329\\
Gaussian $z \geq 0.3$ & $6.567\pm{0.176}$ & $0.235\pm{0.177}$ & $0.089\pm{0.072}$ & 39\\
Gaussian $0.15 \leq z < 0.3$ & $6.456\pm{0.048}$ & $0.378\pm{0.093}$ & $0.064\pm{0.050}$& 130\\
Gaussian $z < 0.15$ & $6.439\pm{0.021}$ & $0.367\pm{0.045}$ & $0.111\pm{0.027}$& 263\\
\hline                  
\end{tabular}
\end{table*}

\begin{table*}
\caption{Summary of regression analysis for X-ray luminosity vs the richness scaling relation $\ln(L_X E^{-1}_z\; \rm ergs^{-1} s)=\alpha + \beta \ln(\lambda E_z / 47.2)+N(0,\sigma)$ for the SCGF sample.}             
\label{table:linmix parameters LX}      
\centering          
\begin{tabular}{ l c c c c }
\hline\hline       
Sample & Intercept $\alpha$ & Slope $\beta$ & Intrinsic scatter $\sigma$ & N clusters\\
\hline
Full & $100.843\pm{0.030}$ & $1.585\pm{0.055}$ & $0.631\pm{0.023}$ & 530\\
Gaussian   & $100.869\pm{0.035}$ & $1.562\pm{0.062}$ & $0.646\pm{0.025}$ & 432\\
Non-Gaussian  & $100.733\pm{0.064}$ & $1.683\pm{0.120}$ & $0.561\pm{0.048}$ &  98 \\
Full $\lambda \geq$ 47.2 & $100.799\pm{0.072}$ & $1.645\pm{0.119}$ & $0.639\pm{0.030}$ & 298\\
Full $\lambda <$ $47.2$ & $100.879\pm{0.075}$ & $1.652\pm{0.159}$ & $0.629\pm{0.035}$&  232\\
Gaussian $\lambda \geq$ 47.2 & $100.838\pm{0.079}$ & $1.605\pm{0.124}$ & $0.637\pm{0.034}$ & 245\\
Gaussian $\lambda <$ 47.2 & $100.888\pm{0.090}$ & $1.595\pm{0.198}$ & $0.659\pm{0.040}$ & 187\\
Non-Gaussian $\lambda \geq$ 47.2 & $100.605\pm{0.176}$ & $1.878\pm{0.318}$ & $0.625\pm{0.072}$ & 53\\
Non-Gaussian $\lambda <$ 47.2 & $100.827\pm{0.124}$ & $1.843\pm{0.271}$ & $0.491\pm{0.072}$ & 45\\
$0.05 < \alpha_{AD} \leq 0.15$ & $100.588\pm{0.077}$ & $1.548\pm{0.133}$ & $0.593\pm{0.060}$  & 73\\
$\alpha_{AD} > 0.15$ & $100.826\pm{0.041}$ & $1.546\pm{0.070}$ & $0.653\pm{0.028}$ & 359\\
%
$\Delta \theta/R_{200c} \geq 0.3$ & $100.660\pm{0.067}$ & $1.595\pm{0.139}$ & $0.694\pm{0.051}$  & 131\\
$\Delta \theta/R_{200c} < 0.3$ & $101.003\pm{0.032}$ & $1.483\pm{0.056}$ & $0.628\pm{0.023}$ & 399\\
%
$z \geq 0.3$ & $101.348\pm{0.296}$ & $1.118\pm{0.293}$ & $0.306\pm{0.091}$ &45\\
$0.15 \leq z < 0.3$ & $101.117\pm{0.116}$ & $1.323\pm{0.201}$ & $0.489\pm{0.053}$& 156\\
$z < 0.15$ & $100.760\pm{0.064}$ & $1.484\pm{0.139}$ & $0.704\pm{0.046}$&329\\
$z \geq 0.3$, $\sigma_v \geq 650$ [km/s] & $101.403\pm{0.309}$ & $1.086\pm{0.308}$ & $0.311\pm{0.101}$  & 40\\
$0.15 \leq z < 0.3$, $\sigma_v \geq 650$ [km/s] & $101.139\pm{0.166}$ & $1.341\pm{0.257}$ & $0.504\pm{0.064}$ & 106\\
$z < 0.15$, $\sigma_v \geq 650$ [km/s] & $100.857\pm{0.120}$ & $1.419\pm{0.260}$ & $0.771\pm{0.084}$& 121\\
Gaussian $z \geq 0.3$, $\sigma_v \geq 650$ [km/s] & $101.512\pm{0.324}$ & $1.019\pm{0.313}$ & $0.312\pm{0.109}$  & 34\\
Gaussian $0.15 \leq z < 0.3$, $\sigma_v \geq 650$ [km/s] & $101.127\pm{0.174}$ & $1.356\pm{0.271}$ & $0.505\pm{0.071}$& 88\\
Gaussian $z < 0.15$, $\sigma_v \geq 650$ [km/s] & $100.921\pm{0.146}$ & $1.331\pm{0.316}$ & $0.808\pm{0.106}$& 89\\
\hline                  
\end{tabular}
\end{table*}

\begin{table*}
\caption{Summary of regression analysis for velocity dispersion  (obtained using the gapper method) vs the X-ray luminosity scaling relation $\ln(\sigma_v\; km^{-1} s)=\alpha + \beta \ln(L_X E^{-1}_z\; 10^{-44} \rm ergs^{-1} s )+N(0,\sigma)$ for the SCGF sample.}  
\label{table:linmix parameters vdisp-LX}      
\centering          
\begin{tabular}{ l c c c c }
\hline\hline       
Sample & Intercept $\alpha$ & Slope $\beta$ & Intrinsic scatter $\sigma$ & N clusters\\
\hline
%
Full sample & $6.555\pm{0.018}$ & $0.168 \pm{0.016}$ & $0.163\pm{0.016}$ & 530\\
$z \geq 0.3$ & $6.599\pm{0.154}$ & $0.172\pm{0.137}$ & $0.073\pm{0.060}$ & 45\\
$0.15 \leq z < 0.3$ & $6.552\pm{0.033}$ & $0.195\pm{0.048}$ & $0.073\pm{0.042}$& 156\\
$z < 0.15$ & $6.526\pm{0.028}$ & $0.148\pm{0.024}$ & $0.191\pm{0.021}$& 329 \\
Gaussian $z \geq 0.3$ & $6.564\pm{0.164}$ & $0.199\pm{0.144}$ & $0.081\pm{0.063}$  & 39\\
Gaussian $0.15 \leq z < 0.3$ & $6.549\pm{0.037}$ & $0.187\pm{0.052}$ & $0.070\pm{0.043}$& 130\\
Gaussian $z < 0.15$ & $6.494\pm{0.028}$ & $0.137\pm{0.023}$ & $0.156\pm{0.023}$& 263\\
Gaussian $z \geq 0.3$, $\lambda \geq 47.2$ & $6.590\pm{0.149}$ & $0.181\pm{0.128}$ & $0.076\pm{0.065}$ & 39\\
Gaussian $0.15 \leq z < 0.3$, $\lambda \geq 47.2$ & $6.563\pm{0.034}$ & $0.187\pm{0.048}$ & $0.062\pm{0.046}$& 116\\
Gaussian $z < 0.15$, $\lambda \geq 47.2$ & $6.589\pm{0.038}$ & $0.091\pm{0.041}$ & $0.213\pm{0.035}$& 90 \\
Gaussian $z \geq 0.3$, $L_X \geq 10^{44}$ [ergs/s] & $6.556\pm{0.168}$ & $0.204\pm{0.136}$ & $0.080\pm{0.071}$ & 39\\
Gaussian $0.15 \leq z < 0.3$, $L_X \geq 10^{44}$ [ergs/s] & $6.557\pm{0.056}$ & $0.174\pm{0.066}$ & $0.073\pm{0.057}$ & 97\\
Gaussian $z < 0.15$, $L_X \geq 10^{44}$ [ergs/s] & $6.505\pm{0.077}$ & $0.099\pm{0.077}$ & $0.168\pm{0.047}$ & 54 \\
\hline                  
\end{tabular}
\end{table*}

\begin{table*}
\caption{Summary of results from the \textit{linmix} routine for the SCGF sample.}
\label{table:linmix results}      
\centering          
\begin{tabular}{ l l c l l }
\hline\hline   
Relation & Test Sample & Result & Baseline sample & Parameter values  \\
\hline
$\sigma_v$ - $\lambda$ & Gaussian & lower $\sigma$ & Full & $0.095\pm{0.021}$ vs $0.138\pm{0.017}$\\
$\sigma_v$ - $\lambda$ & non-Gaussian & higher $\alpha$ & Gaussian  &  $6.524\pm{0.047}$ vs $6.446\pm{0.017}$ \\
$\sigma_v$ - $\lambda$ & non-Gaussian & higher $\sigma$ & Gaussian  & $0.231\pm{0.040}$ vs $0.095\pm{0.021}$ \\
$\sigma_v$ - $\lambda$ & Gaussian $\lambda \geq 47.2$ & lower $\sigma$ & Gaussian $\lambda < 47.2$ & $0.081\pm{0.029}$  vs $0.118\pm{0.034}$  \\
$\sigma_v$ - $\lambda$ &$0.15\leq z < 0.3$ & lower $\sigma$ &  $z < 0.15$ & $0.067\pm{0.045}$ vs $0.164\pm{0.020}$ \\
$\sigma_v$ - $\lambda$ & Gaussian $0.15\leq z < 0.3$ & lower $\sigma$ & Gaussian $z < 0.15$ & $0.064\pm{0.050}$ vs $0.111\pm{0.027}$\\
$L_X$ - $\lambda$ & Non-Gaussian & lower $\alpha$ & Gaussian  & $100.733\pm{0.064}$ vs  $100.869\pm{0.035}$ \\
$L_X$ - $\lambda$ & $0.05 < \alpha_{AD} \leq 0.15$ & lower $\alpha$ &  $\alpha_{AD} < 0.05$  & $100.588\pm{0.077}$ vs $100.869\pm{0.035}$ \\
$L_X$ - $\lambda$ & $\Delta \theta/R_{200c} \geq 0.3$ & lower $\alpha$ & $\Delta \theta/R_{200c} < 0.3$ & $100.660\pm{0.067}$ vs  $101.003\pm{0.032}$ \\
$L_X$ - $\lambda$ & $z < 0.15$ & higher $\sigma$ & $0.15\leq z < 0.3$ & $0.704\pm{0.046}$ vs $0.489\pm{0.053}$ \\
$L_X$ - $\lambda$ & $z < 0.15$, $\sigma_v \geq 650$ [km/s] & higher $\sigma$ & $0.15\leq z < 0.3$, $\sigma_v \geq 650$ [km/s]  & $0.771\pm{0.084}$ vs $0.504\pm{0.064}$ \\
$\sigma_v$ - $L_X$ & $0.15\leq z < 0.3$ & lower $\sigma$ &$z < 0.15$ & $0.073\pm{0.042}$  vs $0.191\pm{0.021}$\\
$\sigma_v$ - $L_X$ & Gaussian $0.15\leq z < 0.3$ & lower $\sigma$ & Gaussian $z < 0.15$ & $0.070\pm{0.043}$ vs $0.156\pm{0.023}$ \\
$\sigma_v$ - $L_X$ & Gaussian $0.15\leq z < 0.3$, $\lambda \geq 47.2$ & lower $\sigma$ & Gaussian $z < 0.15$ $, \lambda \geq 47.2$ &$0.062\pm{0.046}$ vs $0.213\pm{0.035}$\\
$\sigma_v$ - $L_X$ & G $0.15\leq z < 0.3$, $L_X \geq 10^{44}$ [ergs/s] & lower $\sigma$ & G $z < 0.15$, $L_X \geq 10^{44}$ [ergs/s] & $0.073\pm{0.057}$ vs $0.168\pm{0.047}$ \\
\hline                  
\end{tabular}
\end{table*}

To make a comparative study of scaling relations, we need to choose a common variable. For our survey, optical richness is a better-measured quantity, compared to X-ray luminosity, while velocity dispersion is a parameter of our study. Therefore, we selected to study the scaling relations against richness. Richness is also a selection variable to ensure the purity and completeness of the cluster selection, and it determines the success of the spectroscopic follow-up. Analysis of properties against the selection variables simplifies the modelling \citep{Kelly2007}.

In performing the analysis of the scaling relations, we take into account the evolution of the scaling relations, which instructs us to use the X-ray luminosity ($L_X$) specified in the rest frame 0.1--2.4 keV, in the form $\ln(L_X E_z^{-1}\; \mathrm{ergs^{-1} s)}$ and richness as $\ln(\lambda E_z$), where $E^2_z=\Omega_{\rm m}(1+z)^3+{\Omega_\Lambda}$. In addition, to avoid a strong degeneracy between the slope and the normalization, we normalize the richness by its median value of 47.2: $\ln(\lambda E_z)-\ln(47.2)$. Velocity dispersion does not need a correction for evolution, so we use $\ln (\sigma_v\; \mathrm{km^{-1} s)}$.  For all the scaling relations, we used the gapper estimate of the velocity dispersion. To verify this scaling, we also performed the analysis in different redshift bins.

\begin{figure}[ht]
    \centering
    \includegraphics[width=\hsize]{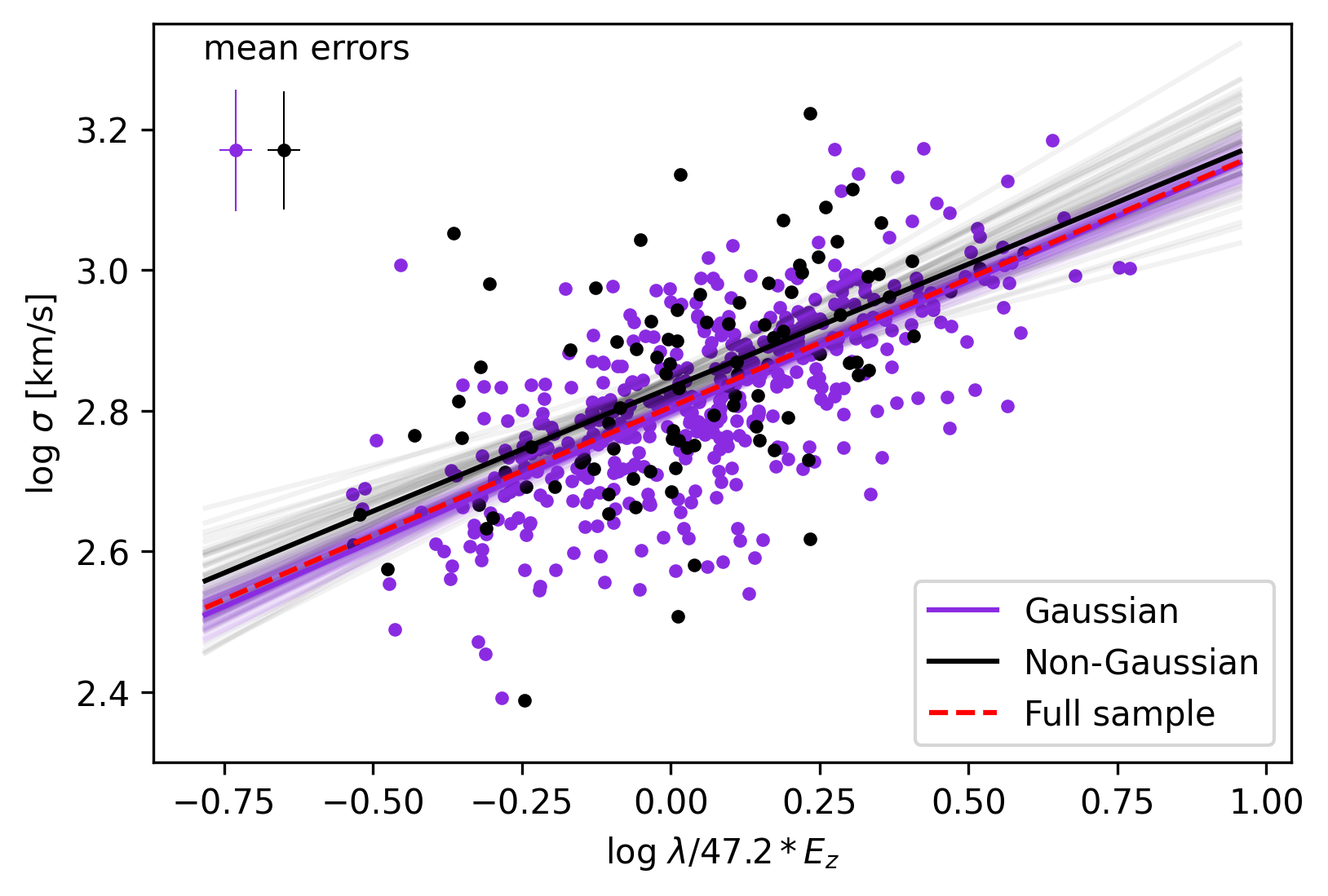}
    \caption{Velocity dispersion vs richness for the SCGF sample with scaling relations overplotted. Data points for Gaussian clusters are marked with purple dots and for non-Gaussian with black dots with 1$\sigma$ errors on both axes. Purple lines show the fits for Gaussian clusters, black lines -- for non-Gaussian clusters and the red dashed line -- for the median of the full sample. The thin lines display 0.5\% of all the {\it linmix} MCMC chains. For the bold lines, we take the median values for each sample.}
    \label{fig:scaling richness vdisp SCGF}
\end{figure}

\begin{figure}[ht]
    \centering
    \includegraphics[width=\hsize]{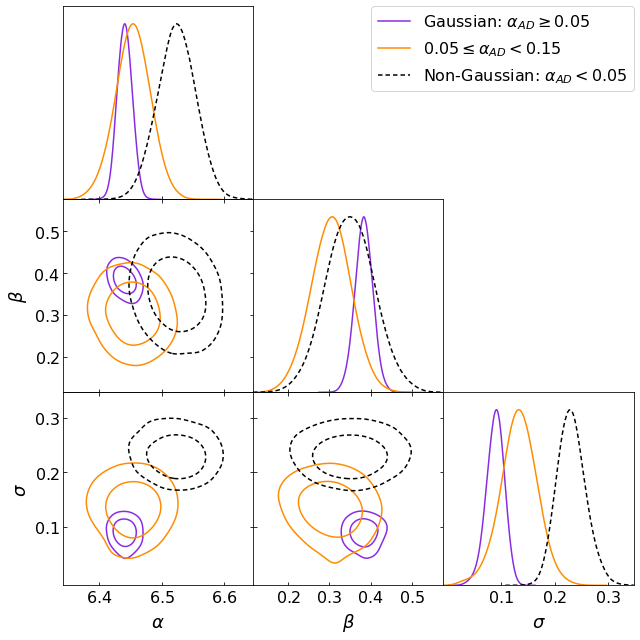}
    \caption{Effect of velocity substructure on the velocity dispersion - richness scaling relation for the SCGF catalogue with a different range of $\alpha_{AD}$ values. $\alpha$ stands for intercept, $\beta$ -- for slope and $\sigma$ -- for the intrinsic scatter. 2D contours show the 68\% and 95\% significance levels. Gaussian clusters are marked with solid purple lines and non-Gaussian clusters with dashed black lines. Solid orange lines show the cluster population with a marginal probability of having a substructure (between 85\% and 95\% confidence level).}
    \label{fig:triangle scaling vdisp alpha SCGF}
\end{figure}

We applied the \textit{linmix}\footnote{https://github.com/jmeyers314/linmix.git} routine by \cite{Kelly2007} for fitting, which performs a Bayesian analysis that accounts for the distribution function of the observed clusters (specified using N=3 Gaussian components). 
There are three regression parameter equations used for calculating line fits in the \textit{linmix} routine:

\begin{equation}\label{eq:linmix eta}
    \eta=\alpha+\beta x_i +\epsilon
,\end{equation}
where $\alpha$ is the intercept, $\beta$ is the slope and $\epsilon$ is intrinsic random scatter about the regression. Here, $\epsilon$ is assumed to be normally distributed with a zero mean and variance $\sigma^2$ ($N(0,\sigma)$).
\begin{equation}\label{eq:linmix x}
    x=x_i + x_{\mathrm {err}} ,
\end{equation}
where $x_i$ are the data points with $x_{\mathrm {err}}$ errors.
\begin{equation}\label{eq:linmix y}
    y_i=\eta+ y_{\mathrm {err}},
\end{equation}
where $y_{\mathrm {err}}$ is the error in $y_i$ (both are data), and
\begin{equation}\label{eq:linmix sigma}
    \sigma^2 = {\rm Var} (\epsilon),
\end{equation}
where $\sigma^2$ is the variance. 
To remove the need to simulate the detection vector, we performed the analysis against the optical richness, which is the selection variable. The results, tabulated in Tables~\ref{table:linmix parameters vdisp} and \ref{table:linmix parameters LX}, are obtained from running 100\,000 Markov chain Monte Carlo steps, rejecting the 1000 initial burn-in steps. We quote the median values of the posterior distribution of the parameters and their 68\% confidence intervals.

\begin{figure}[ht]
    \centering
    \includegraphics[width=\hsize]{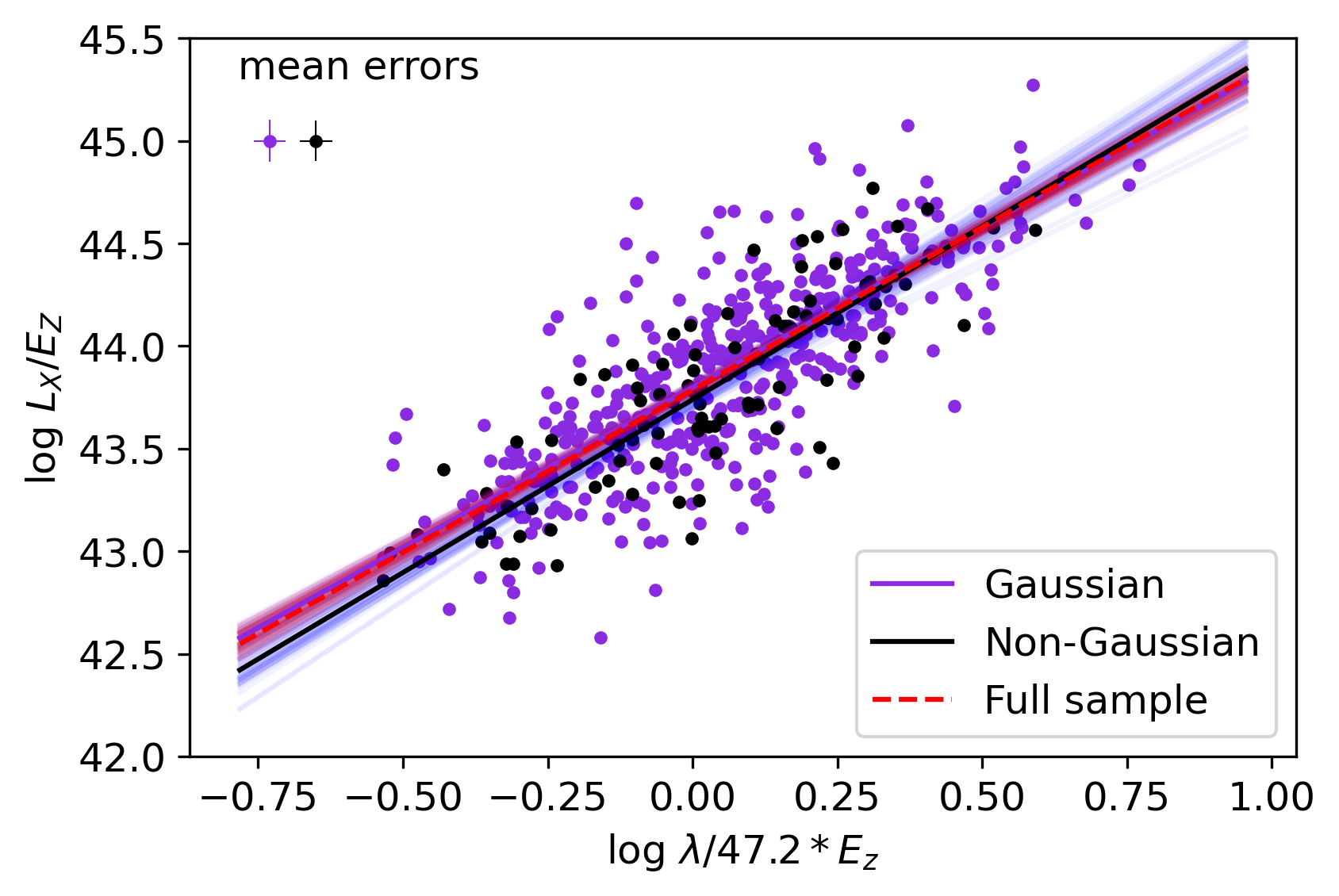}
    \caption{X-ray luminosity vs richness scaling relations for the SCGF sample, re-scaled to redshift. Details are the same as in Fig. \ref{fig:scaling richness vdisp SCGF}.}
    \label{fig:scaling richness LX SCGF}
\end{figure}

\begin{figure}[ht]
    \centering
    \includegraphics[width=\hsize]{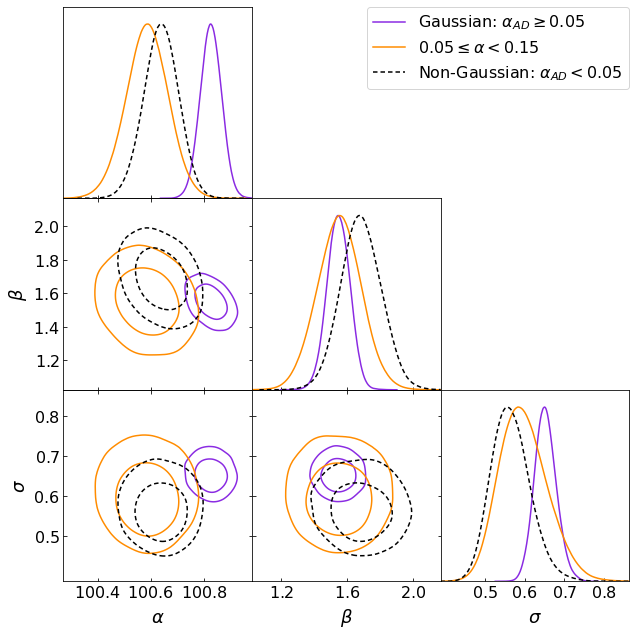}
    
    \caption{Effect of velocity substructure on the X-ray luminosity -- richness scaling relation. Details are the same as in Fig. \ref{fig:triangle scaling vdisp alpha SCGF}.} 
    \label{fig:triangle scaling LX alpha SCGF}
\end{figure}

In the following, we mainly discuss the results of the SCGF sample, while presenting the results for the SPIDERS sample in Appendix \ref{SPIDERS appendix}, with the parameters of all fits summarized in Tables~\ref{table:linmix parameters vdisp} and \ref{table:linmix parameters LX}. We do not find any significant differences in the results between the two samples. 
The scaling relations of velocity dispersion and X-ray luminosity against richness are presented in Figs.\ref{fig:scaling richness vdisp SCGF} to \ref{fig:scaling richness LX SCGF}, showing the data split based on the substructure test. It is already noticeable in those figures that clusters with substructure more often exhibit high-velocity dispersion for a given richness but do not show differences in X-ray luminosities. 
We fitted the relation for the full sample and clusters with and without substructure. 
The fits show an overall agreement in the scaling relations between the clusters with or without substructure. We detail this comparison in Figs. \ref{fig:triangle scaling vdisp alpha SCGF} and \ref{fig:triangle scaling LX alpha SCGF}. A 2D KS test on velocity dispersion and richness between Gaussian and non-Gaussian samples gives a p-value of 0.10. For the X-ray luminosity and richness, the p-value is 0.52. Our main result is the detection of significantly large (2.5 higher) intrinsic scatter in the scaling relations involving velocity dispersion for non-Gaussian clusters, and a somewhat elevated normalization  ($7\pm3$\%). At the same time, there is a marginal decrease in scatter and a  ($17\pm6$\%) decrease in the normalization of the scaling relation between richness and X-ray luminosity. All these differences are expected for merging clusters, which on one hand have higher dynamical disturbance but on the other have weaker cool cores. Disruption of cool cores reduces the total X-ray luminosity but also improves the self-similarity of clusters. Given that our sample of Gaussian clusters contains some contamination from the non-Gaussian clusters, we performed an analysis of the marginally non-Gaussian clusters by considering the interval of $0.05<\alpha_{AD}<0.15$ in Fig. \ref{fig:triangle scaling vdisp alpha SCGF}. As it can be seen from the plots, the border sample already behaves like a Gaussian sample in the velocity dispersion versus richness scaling relations, while it is closer to a non-Gaussian sample on the $L_X$--richness plane. The cleaner definition of the Gaussian sample using $\alpha>0.15$ does not result in any differences in the scaling relations.

\begin{figure}[ht]
    \centering
    \includegraphics[width=\hsize]{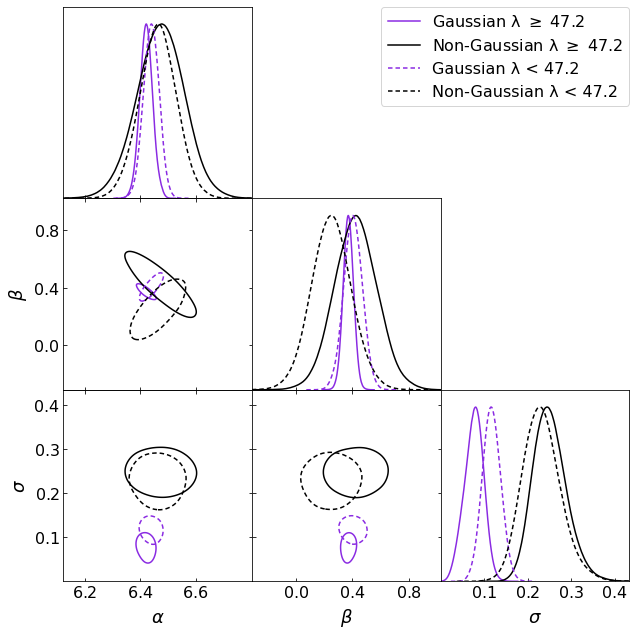}
    \caption{ Reducing the effect of the complexity of scaling relations on the inferred scatter between velocity dispersion and richness. Contours show 68\% confidence levels. Gaussian clusters are marked with purple lines and non-Gaussian clusters with black lines. The high-richness sample is marked with solid lines and the low-richness sample with dashed lines. Other details are the same as in Fig. \ref{fig:triangle scaling vdisp alpha SCGF}.}
    \label{fig:triangle vdisp richness cut}
\end{figure}

\begin{figure}[ht]
    \centering
    \includegraphics[width=\hsize]{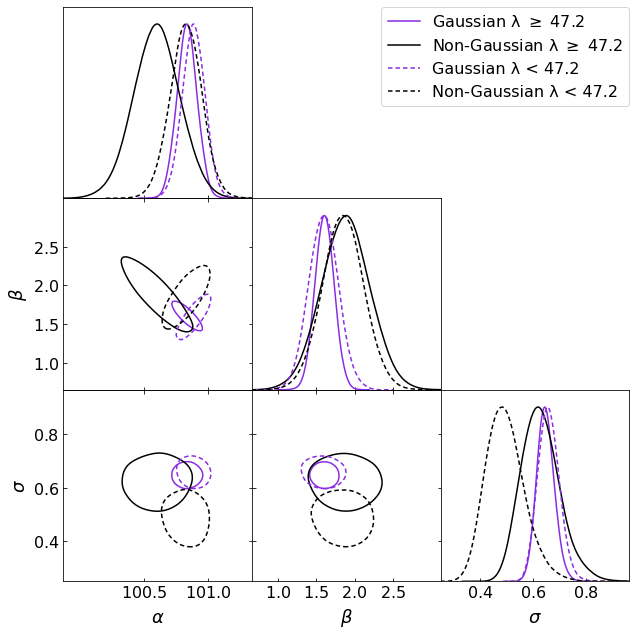}
    \caption{Effect of velocity substructure on the richness--X-ray luminosity scaling relation vs complexity of scaling relations for the SCGF catalogue with a richness cut at $\lambda = 47.2$.  Details are the same as in Fig. \ref{fig:triangle vdisp richness cut}.} 
    \label{fig:triangle LX richness cut}
\end{figure}

As next, we test whether the scatter in the scaling relations is due to a deviation of the shape of the scaling relations from a simple power law form. For that, we split the sample in two at the median value of richness (47.2). The results of this analysis, which are reported in Figs. \ref{fig:triangle vdisp richness cut} and \ref{fig:triangle LX richness cut},  reveal the same parameters of the scaling relations as seen in the analysis of the whole sample.  We show that the slopes and normalizations of the scaling relations for low- and high-richness clusters agree, allowing us to exclude the change in the scaling relations as a putative explanation for the increased scatter. We note a marginally lower scatter for rich Gaussian clusters, compared to low-richness Gaussian clusters. Improvements in cluster identification at the low richness, achieved with the SCGF sample, allows us to significantly reduce the uncertainties of scaling relations at the low mass end, which are particularly important for the tests of the consistency of scaling relations.

\begin{figure}[ht]
    \centering
    \includegraphics[width=\hsize]{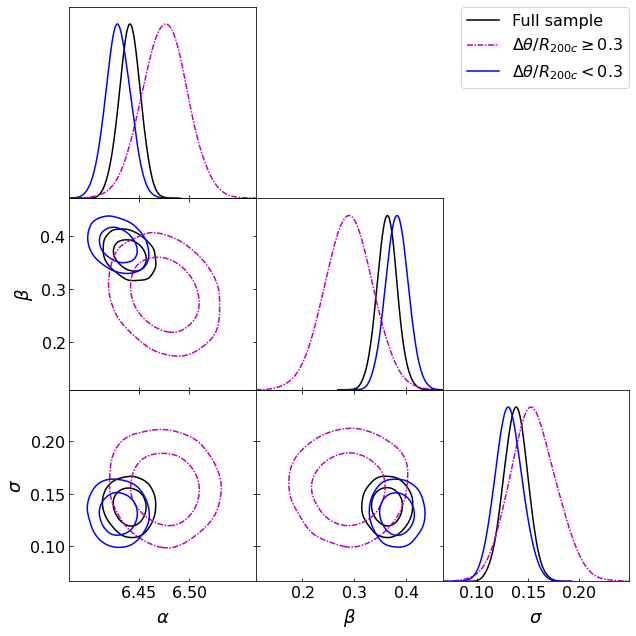}
    \caption{Effect of the offset between X-ray and optical centres on the velocity dispersion - richness scaling relation for the SCGF Gaussian subsample.  The parameters of the scaling relations for the small offset clusters are shown using solid blue lines and for the large offset clusters - with dash-dotted magenta lines. The separation is done at 0.3R$_{200c}$. For reference, we also present the parameters describing the scaling relations for the full sample (all clusters) with solid black lines. Other details are the same as in Fig. \ref{fig:triangle scaling vdisp alpha SCGF}.}
    \label{fig: triangle separation richness vdisp SCGF}
\end{figure}

\begin{figure}[ht]
    \centering
    \includegraphics[width=\hsize]{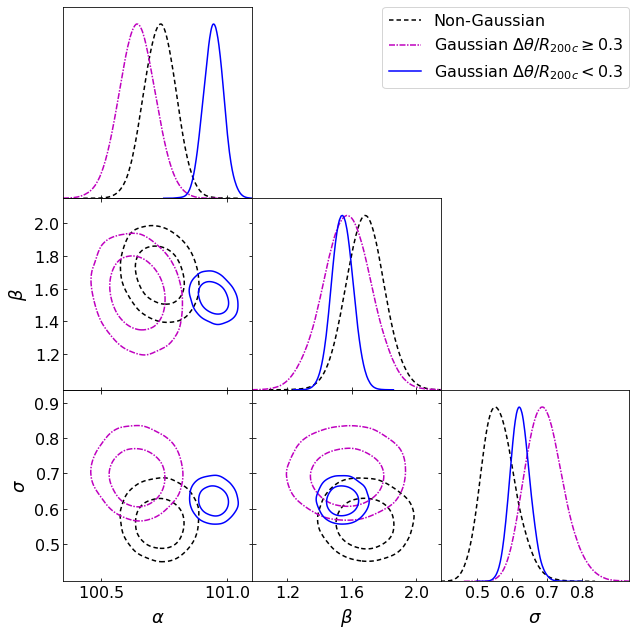}
    \caption{Effect of the offset between X-ray and optical centres on the X-ray luminosity vs richness scaling relation for the SCGF Gaussian subsample. The parameters of the scaling relations for the non-Gaussian sample are shown in dashed black lines, small offset clusters - in solid blue lines and large offset clusters - in dash-dotted magenta lines. The separation is done at 0.3R$_{200c}$. Other details are the same as in Fig. \ref{fig:triangle scaling vdisp alpha SCGF}.}
    \label{fig: triangle separation richness LX SCGF}
\end{figure}

In Fig. \ref{fig: triangle separation richness vdisp SCGF} we test whether separating the Gaussian sample based on the value of the offset between the optical and X-ray centre results in a change in the scaling relations, concluding that the obtained values are consistent, and the scatter in the $\sigma_v-\lambda$ relation of the offset clusters is in between the scatter of non-Gaussian and Gaussian clusters.  Performing a similar analysis of the $L_X-\lambda$ relation in Fig. \ref{fig: triangle separation richness LX SCGF} we find that offset clusters exhibit $25\pm7$\% lower normalization. This offset is consistent with the normalization for the non-Gaussian clusters. Such changes would be expected from the plane of the sky mergers, where projection effects reduce the scatter in the dynamical properties, a famous example of such clusters is A3667 \citep{melanie08,finoguenov10}.  The observed reduction of $L_X$ also rules out chance contamination of $L_X$ as a possible explanation for the off-centring, as one would expect a higher $L_X$ for contaminated systems, contrary to what is observed.

\begin{figure}[ht]
    \centering
    \includegraphics[width=\hsize]{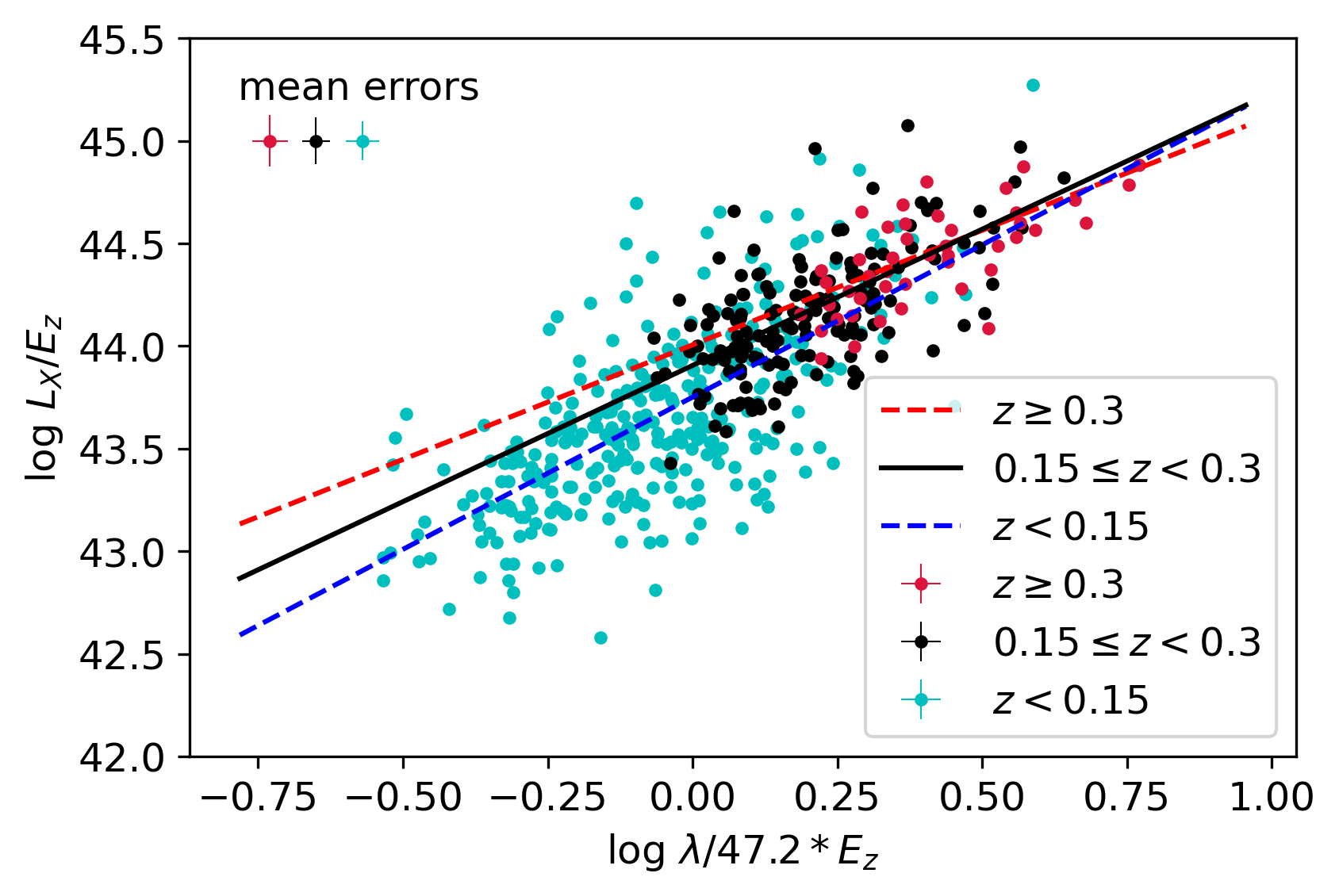}
    \caption{Redshift evolution of X-ray luminosity vs richness scaling relations for the SCGF sample. The $z \geq 0.3$ sample is marked with red, the $0.15 \leq z < 0.3$ sample with black, and the $z < 0.15$ sample with light blue.}
    \label{fig:scaling richness LX SCGF redshift cut}
\end{figure}

\begin{figure}[ht]
    \centering
    \includegraphics[width=\hsize]{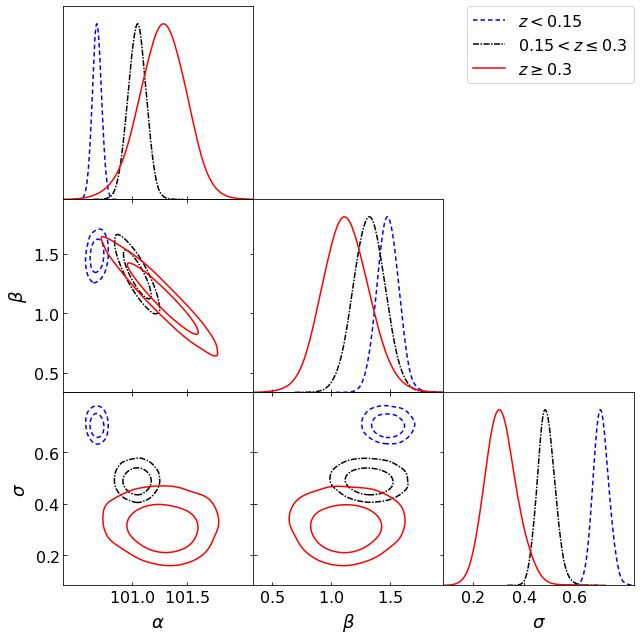}
    \caption{Marginal distributions and covariances of the parameters of the X-ray luminosity vs richness scaling relation in a range of redshift bins. The redshift sample $z < 0.15$ is marked with dashed blue, $0.15 \leq z < 0.3$ with dash-dotted black, and $z \geq 0.3$ with solid red lines. Other details are the same as in Fig. \ref{fig:triangle scaling vdisp alpha SCGF}.}
    \label{fig:triangle LX redshift cut SCGF}
\end{figure}

\begin{figure}[ht]
    \centering
    \includegraphics[width=\hsize]{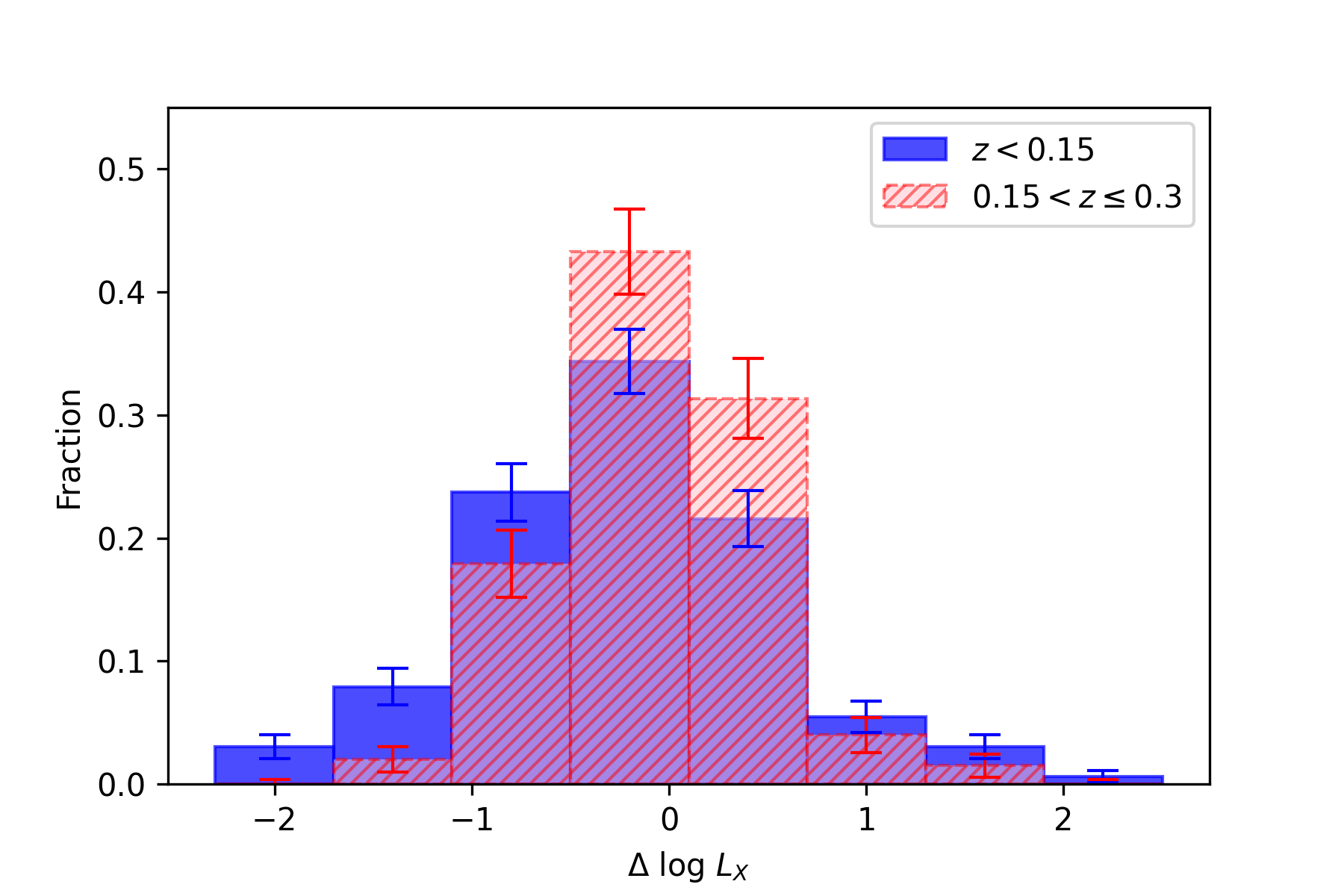}
    \caption{Deviations from the median X-ray luminosity--richness fits. The lower-redshift sample ($z<0.15$) is marked with blue and the higher-redshift sample ($0.15 < z \leq 0.3$) with hatched red.}
    \label{fig:hist deviations LX redshift cut SCGF}
\end{figure}

\begin{figure}[ht]
    \centering
    \includegraphics[width=\hsize]{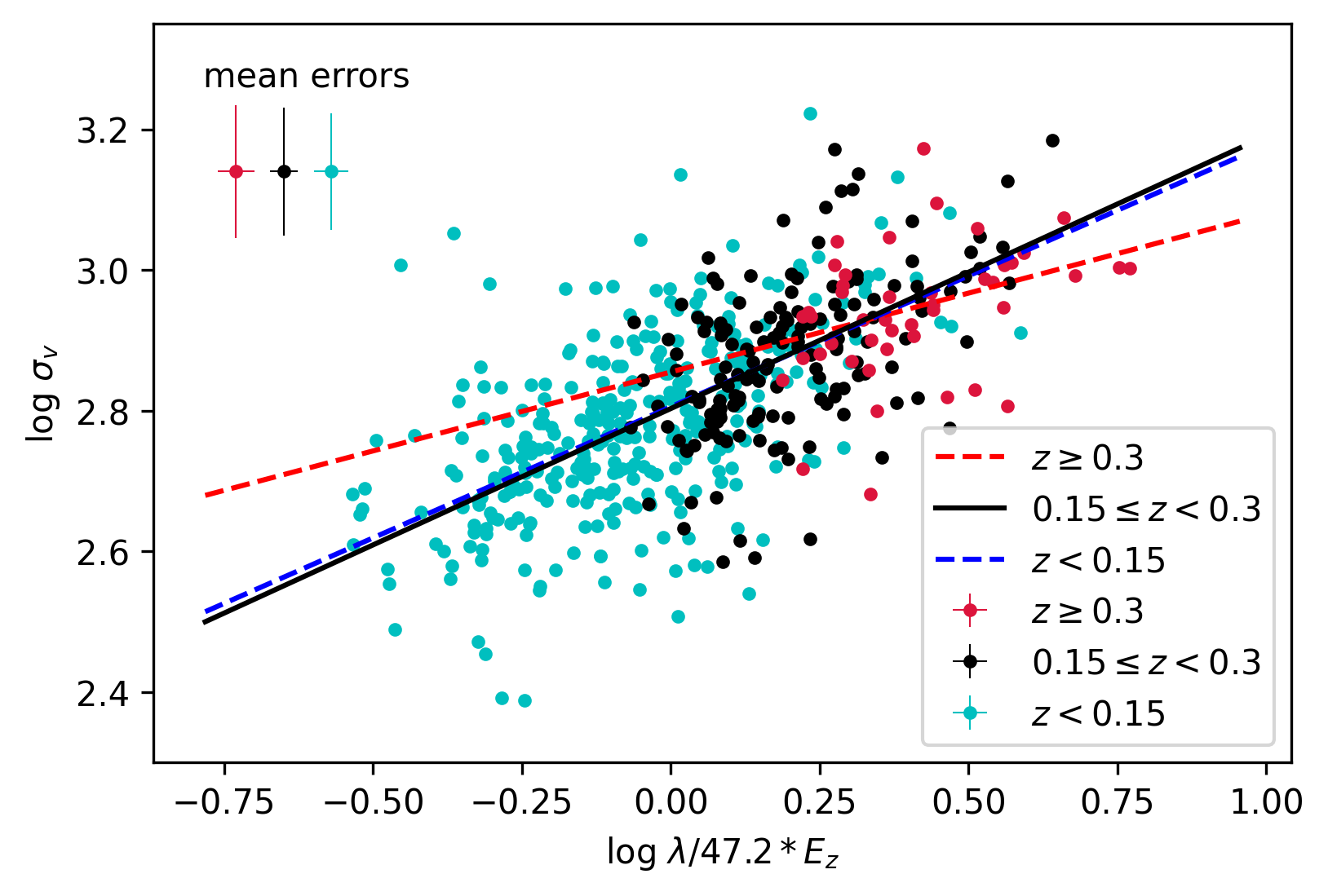}
    \caption{Redshift evolution of velocity dispersion vs richness scaling relations for the SCGF sample. Details are the same as in Fig. \ref{fig:scaling richness LX SCGF redshift cut}. }
    \label{fig:scaling richness vdisp SCGF redshift cut}
\end{figure}

\begin{figure}[ht]
    \centering
    \includegraphics[width=\hsize]{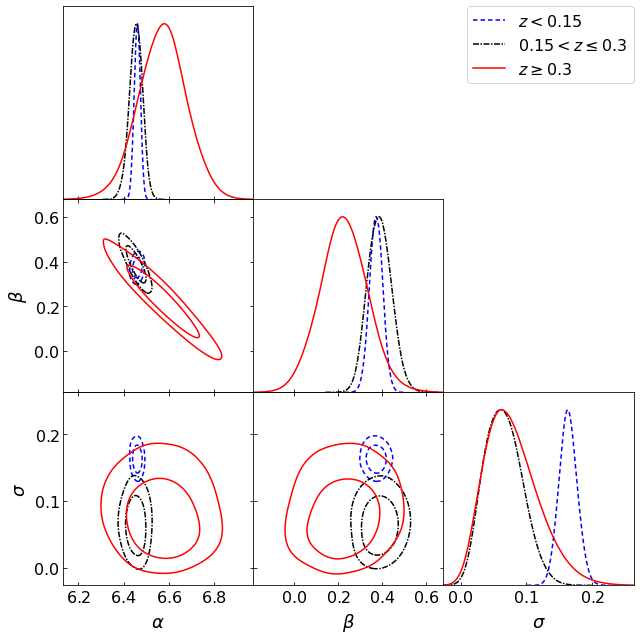}
    \caption{Same as Fig.\ref{fig:triangle LX redshift cut SCGF}, but for velocity dispersion vs richness.}
    \label{fig:triangle vdisp redshift cut SCGF}
\end{figure}

\begin{figure}[ht]
    \centering
    \includegraphics[width=\hsize]{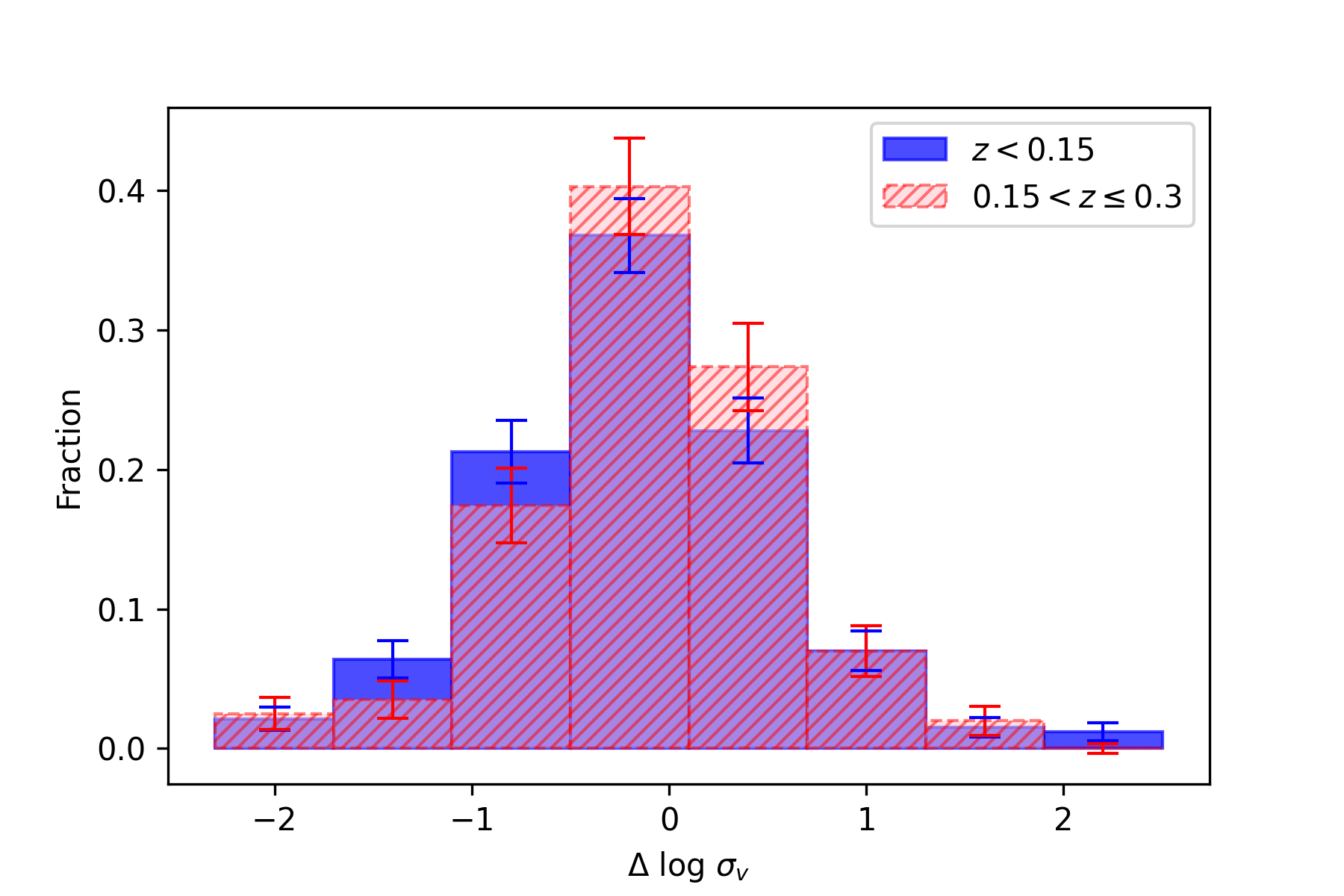}
    \caption{Deviations from the median velocity dispersion - richness fits. The lower-redshift sample ($z<0.15$) is marked with blue and the higher-redshift sample ($0.15 < z \leq 0.3$) with hatched red.}
    \label{fig:hist deviations vdisp redshift cut SCGF}
\end{figure}

\begin{figure}[ht]
    \centering
    \includegraphics[width=\hsize]{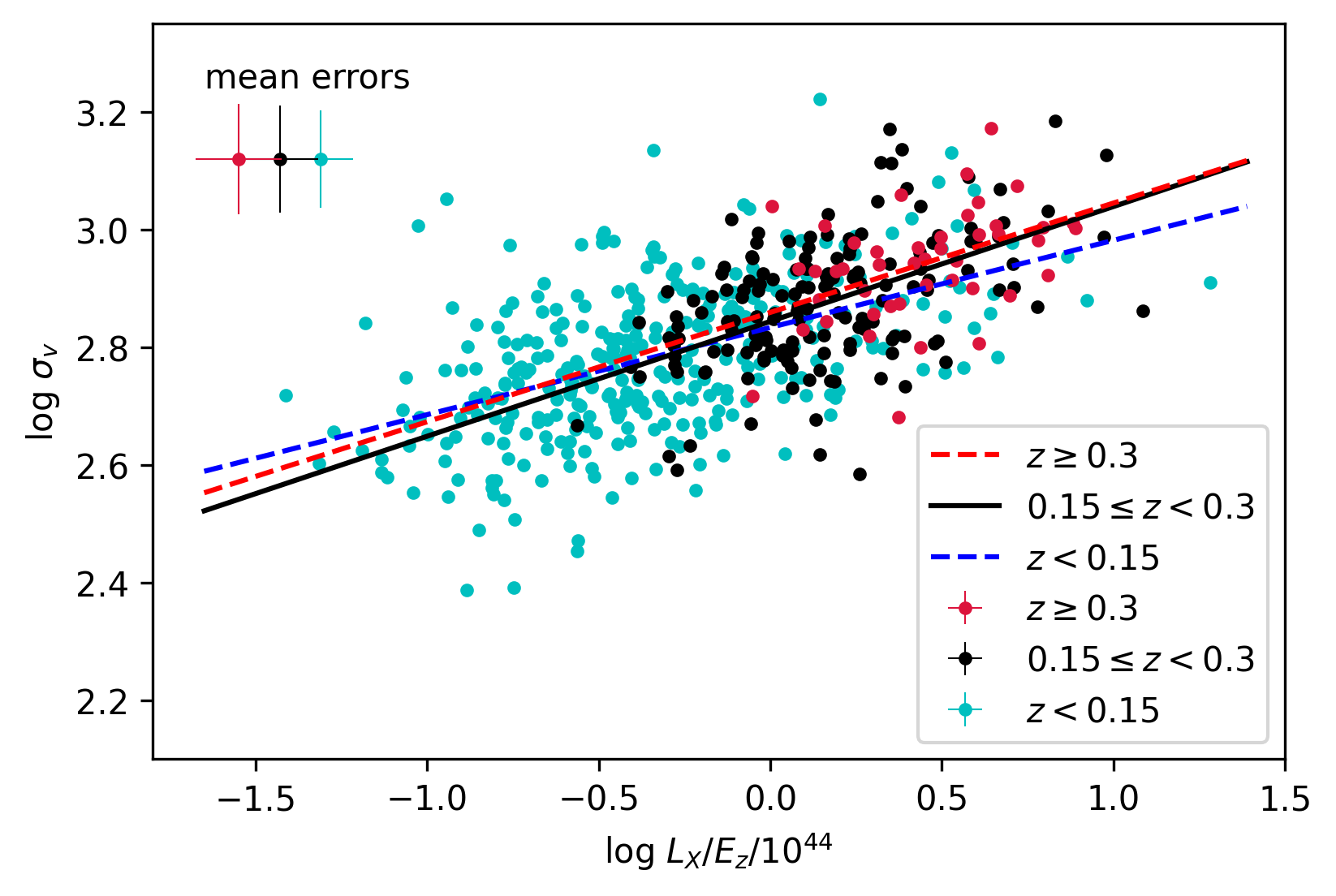}
    \caption{Redshift evolution of velocity dispersion vs X-ray luminosity relations for the SCGF sample. Details are the same as in Fig. \ref{fig:scaling richness LX SCGF redshift cut}. }
    \label{fig:scaling vdisp-LX SCGF redshift cut}
\end{figure}

\begin{figure}[ht]
    \centering
    \includegraphics[width=\hsize]{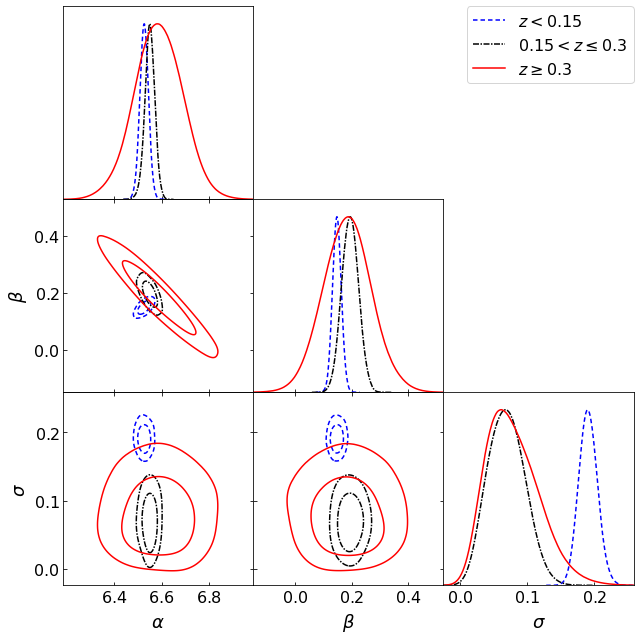}
    \caption{Same as Fig.\ref{fig:triangle LX redshift cut SCGF}, but for velocity dispersion vs X-ray luminosity.}
    \label{fig:triangle vdisp-LX redshift ranges SCGF}
\end{figure}

\begin{figure}[ht]
    \centering
    \includegraphics[width=\hsize]{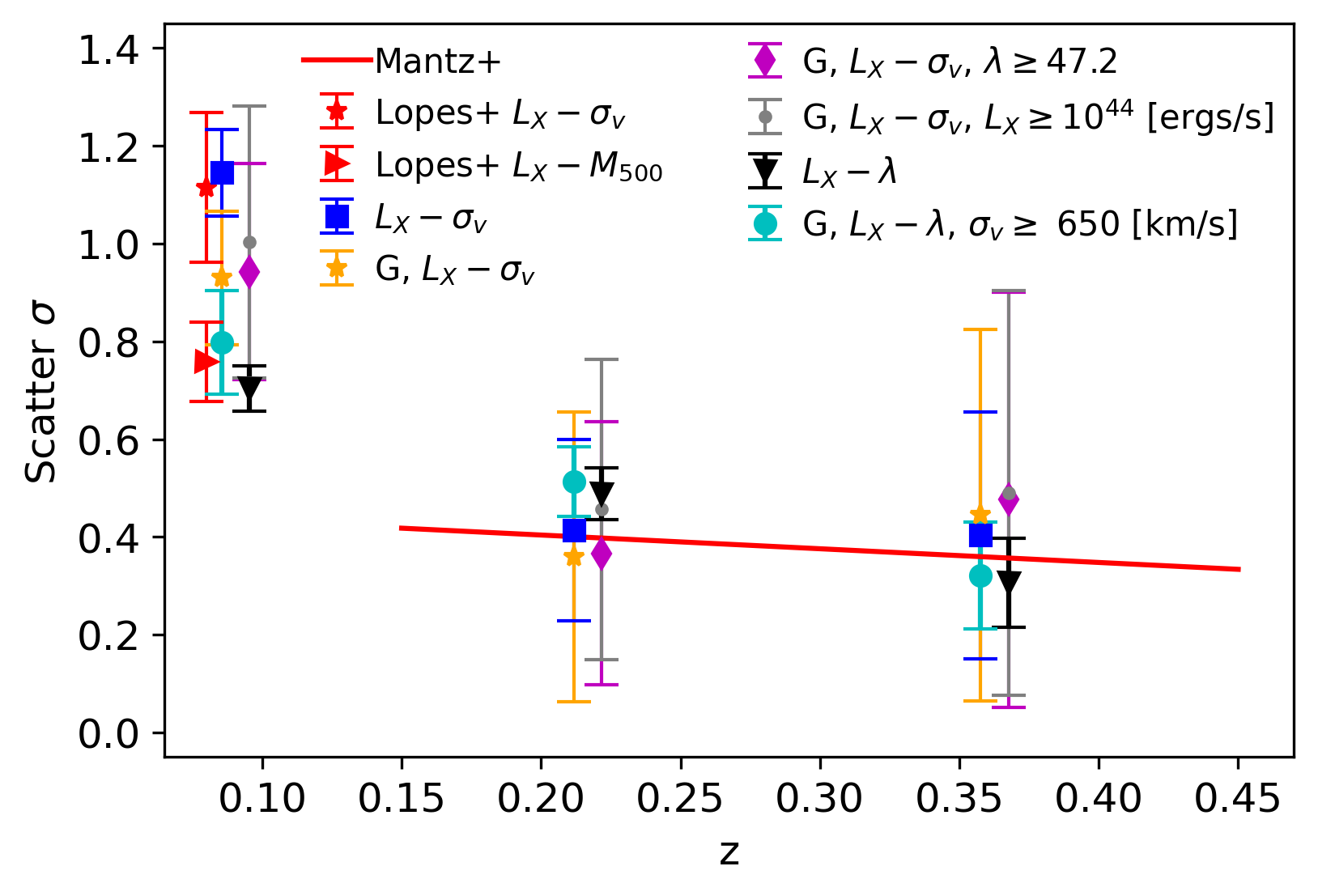}
    \caption{Redshift evolution of the intrinsic scatter of X-ray luminosity in scaling relations against optical mass proxies. The scatter of X-ray luminosity vs velocity dispersion for the full sample is marked with blue squares, for the Gaussian sample with orange stars, for the Gaussian high-richness sample ($\lambda \geq 47.2$) with magenta diamonds and the Gaussian high $L_X$ sample ($L_X \geq 10^{44}$ ergs s$^{-1}$) with small grey circles. The scatter of the X-ray luminosity versus richness for the full sample is marked with black triangles and for the Gaussian high-velocity dispersion sample ($\sigma_v \geq 650$ km s$^{-1}$) with light blue circles. For comparison, we show the linear fit to the evolution of scatter in $L_x$ obtained using high-quality X-ray data on relaxed clusters from \cite{mantz2016}  with a red line. A larger scatter seen in our data at low redshifts is also seen in the analysis of \cite{lopes09}. \cite{lopes09} X-ray luminosity versus velocity dispersion relation at redshift $z=0.08$ is marked with a red star, and their X-ray luminosity versus $M_{500}$ at the same redshift is marked with a red triangle.}
    \label{fig:scatter LX redshift ranges SCGF}
\end{figure}

In Figs. \ref{fig:scaling richness LX SCGF redshift cut} and \ref{fig:triangle LX redshift cut SCGF} we study the redshift evolution of the X-ray luminosity-richness scaling relation. We see a substantial (a factor of 2) reduction of scatter in the scaling relation with increasing redshift.  A 2D KS test between the z < 0.15 and 0.15 < z $\leq$ 0.3 samples yields a p-value of $10^{-30}$. A sharp transition in the behaviour of scatter with redshift, using a
consistent and complete dataset, is presented for the first time. To highlight the differences in scatter, we show the residuals from the different redshift fits (z < 0.15 and 0.15 $\leq$ z < 0.3) in Fig. \ref{fig:hist deviations LX redshift cut SCGF}. The figure shows that at $z<0.15,$ deviations in $L_X$ are larger, which can be seen from the broader wings of the distribution as well as a depression in the centre. Shuffling on the standard deviation and KS test p-values between the samples are p<0.0001 and p=$6.85\times10^{-5}$, respectively. Previous results on the redshift evolution of scatter have suggested only a mild evolution of scatter \citep{mantz2016}, and we have embarked on a detailed study of the possible differences.
Using the large size of our sample, we were able to reduce the redshift bin size, finding that the high scatter is localized to z<0.15. This redshift range did not have weak lensing mass measurements in \cite{mantz2016} and so, their relation for the scatter has not been based on $z<0.15$ data.  To arrive at a full picture we also tested the behaviour of scatter in richness and velocity dispersion. The scatter is marginally smaller at high z for the velocity dispersion -- richness scaling relation, as illustrated in Figs. \ref{fig:scaling richness vdisp SCGF redshift cut} and \ref{fig:triangle vdisp redshift cut SCGF} where we display the sample and explore the parameters of the scaling relations. A 2D KS test p-value between the z < 0.15 and 0.15 < z $\leq$ 0.3 samples is $1.74\times10^{-23}$. The residuals for these redshift fits are shown in Fig. \ref{fig:hist deviations vdisp redshift cut SCGF}. We see that the marginally larger scatter at $z<0.15$ is caused by an increase in the fraction of systems with lower velocity dispersion for a given richness, which is indicative of projection effects, anticipated for redMaPPer performance at low z. These projection effects lead to a preferential upscatter of low-mass clusters in richness, while the measurement of their velocity dispersion is less affected. Shuffling on standard deviations and KS test p-values for the residuals are 0.06 and 0.20, respectively.
In order to remove the extra scatter due to projection effects on richness at $z<0.15$, we included an analysis of $\sigma_v-L_X$ relation. Doing this analysis allows us to fully exclude any residual sensitivity to the cluster selection. In presenting the results as the scatter in $L_X$, we have translated the measured scatter to that of $L_X$ by normalizing the former by the slope of the relation. To remove the effect of the noise of individual slope measurement, when inferring the scatter in $L_X$ we use the same value of the slope, obtained for the full sample ($\beta=0.168$). The results of $\sigma_v-L_X$ analysis are given in Table \ref{table:linmix parameters vdisp-LX} and the full details of the scaling relation analysis are shown in Figs. \ref{fig:scaling vdisp-LX SCGF redshift cut} and \ref{fig:triangle vdisp-LX redshift ranges SCGF}. The 2D KS test p-value for the z < 0.15 and 0.15 < z $\leq$ 0.3 samples is $8.12\times10^{-23}$. The results on the $L_X$ scatter are summarized in Fig. \ref{fig:scatter LX redshift ranges SCGF}, including a comparison to literature values, where in addition to \cite{mantz2016}, we presented the low-z results from \cite{lopes09}. While our results are consistent with literature values in each redshift bin, extrapolation of the trend published by \cite{mantz2016} to lower redshifts would result in an underestimation of the scatter. Also, the \cite{mantz2016} results strictly apply only to relaxed clusters, and so differences to other results could have been ignored based on that. But, as we show, the scatter at $z>0.15$ is low for all the X-ray-selected clusters. The enhancement of scatter in $\sigma_v-\lambda$ scaling relation we report can be compared to a similar difference between the scatter in $\sigma_v-L_X$ versus $M_{500c}-L_X$, presented in \cite{lopes09}, with the explanation that dynamical modelling reduces the scatter intrinsic to $\sigma_v$. We see a good agreement in the inferred scatter between our $L_X-\lambda$ relation and $M_{500}-L_X$ of \cite{lopes09} and for the same scaling relation $\sigma_v-L_X$ in both works, with all the results shown in Fig.\ref{fig:scatter LX redshift ranges SCGF}. Following our results on the reduced scatter using Gaussian clusters, we added the analysis of $\sigma_v-L_X$, which is more consistent with the inferred scatter with the results of $L_X-\lambda$. We have also demonstrated that selecting Gaussian clusters and removing low-$L_X$, low-richness, or low-$\sigma_v$ clusters does not lead to a reduction of scatter at $z<0.15$. 
  
High scatter at $z<0.15$ could be partially due to our flux correction procedure, which changes from a median value of 20\% at high z (with a median ratio of the aperture to $R_{500c}$ of 0.7) to 60\% at a redshift of 0.1 (with the median ratio of the aperture to $R_{500c}$ of 0.4). Large deviations, of the order of 100\% in the behaviour of the central part ($<0.2R_{500c}$) of the cluster surface brightness profile have been reported by the XCOP \citep{Ghirardini19} and REXCESS projects \citep{Croston08}. However, given that similar results are obtained integrating the X-ray surface brightness to $R_{500}$ by \cite{lopes09}, a conclusion on the level of scatter at low redshifts appears robust.
The milder part of the redshift evolution of scatter is associated with decreasing contribution of cluster cores to the X-ray luminosity at high redshifts \citep{McDonald16}. Their explanation consists of the similar luminosity of the cool core at a given mass, but much larger total X-ray luminosity, as predicted by the evolution of the scaling relations. However, our results cannot be explained by this. In terms of cool core, it would require a much larger contribution of cool core to the total $L_X$, not a constant one, suggested by \cite{McDonald16}, whose data barely sample $z<0.15$, due to the survey construction.

\section{Discussion and conclusions}
\label{Discussion}

Using the large spectroscopic dataset of the CODEX clusters, we show that the presence of velocity substructure leads to a noticeable increase in scatter in $\sigma_v$ (Fig. \ref{fig:triangle scaling vdisp alpha SCGF}), which, as we show in Fig. \ref{fig:triangle vdisp richness cut}, is not associated with richness-dependent deviations from a single power law.  The use of different velocity dispersion estimators does not lead to a reduction in scatter.
Diversity in the intrinsic scatter in the scaling relations is important for using velocity functions in cosmology (${\rm d}N/{\rm d}\sigma_v {\rm d} V$), which currently sets the goal for understanding cluster systematics to a per cent level. On the positive side, the scatter in the velocity dispersion-- richness scaling relation is very small ($8\pm2$\% for rich Gaussian clusters and  $12\pm2$\% for poor Gaussian ones; Table~\ref{table:linmix parameters vdisp}). In rich clusters, the contribution of scatter in richness at fixed mass has been estimated to amount to $8\pm2$\% \citep{Mulroy2019}, so the contribution of the scatter in $\sigma_v$ is $26\pm5$\% for the inferred total mass, using a power law index of 3 (respectively of 1) to rescale the scatter in $\sigma_v$ (respectively in lambda) at fixed mass. An increase in the scatter towards low-richness clusters might be associated with a corresponding increase in the richness scatter. Thus, adding an AD test to the characterization of clusters allows one to associate the correct uncertainty on the inferred mass with $z>1$ clusters as well as with low-mass groups, where baryonic mass proxies still require calibration. We also provide the scaling relations for the non-Gaussian clusters, which allows one to correct for the bias and to estimate the systematic error associated with mass estimates based on velocity dispersion in this case (which is $69\pm8$\%). One can also see that, after the removal of the contribution of scatter in richness, the ratio in the $\sigma_v$ scatter between Gaussian and non-Gaussian clusters at fixed mass is 2.7. Our results on the scatter in the mass estimate based on the velocity dispersion compare well with the methods outlined in \cite{old18}. While the scatter of the non-Gaussian clusters is typical to the outcome of applying \textit{Clean} to simulated data \citep[see Fig. 2 in][and account for their use of dex units]{old18}, the scatter on Gaussian clusters performs better than most estimates and is matched only by methods that use abundance matching and richness to infer the mass. Although we do not have access to the true mass in our tests,  we back up our claim on the low scatter in $\sigma_v$ using richness, which shows a low scatter in the tests against the true mass in \cite{old18}.

An important aspect of our work is that we cover the full range of cluster richness, sampling a great deal of clusters of 20 galaxies, adopted as the low-richness threshold for DES cluster cosmology papers \citep{Costanzi21}. We provide one of the first consistency checks for scaling relations between low- and high-richness clusters, concluding that, within a single power law description of the dataset, the normalization, slope, and scatter parameters of the model are the same among rich and poor X-ray clusters.
If we substitute $\sigma_v$ with total mass, we obtain an $M-\lambda$ slope of $0.93\pm0.05$, which is consistent with the results of our dynamical mass calibration \citep{2019Capasso}.

As we are entering a new era of large sample sizes of clusters, the errors on the derived parameters of the scaling relations reveal the intrinsic inhomogeneities of a sample. For example, the constraints on the $L_X-\lambda$ relation improve for the reduced sample that contains only clusters with small offsets in the X-ray-to-optical centres. \cite{seppi2023} demonstrate the importance of cluster offsets for eROSITA \citep{predehl21} and their link to a cluster's dynamical state and feedback through a comparison to simulations. Our results on the change in the scaling relations for the large-offset clusters might also be linked to the poorly understood phenomenon of X-ray-underluminous clusters in \citet{Andreon22}, who show that underluminous clusters populate the low concentration of dark matter end of the distribution for a given mass. To explain the large offset with X-rays, there should be a link between the concentration and formation of the cool core. While the nature of underluminous clusters might be more complicated,  the lack of dynamical disturbance, which follows from no change in the $\sigma_v-\lambda$ relation, supports the association of this subsample with low-concentration clusters.

Finally, we find that intrinsic scatter in the scaling relations is reduced at high redshifts. By matching our redshift binning to literature studies, we find consistency between our results and those of both \cite{mantz2016} and \cite{lopes09}. Our main result consists of a sharp change in the scatter followed by a much milder evolution. Our results call for a more sophisticated evolution of the scaling relation compared to the functional form explored currently. We also extend the results of \cite{mantz2016} to the full population of clusters. There is an increased scatter at $z<0.15$ in the $\sigma_v$--richness scaling relation, which can be reduced by considering Gaussian clusters in our sample or by replacing $\sigma_v$ with the results of the full dynamical analysis in \cite{lopes09}. For $L_X$, a reduction in scatter is attributed to the reduced role of cool cores \citep{McDonald16}; however, \cite{McDonald16} did not sample the clusters at $z<0.1$ well. Our results shed new light on the discussion of low-z cluster cosmology sparked by \cite{stanek06}, who presented results on the scatter that are in agreement with ours and showed a consistency in cosmology between the studies using the Cosmic Microwave Background and local galaxy cluster abundance once these results are taken into account. The sharp transition in the scatter would not appear as sharp if we used cosmic time. Indeed, while the evolution of scaling relations as well as the growth of structure slows down at low redshifts, the time available to develop cooling and feedback is several gigayears. This supports the idea that cooling and feedback processes play a primary role in the physical explanation. Our results support a picture in which not only overluminous but also underluminous systems are produced, which would imply that the associated active galactic nucleus feedback operating in cool cluster cores is capable of affecting the scaling relations of the clusters. Another source of variations is linked to the importance of cooling and feedback in the infalling substructure, which increases the diversity. At lower cluster masses there is the additional effect of a difference in the efficiency of baryonic conversion to stars, which enhances the scatter of the hot intracluster medium. This effect is not limited to low z; it has also been seen in serendipitous \textit{XMM-Newton} data on redMaPPer clusters at high z \citep{giles22}.

In the future, an expansion of this study will be feasible thanks to the performance of the 4MOST eROSITA cluster follow-up programme \citep{finoguenov2019}, detailed cluster studies by 4MOST CHANCES \citep{Haines23}, and blind spectroscopic surveys, such as WAVES \citep{WAVES} and 4HS \citep{4HS} in the Southern Hemisphere and the DESI bright galaxy survey \citep{DESI-BGS} and WEAVE \citep{WEAVE,WEAVE-clusters} in the Northern Hemisphere.

\begin{acknowledgements}
S. Damsted has received funding from the Finnish
      Emil Aaltonen Foundation. Funding for the Sloan Digital Sky Survey IV has been provided by the Alfred P. Sloan Foundation, the U.S. Department of
Energy Office of Science, and the Participating Institutions. SDSS-IV acknowledges support and resources from the Center for High Performance Computing at the University of Utah. The SDSS web
site is www.sdss.org.
SDSS-IV is managed by the Astrophysical Research Consortium for the Participating Institutions of the SDSS Collaboration
including the Brazilian Participation Group, the Carnegie Institution for Science, Carnegie Mellon University, the Chilean Participation Group, the French Participation Group, Harvard-Smithsonian
Center
 for Astrophysics, Instituto de Astrofísica de Canarias, The
Johns Hopkins University, Kavli Institute for the Physics and Mathematics of the Universe (IPMU) / University of Tokyo, the Korean Participation Group, Lawrence Berkeley National Laboratory,
Leibniz Institut für Astrophysik Potsdam (AIP), Max-Planck-Institut
für Astronomie (MPIA Heidelberg), Max-Planck-Institut für Astrophysik (MPA Garching), Max-Planck-Institut für Extraterrestrische
Physik (MPE), National Astronomical Observatories of China, New
Mexico State University, New York University, University of Notre
Dame, Observatário Nacional / MCTI, The Ohio State University,
Pennsylvania State University, Shanghai Astronomical Observatory, United Kingdom Participation Group, Universidad Nacional
Autónoma de México, University of Arizona, University of Colorado Boulder, University of Oxford, University of Portsmouth,
University of Utah, University of Virginia, University of Washington, University of Wisconsin, Vanderbilt University, and Yale
University. 
The data presented here were obtained in part with ALFOSC, which is provided by the Instituto de Astrofisica de Andalucia (IAA) under a joint agreement with the University of Copenhagen and the Nordic Optical Telescope. 
\end{acknowledgements}

\bibliographystyle{aa}
\bibliography{references}

\begin{appendix}
\section{Catalogues}\label{catalogs}

In performing this work we produced spectroscopic cluster member galaxy catalogues, applied the cleaning of the catalogues, updated the cluster properties and performed the measurement of the substructure parameter. In this subsection we describe the structure of the data release. The spectroscopic membership catalogues and the substructure analysis of clusters are released for the first time.\footnote{The data described in Tables A.1--A.4 are only available in electronic form at the CDS via anonymous ftp to cdsarc.cds.unistra.fr (130.79.128.5) or via https://cdsarc.cds.unistra.fr/cgi-bin/qcat?J/A+A/}

\begin{table}
\caption{CODEX red-sequence galaxy to SCGF group assignment.}
\label{table:SCGF full catalogue}      
\centering          
\begin{tabular}{ l l r }
\hline\hline       
Keyword & Description  \\
\hline
SPIDERS\_ID & SPIDERS ID\\
SCGF\_ID & SCGF group ID \\
RA\_GAL & Galaxy right ascension (FK5)    \\
DEC\_GAL & Galaxy declination   (FK5)\\
GAL\_SPECZ & Galaxy spectroscopic redshift \\
P\_SAT & Probability being a satellite \\ 
N\_SAT & Number of satellites \\
\hline                  
\end{tabular}
\end{table}

In Table~\ref{table:SCGF full catalogue} we provide the SCGF cluster membership catalogue for all 66k target red-sequence galaxies in Table~\ref{table:SCGF full catalogue}, which provides spectroscopic counterparts to  5024 clusters down to 3 members. The cross-matching of the SCGF group ID to the SPIDERS ID is done using SCGF member galaxies. The unique match between SPIDERS and CODEX ID is presented in \citep{finoguenov2020}. We list galaxy coordinates in the FK5 system (RA\_GAL, DEC\_GAL), and redshift (GAL\_SPECZ). We include the output of SCGF (P\_SAT, N\_SAT, SCGF\_ID). P\_SAT is the probability that this galaxy is a satellite. If P\_SAT$>0.5$, it is assumed to be a satellite and will be assigned to a group with a different central galaxy, and if the galaxy is a central P\_SAT$<0.5$. N\_SAT provides the number of satellites of the group. SCGF\_ID keyword stands for the SCGF group id.
SCGF often finds several components per cluster. These cases are identified by having several SCGF group IDs for the same SPIDERS ID. In the SPIDERS catalogue release, only the primary component was reported \citep{Kirkpatrick2021}, but the removal of the projected components has been performed.

\begin{table*}[ht]
\caption{Description of the clean CODEX spectroscopic cluster catalogue, spectroscopically identified with SCGF.}
\label{table:SCGF cluster catalogue}   
\centering          
\begin{tabular}{ l l r }
\hline\hline       
Keyword & Description  \\
\hline
SPIDERS\_ID & SPIDERS ID \\
SCGF\_ID & SCGF group ID \\
NGAL\_CLEAN & Number of clean galaxies \\
CLU\_SPECZ & Spectroscopic cluster redshift \\
CLUVDISP\_GAP & Cluster Gapper velocity dispersion  \\
CLUVDISP\_GAP\_E & Gapper velocity dispersion error  \\
CLU\_R200C & $R_{200c}$ in kpc accounting for aperture effects in velocity dispersion estimate  \\
ALPHA & AD test statistic \\
NGFLAG & Gaussianity flag  \\
RA\_XRAY & CODEX X-ray centre right ascension &  \\
DEC\_XRAY & CODEX X-ray centre declination \\
LAMBDA\_CHISQ\_OPT & Optical richness &  \\
LAMBDA\_CHISQ\_OPT\_E & Optical richness error \\
RA\_OPT & SCGF group centre right ascension \\
DEC\_OPT & SCGF group centre declination  \\
LX0124 & X-ray luminosity, ergs s$^{-1}$,  0.1--2.4 keV\\
LX0124\_E & X-ray luminosity error, 68\% c.l. \\
SAMPLING & Spectroscopic sampling of red-sequence members inside CLU\_R200C \\
\hline                  
\end{tabular}
\end{table*}

\begin{table*}[ht]
\caption{Description of the clean CODEX spectroscopic cluster catalogue, spectroscopically confirmed using SPIDERS manual inspection.}
\label{table:SPIDERS cluster catalogue}   
\centering          
\begin{tabular}{ l l r }
\hline\hline       
Keyword & Description  \\
\hline
SPIDERS\_ID & SPIDERS ID \\
NGAL\_CLEAN & Number of clean galaxies  \\
CLU\_SPECZ & Spectroscopic cluster redshift \\
CLUVDISP\_GAP & Cluster Gapper velocity dispersion  \\
CLUVDISP\_GAP\_E & Gapper velocity dispersion error  \\
CLU\_R200C & $R_{200c}$ in kpc   \\
ALPHA & AD test statistic \\
NGFLAG & Gaussianity flag  \\
RA\_XRAY & CODEX X-ray centre right ascension &  \\
DEC\_XRAY & CODEX X-ray centre declination \\
LAMBDA\_CHISQ\_OPT & Optical richness &  \\
LAMBDA\_CHISQ\_OPT\_E & Optical richness error \\
RA\_OPT & redMaPPer centre, right ascension \\
DEC\_OPT & redMaPPer centre, declination  \\
LX0124 & X-ray luminosity, ergs s$^{-1}$,  0.1--2.4 keV\\
LX0124\_E & X-ray luminosity error, 68\% c.l. \\
SAMPLING & Spectroscopic sampling of red-sequence members inside CLU\_R200C \\
INPUT\_CLUVDISP & Cluster velocity dispersion from SPIDERS manual inspection  \\
INPUT\_CLUZ & Cluster redshift from SPIDERS manual inspection  \\
\hline                  
\end{tabular}
\end{table*}

In addition, we release the catalogue of clusters (Table~\ref{table:SCGF cluster catalogue} and \ref{table:SPIDERS cluster catalogue}) and their membership (Table~\ref{table:member catalogue}) for clusters having at least 15 members, after performing the membership acquisition through SCGF or SPIDERS and running {\it Clean}. Only a single spectroscopic component per cluster remained at this step.

Tables~\ref{table:SCGF cluster catalogue} and \ref{table:SPIDERS cluster catalogue} show the keywords of our substructure catalogue and a short description of the keywords in the catalogue. The virial radius $R_{200c}$ is estimated by {\it Clean}, which takes into account the spatial extent of galaxy sampling and makes a correction to observed velocity dispersion based on the caustic profile. The X-ray luminosity is reported for the 0.1--2.4 keV rest-frame band and is recomputed for the updated spectroscopic redshift of the cluster. 
Input values from manual SPIDERS cluster inspection outside the area covered by \cite{Kirkpatrick2021}, is released for the first time.
 
Table~\ref{table:member catalogue} shows the keywords and short descriptions for the catalogue of cluster member galaxies. Keywords include the cluster ID in the SPIDERS catalogue; flags of the source catalogue; the member galaxy Right Ascension and Declination; the member galaxy spectroscopic redshift.

\begin{table}
\caption{Description of the clean cluster member galaxy catalogue.}
\label{table:member catalogue}      
\centering          
\begin{tabular}{ l l r }
\hline\hline       
Keyword & Description  \\
\hline
SPIDERS\_ID & SPIDERS ID \\
SCGF\_ID & SCGF group ID \\
RA\_GAL & Galaxy right ascension (FK5)    \\
DEC\_GAL & Galaxy declination   (FK5)\\
GAL\_SPECZ & Galaxy spectroscopic redshift \\
IN\_SPIDERS & True when in SPIDERS \\
IN\_SCGF & True when in SCGF \\
\hline                  
\end{tabular}
\end{table}

\section{Scaling relations for the SPIDERS sample}\label{SPIDERS appendix}

In the paper we have defined two different ways of determining galaxy membership while leading to two ways of producing the measurements of velocity dispersion and a corresponding classifier of the substructure. The main method for cluster member acquisition used in this paper is consistent with the methods used in cluster literature and shall be preferred. Here we study whether these differences are important in deriving our conclusions on the behaviour of scaling relations. We repeat here the analysis, performed in the paper for the SCGF sample, using the SPIDERS sample constructed from the sample of \cite{Kirkpatrick2021}, as explained in Sect. \ref{data}, and refer to a counterpart plot in the main body of the paper. The results are summarized in Tables~\ref{table:linmix parameters vdisp SPIDERS}, \ref{table:linmix parameters LX SPIDERS}, \ref{table:linmix parameters vdisp-LX SPIDERS}, and \ref{table:linmix results SPIDERS}. No statistically significant differences to the main results have been found, though one statement, on the lower normalization of the $L_x-\lambda$ relation for the sample at marginal substructure detection, was not found as significant. 

\begin{figure}[ht]
    \centering
    \includegraphics[width=\hsize]{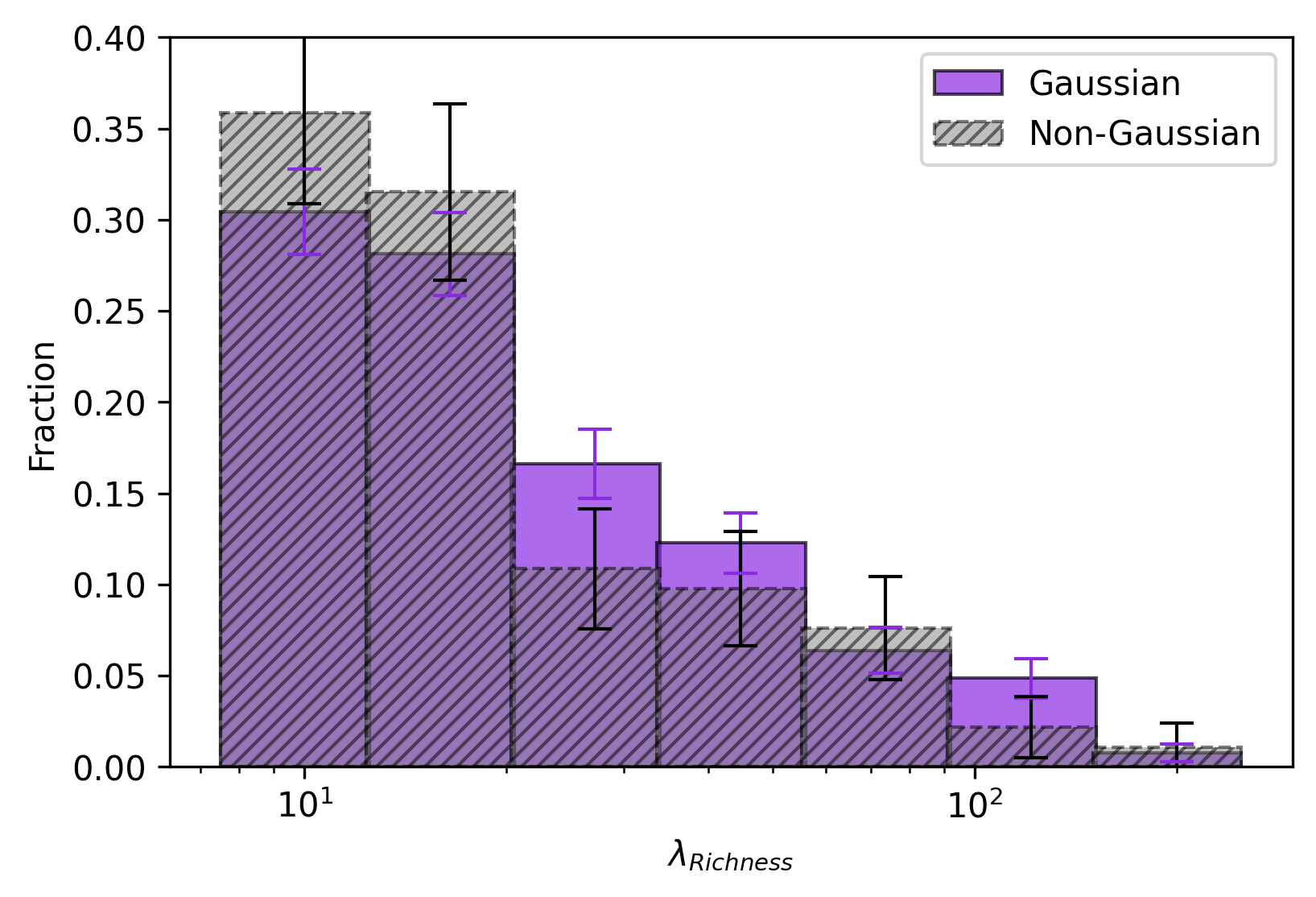}
    \caption{Normalized fraction of Gaussian and non-Gaussian clusters at different richnesses for the SPIDERS sample. Gaussian clusters are marked with purple and non-Gaussian clusters with hatched grey. Similar to Fig. \ref{fig:hist fraction of G NG richness SCGF}.
    }
    \label{fig:hist fraction of G NG richness SPIDERS}
\end{figure}

Figure \ref{fig:hist fraction of G NG richness SPIDERS} shows the normalized distributions of richnesses for Gaussian and non-Gaussian fractions of the SPIDERS sample (similar to Fig. \ref{fig:hist fraction of G NG richness SCGF}). The shuffling test on medians and KS test between the samples have p-values of 0.17 and 0.37, respectively.

\begin{figure}[ht]
    \centering
    \includegraphics[width=\hsize]{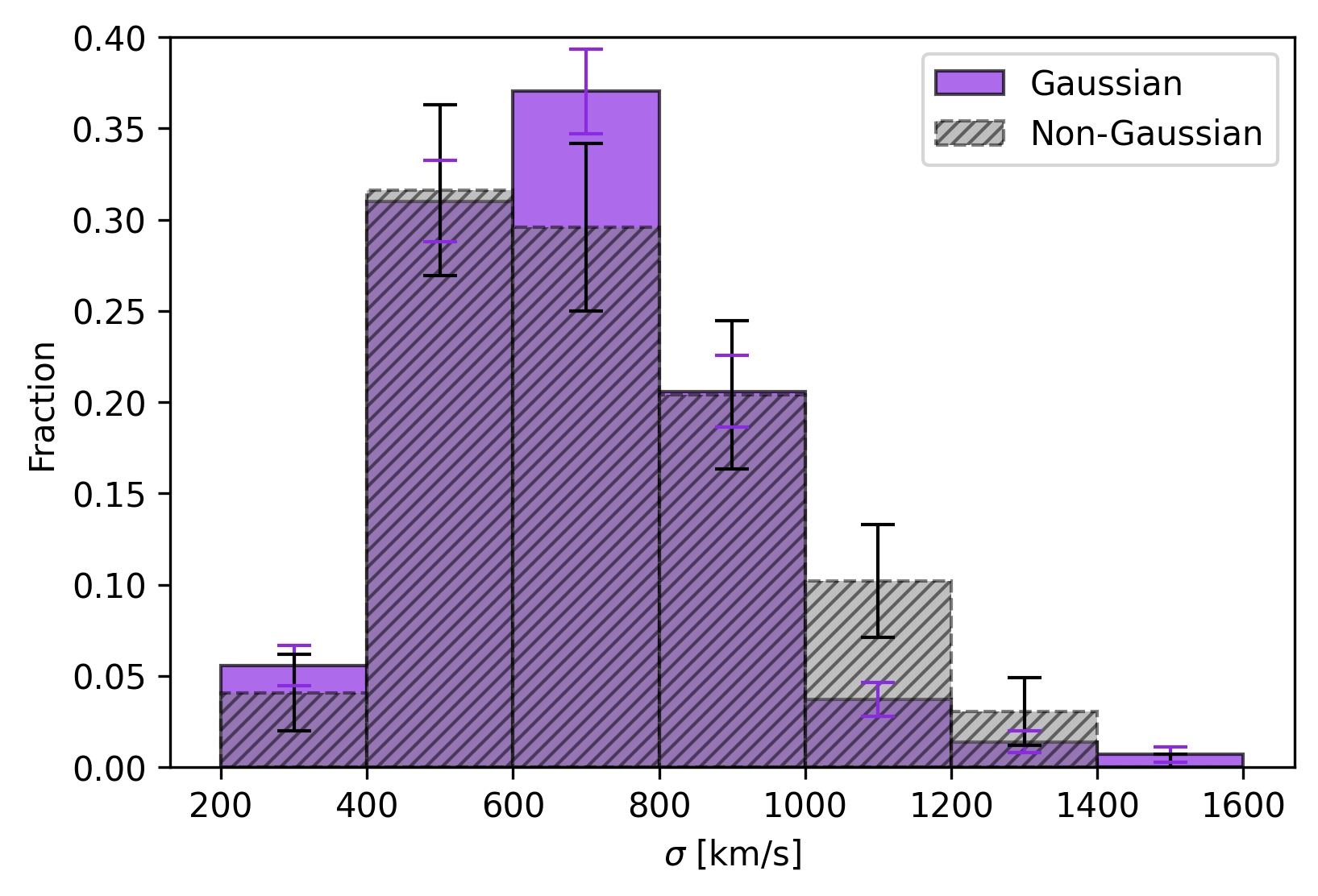}
    \caption{Normalized fractions of Gaussian and non-Gaussian clusters at different velocity dispersions for the SCGF sample. Gaussian clusters are marked with purple and non-Gaussian clusters with hatched grey. Similar to Fig. \ref{fig:hist fraction of G NG vdisp SCGF}.}
    \label{fig:hist fraction of G NG vdisp SPIDERS}
\end{figure}

Figure \ref{fig:hist fraction of G NG vdisp SPIDERS} shows the normalized fractions of Gaussian and non-Gaussian clusters
at different velocity dispersions (similar to Fig. \ref{fig:hist fraction of G NG vdisp SCGF}). A shuffling test between the samples on median and standard deviation gives p-values of 0.99 and 0.98, respectively, and the KS test yields a p-value of 0.11.

\begin{figure}[ht]
    \centering
    \includegraphics[width=\hsize]{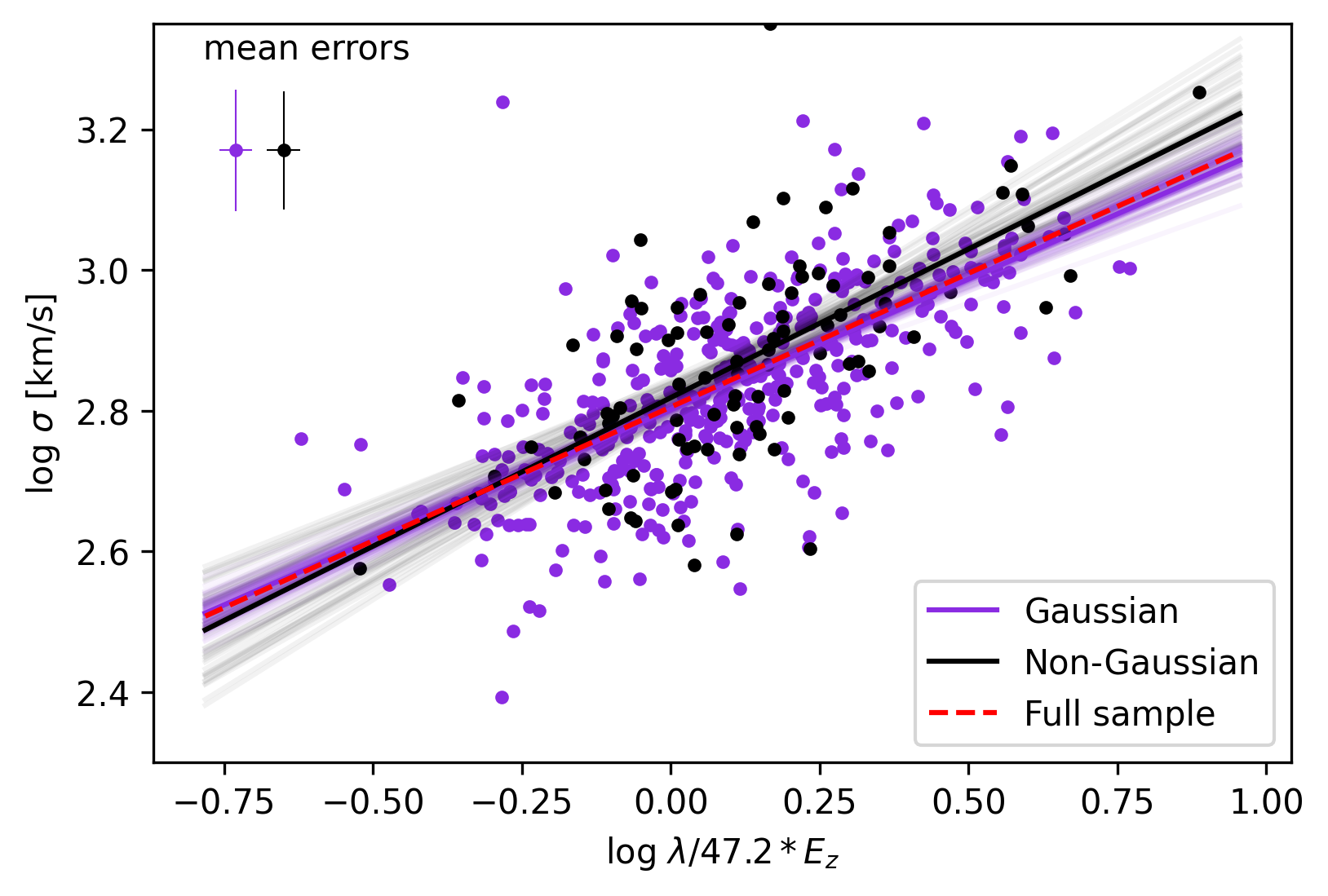}
    \caption{Velocity dispersion vs richness scaling relation for the SPIDERS sample. Similar to Fig. \ref{fig:scaling richness vdisp SCGF}.}
    \label{fig:scaling richness vdisp SPIDERS}
\end{figure}

\begin{figure}[ht]
    \centering
    \includegraphics[width=\hsize]{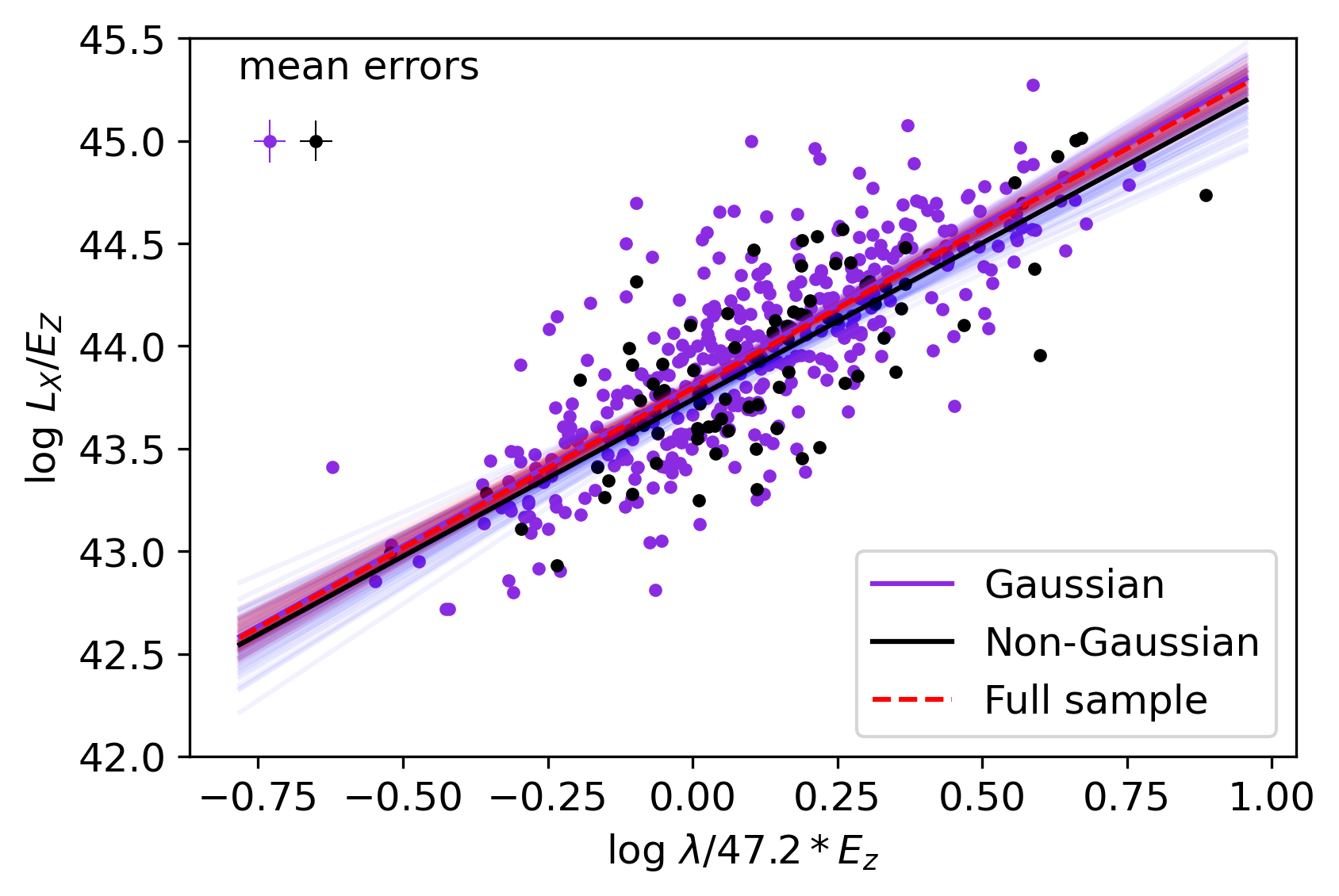}
    \caption{X-ray luminosity vs richness scaling relation for the SPIDERS sample. Similar to Fig. \ref{fig:scaling richness LX SCGF}.}
    \label{fig:scaling richness LX SPIDERS}
\end{figure}

Figures \ref{fig:scaling richness vdisp SPIDERS} and \ref{fig:scaling richness LX SPIDERS} show the scaling relations obtained from the \textit{linmix} routine for the velocity dispersion versus richness and X-ray luminosity versus richness (similar to Figs.\ref{fig:scaling richness vdisp SCGF} and \ref{fig:scaling richness LX SCGF}). A 2D KS test between the two samples has a p-value of 0.17 for velocity dispersion versus richness and 0.17 for X-ray luminosity versus richness.

\begin{figure}[ht]
    \centering
    \includegraphics[width=\hsize]{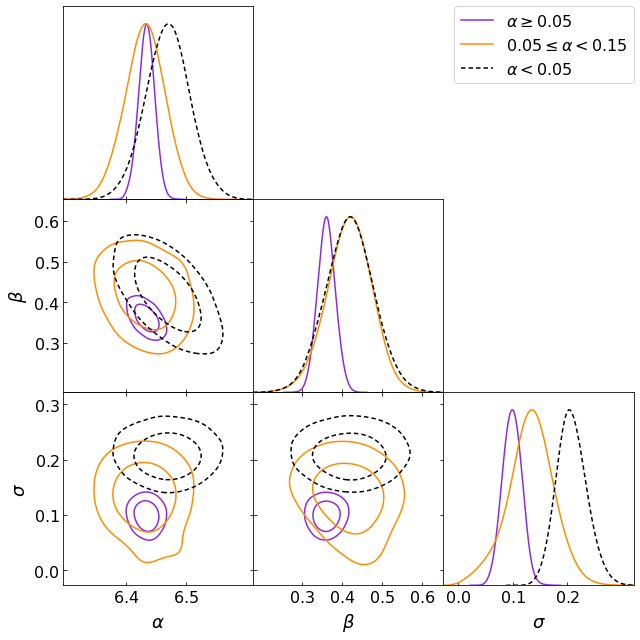}
    \caption{Marginal distributions and covariances for velocity dispersion vs richness scaling relation for the SPIDERS catalogue, split into different values of $\alpha_{AD}$. Similar to Fig. \ref{fig:triangle scaling vdisp alpha SCGF}.}
    \label{fig:triangle scaling vdisp alpha SPIDERS}
\end{figure}

\begin{figure}[ht]
    \centering
    \includegraphics[width=\hsize]{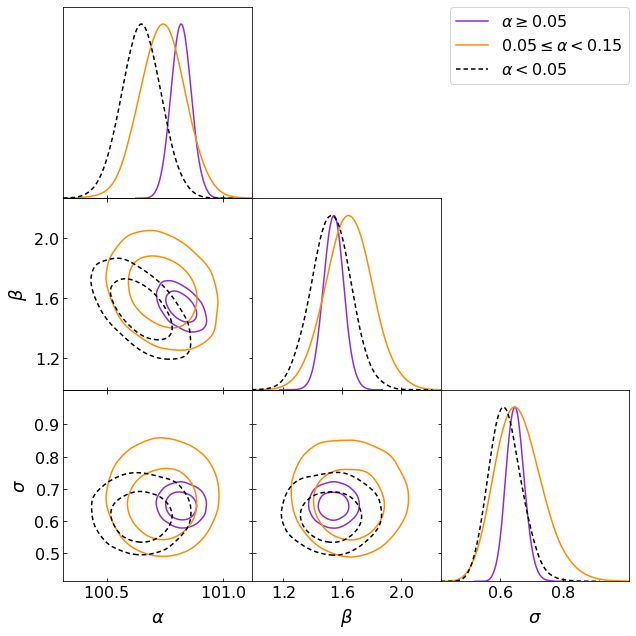}
    \caption{Marginal distributions and covariances of parameters of X-ray luminosity vs richness scaling relation for the SPIDERS catalogue, split into different values of $\alpha_{AD}$. Similar to Fig. \ref{fig:triangle scaling LX alpha SCGF}.}
    \label{fig:triangle scaling LX alpha SPIDERS}
\end{figure}

Figure \ref{fig:triangle scaling vdisp alpha SPIDERS} shows the marginal distributions and covariances of the \textit{linmix} parameters for intercept, slope and intrinsic scatter for velocity dispersion versus richness and Fig. \ref{fig:triangle scaling LX alpha SPIDERS} shows the same for X-ray luminosity versus richness. The figures also show results for possible substructure at $0.05 \leq \alpha_{AD} < 0.15$ (similar to Figs. \ref{fig:triangle scaling vdisp alpha SCGF} and \ref{fig:triangle scaling LX alpha SCGF}).

\begin{figure}[ht]
    \centering
    \includegraphics[width=\hsize]{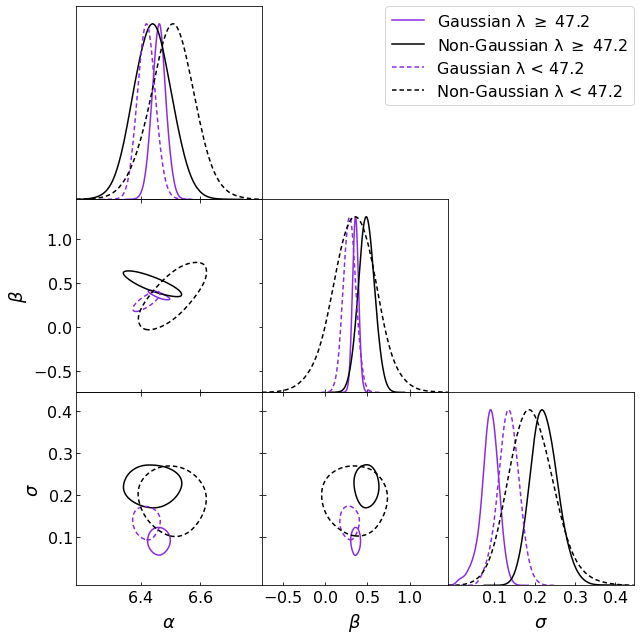}
    \caption{Effect of velocity substructure on scaling relations vs complexity of scaling relations in velocity dispersion vs richness. Similar to Fig. \ref{fig:triangle vdisp richness cut}.}
    \label{fig:triangle vdisp richness cut comparison SPIDERS}
\end{figure}

\begin{figure}[ht]
    \centering
    \includegraphics[width=\hsize]{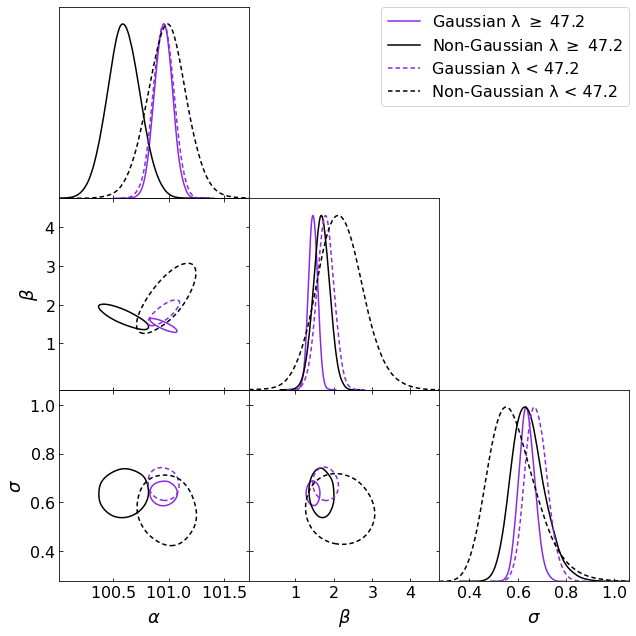}
    \caption{Effect of velocity substructure on scaling relations vs complexity of scaling relations in X-ray luminosity vs richness. Similar to Fig. \ref{fig:triangle LX richness cut}.}
    \label{fig:triangle LX richness cut comparison SPIDERS}
\end{figure}

Figures \ref{fig:triangle vdisp richness cut comparison SPIDERS} and \ref{fig:triangle LX richness cut comparison SPIDERS} show Gaussian and non-Gaussian samples with a richness cut at $\lambda = 47.2$. The first is for velocity dispersion versus richness and the second for X-ray luminosity versus richness scaling relations (similar to Figs. \ref{fig:triangle vdisp richness cut} and \ref{fig:triangle LX richness cut}).

\begin{figure}[ht]
    \centering
    \includegraphics[width=\hsize]{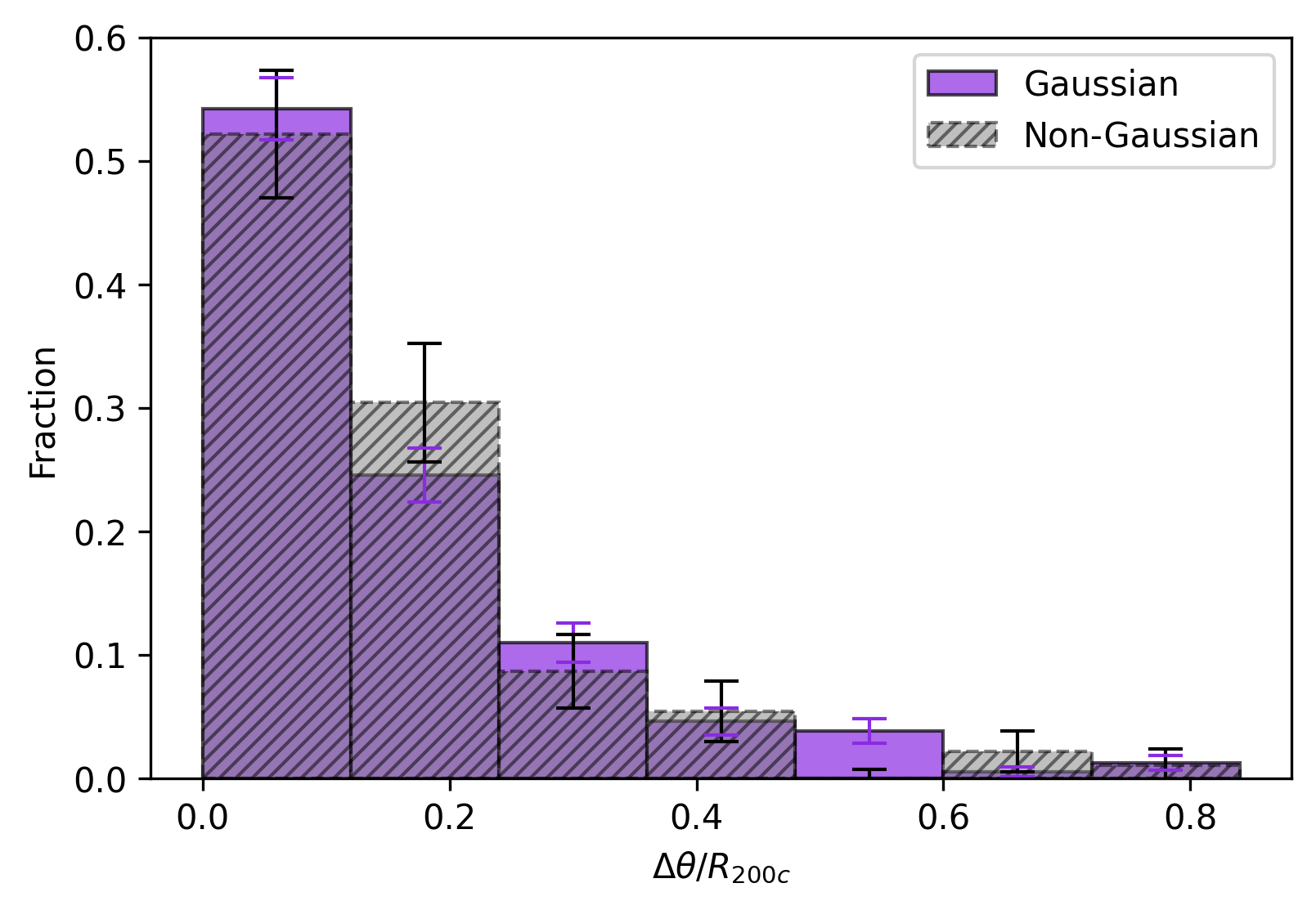}
    \caption{Normalized distributions for angular separation of SPIDERS optical and X-ray centres as a fraction of the virial radius of clusters. Similar to Fig. \ref{fig:center separation dist SCGF}.}
    \label{fig:center separation dist SPIDERS}
\end{figure}

\begin{figure}[ht]
    \centering
    \includegraphics[width=\hsize]{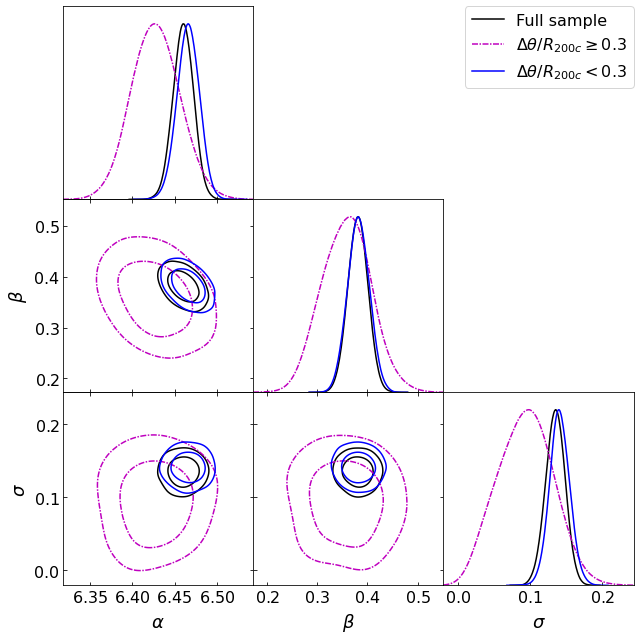}
    \caption{Effect of the offset between X-ray and optical centres on the velocity dispersion--richness scaling relation for the SPIDERS Gaussian subsample. The full sample is shown in solid black, small offset clusters in solid blue, and large offset clusters in dash-dotted magenta lines. The separation is done at 0.3R$_{200c}$. Equivalent to Fig. \ref{fig: triangle separation richness vdisp SCGF}.}
    \label{fig: triangle separation richness vdisp SPIDERS}
\end{figure}

\begin{figure}[ht]
    \centering
    \includegraphics[width=\hsize]{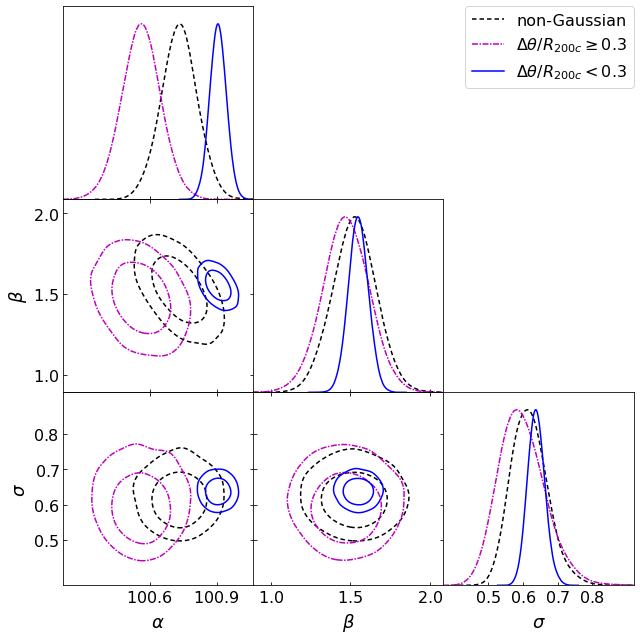}
    \caption{Effect of the offset between X-ray and optical centres on the X-ray luminosity - richness scaling relation for the SPIDERS Gaussian subsample. The non-Gaussian sample is shown in dashed black, small offset clusters - in solid blue and large offset clusters - in dash-dotted magenta lines. The separation is done at 0.3R$_{200c}$. Similar to Fig. \ref{fig: triangle separation richness LX SCGF}.}
    \label{fig: triangle separation richness LX SPIDERS}
\end{figure}

Figure \ref{fig:center separation dist SPIDERS} shows the normalized distributions of the angular separation of X-ray and optical centres for Gaussian and non-Gaussian clusters (the equivalent of Fig. \ref{fig:center separation dist SCGF}). Shuffling and KS tests between the samples have p-values of 0.29 and 0.65. Figures \ref{fig: triangle separation richness vdisp SPIDERS} and \ref{fig: triangle separation richness LX SPIDERS} show the marginal distributions and covariances for the same separations (similar of Figs. \ref{fig: triangle separation richness vdisp SCGF} and \ref{fig: triangle separation richness LX SCGF}).

\begin{figure}[ht]
    \centering
    \includegraphics[width=\hsize]{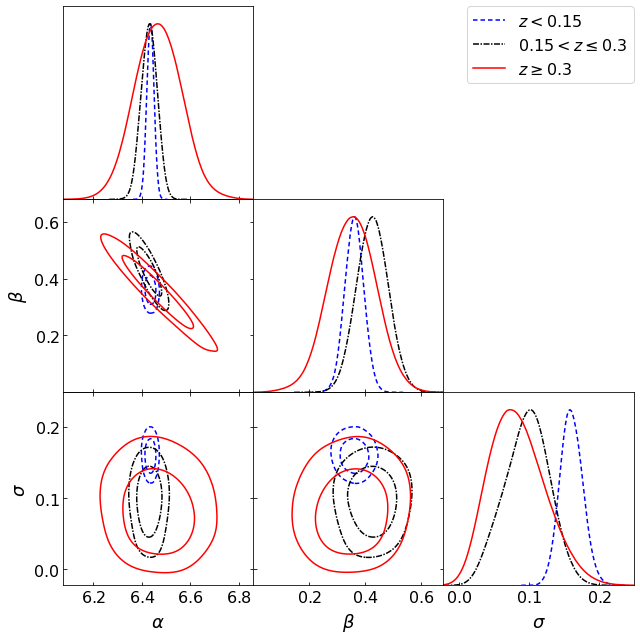}

    \caption{Marginal distributions and covariances of the parameters of velocity dispersion vs richness scaling relation in a range of redshift bins. The redshift sample $z < 0.15$ is marked with dashed blue, $0.15 \leq z < 0.3$ with dash-dotted black, and $z \geq 0.3$ with solid red lines. Similar to Fig. \ref{fig:triangle vdisp redshift cut SCGF}.}
    \label{fig:triangle vdisp redshift cut SPIDERS}
\end{figure}

\begin{figure}[ht]
    \centering
    \includegraphics[width=\hsize]{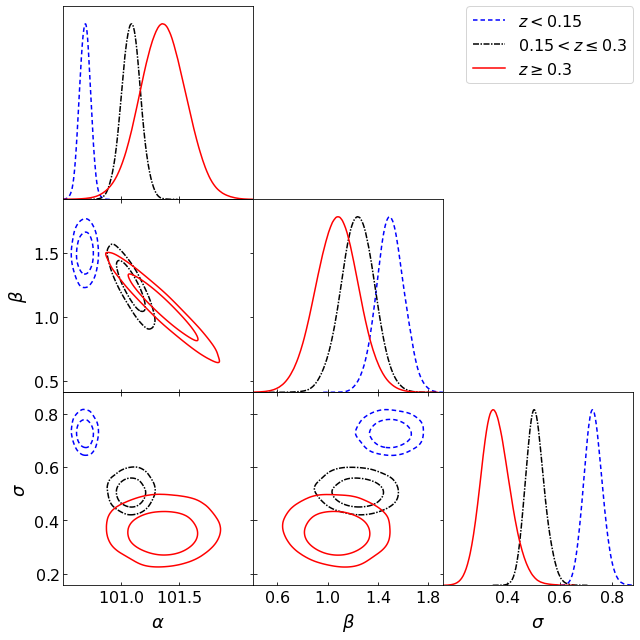}

    \caption{Same as Fig. \ref{fig:triangle vdisp redshift cut SPIDERS}, but for X-ray luminosity vs richness. Similar to Fig. \ref{fig:triangle LX redshift cut SCGF}.}
    \label{fig:triangle LX redshift cut SPIDERS}
\end{figure}

\begin{figure}[ht]
    \centering
    \includegraphics[width=\hsize]{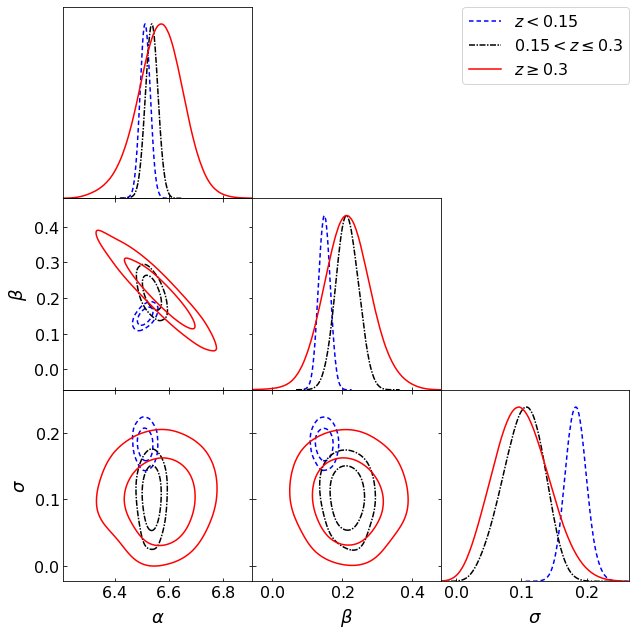}

    \caption{Same as Fig. \ref{fig:triangle vdisp redshift cut SPIDERS}, but for velocity dispersion vs X-ray luminosity. Similar to Fig. \ref{fig:triangle vdisp-LX redshift ranges SCGF}.}
    \label{fig:triangle vdisp-LX redshift ranges SPIDERS}
\end{figure}

Figure \ref{fig:triangle vdisp redshift cut SPIDERS} shows the distributions of the \textit{linmix} parameters for velocity dispersion versus richness using three redshift bins: $z<0.15$, $0.15<z<0.3$ and $z>0.3$ (similar to Fig. \ref{fig:triangle vdisp redshift cut SCGF}). Figure \ref{fig:triangle LX redshift cut SPIDERS} shows the same but for  X-ray luminosity versus richness (similar to Fig. \ref{fig:triangle LX redshift cut SCGF}), and Fig.\ref{fig:triangle vdisp-LX redshift ranges SPIDERS} shows the results for the $\sigma_v-L_x$ relation (similar to Fig. \ref{fig:triangle vdisp-LX redshift ranges SCGF}).

We conclude that our main results on the change in the scatter of scaling relations for non-Gaussian clusters, X-ray and optical centre separation, and lower scatter for the high-redshift sample, are robust against the cluster membership definition. Different scatter leads to stronger up-scattering of the non-Gaussian cluster in the velocity dispersion, and in \cite{Kirkpatrick2021} the scatter of the total sample, weighed by the shape of the mass function, is found to be closer to the scatter for non-Gaussian clusters.

\begin{table*}
\caption{Summary of regression analysis for velocity dispersion vs the richness scaling relation $\ln(\sigma_v\; {\rm km^{-1} s)}=\alpha + \beta \ln(\lambda E_z / 47.2)+N(0,\sigma).$
}             
\label{table:linmix parameters vdisp SPIDERS}      
\centering          
\begin{tabular}{ l c c c c}
\hline\hline       
Sample & Intercept $\alpha$ & Slope $\beta$ & Intrinsic scatter $\sigma$ & N clusters\\
\hline
Full  & $6.460\pm{0.012}$ & $0.380\pm{0.021}$ & $0.134\pm{0.014}$ & 483\\
Gaussian & $6.453\pm{0.013}$ & $0.371\pm{0.021}$ & $0.103\pm{0.017}$ & 391\\
Non-Gaussian  & $6.491\pm{0.036}$ & $0.422\pm{0.060}$ & $0.207\pm{0.028}$& 92 \\
Full $\lambda \geq$ 47.2 & $6.455\pm{0.024}$ & $0.392\pm{0.036}$ & $0.133\pm{0.016}$& 302 \\
Full $\lambda <$ 47.2 & $6.439\pm{0.029}$ & $0.315\pm{0.072}$ & $0.145\pm{0.022}$ & 181\\
Gaussian $\lambda \geq$ 47.2 & $6.462\pm{0.024}$ & $0.362\pm{0.036}$ & $0.090\pm{0.022}$ & 239\\
Gaussian $\lambda <$ 47.2 & $6.420\pm{0.030}$ & $0.291\pm{0.075}$ & $0.136\pm{0.026}$ & 152\\
Non-Gaussian $\lambda \geq$ 47.2 & $6.440\pm{0.065}$ & $0.490\pm{0.096}$ & $0.223\pm{0.034}$ & 63\\
Non-Gaussian $\lambda <$ 47.2 & $6.508\pm{0.075}$ & $0.352\pm{0.255}$ & $0.191\pm{0.056}$& 29\\
$0.05 < \alpha_{AD} \leq 0.15$ & $6.433\pm{0.050}$ & $0.420\pm{0.083}$ & $0.134\pm{0.061}$& 58\\
$\alpha_{AD} > 0.15$ & $6.435\pm{0.020}$ & $0.361\pm{0.034}$ & $0.099\pm{0.027}$&  333\\
$\Delta \theta/R_{200c} \geq 0.3$ & $6.427\pm{0.044}$ & $0.359\pm{0.072}$ & $0.094\pm{0.059}$& 64\\
$\Delta \theta/R_{200c} < 0.3$ & $6.466\pm{0.020}$ & $0.382\pm{0.033}$ & $0.141\pm{0.021}$& 419\\
$z \geq 0.3$ & $6.466\pm{0.146}$ & $0.354\pm{0.128}$ & $0.080\pm{0.062}$ & 64\\
$0.15 \leq z < 0.3$ & $6.431\pm{0.051}$ & $0.426\pm{0.085}$ & $0.098\pm{0.048}$ & 159\\
$z < 0.15$ & $6.435\pm{0.022}$ & $0.361\pm{0.051}$ & $0.160\pm{0.024}$& 260\\
Gaussian $z \geq 0.3$ & $6.538\pm{0.169}$ & $0.291\pm{0.161}$ & $0.097\pm{0.068}$  & 52 \\
Gaussian $0.15 \leq z < 0.3$ & $6.459\pm{0.050}$ & $0.395\pm{0.091}$ & $0.089\pm{0.049}$& 131\\
Gaussian $z < 0.15$ & $6.440\pm{0.024}$ & $0.342\pm{0.050}$ & $0.122\pm{0.030}$& 208\\
\hline                  
\end{tabular}
\end{table*}

\begin{table*}
\caption{Summary of the regression analysis for X-ray luminosity vs the richness scaling relation $\ln(L_X E^{-1}_z\; \rm ergs^{-1} s)=\alpha + \beta \ln(\lambda E_z / 47.2)+N(0,\sigma)$ for the SPIDERS sample.}             
\label{table:linmix parameters LX SPIDERS}      
\centering          
\begin{tabular}{ l c c c c }
\hline\hline       
Sample & Intercept $\alpha$ & Slope $\beta$ & Intrinsic scatter $\sigma$ & N clusters\\
\hline
Full  & $100.862\pm{0.034}$ & $1.553\pm{0.059}$ & $0.640\pm{0.024}$ & 483\\
Gaussian   & $100.890\pm{0.034}$ & $1.566\pm{0.063}$ & $0.645\pm{0.027}$ & 391\\
Non-Gaussian  & $100.731\pm{0.081}$ & $1.526\pm{0.137}$ & $0.615\pm{0.053}$  & 92\\
Full $\lambda \geq$ 47.2 & $100.872\pm{0.071}$ & $1.519\pm{0.104}$ & $0.643\pm{0.030}$ & 302\\
Full $\lambda <$ $47.2$ & $100.956\pm{0.081}$ & $1.824\pm{0.201}$ & $0.652\pm{0.040}$ & 181\\
Gaussian $\lambda \geq$ 47.2 & $100.953\pm{0.080}$ & $1.463\pm{0.115}$ & $0.633\pm{0.032}$ &239 \\
Gaussian $\lambda <$ 47.2 & $100.954\pm{0.094}$ & $1.789\pm{0.223}$ & $0.675\pm{0.045}$ & 152\\
Non-Gaussian $\lambda \geq$ 47.2 & $100.591\pm{0.148}$ & $1.689\pm{0.214}$ & $0.639\pm{0.068}$ & 63\\
Non-Gaussian $\lambda <$ 47.2 & $100.984\pm{0.175}$ & $2.146\pm{0.574}$ & $0.576\pm{0.099}$ & 29\\
$0.05 < \alpha_{AD} \leq 0.15$ & $100.741\pm{0.098}$ & $1.651\pm{0.160}$ & $0.654\pm{0.075}$ &  58\\
$\alpha_{AD} > 0.15$ & $100.819\pm{0.045}$ & $1.546\pm{0.070}$ & $0.648\pm{0.029}$& 333\\
 $\Delta \theta/R_{200c} \geq 0.3$ & $100.564\pm{0.089}$ & $1.475\pm{0.145}$ & $0.593\pm{0.066}$  & 64\\
 $\Delta \theta/R_{200c} < 0.3$ & $100.906\pm{0.038}$ & $1.552\pm{0.062}$ & $0.637\pm{0.025}$ & 419\\
 $z \geq 0.3$ & $101.421\pm{0.277}$ & $1.072\pm{0.254}$ & $0.355\pm{0.082}$  & 64\\
 $0.15 \leq z < 0.3$ & $101.151\pm{0.116}$ & $1.240\pm{0.203}$ & $0.505\pm{0.055}$& 159\\
 $z < 0.15$ & $100.767\pm{0.073}$ & $1.500\pm{0.162}$ & $0.727\pm{0.054}$& 260\\
Gaussian $z \geq 0.3$, $\sigma_v \geq 650$ [km/s] & $101.630\pm{0.334}$ & $0.908\pm{0.300}$ & $0.331\pm{0.093}$ & 52\\ 
Gaussian $0.15 \leq z < 0.3$, $\sigma_v \geq 650$ [km/s] & $101.210\pm{0.182}$ & $1.262\pm{0.283}$ & $0.527\pm{0.075}$& 87\\ 
Gaussian $z < 0.15$, $\sigma_v \geq 650$ [km/s] & $100.965\pm{0.170}$ & $1.391\pm{0.367}$ & $0.844\pm{0.117}$& 74\\
\hline                  
\end{tabular}
\end{table*}

\begin{table*}
\caption{Summary of regression analysis for velocity dispersion  (obtained using the gapper method) vs the X-ray luminosity scaling relation $\ln(\sigma_v\; km^{-1} s)=\alpha + \beta \ln(L_X E^{-1}_z\; 10^{-44} \rm ergs^{-1} s )+N(0,\sigma)$ for the SPIDERS sample.}             
\label{table:linmix parameters vdisp-LX SPIDERS}      
\centering          
\begin{tabular}{ l c c c c }
\hline\hline       
Sample & Intercept $\alpha$ & Slope $\beta$ & Intrinsic scatter $\sigma$ & N clusters\\
\hline
Full sample & $6.569\pm{0.018}$ & $0.180 \pm{0.017}$ & $0.159\pm{0.019}$ & 483\\
$z \geq 0.3$ & $6.584\pm{0.135}$ & $0.220 \pm{0.101}$ & $0.090\pm{0.068}$ &64\\
$0.15 \leq z < 0.3$ & $6.556\pm{0.034}$ & $0.218\pm{0.049}$ & $0.109\pm{0.045}$& 159\\
$z < 0.15$ & $6.529\pm{0.028}$ & $0.147\pm{0.025}$ & $0.179\pm{0.023}$& 260\\
Gaussian $z \geq 0.3$ & $6.512\pm{0.159}$ & $0.264\pm{0.128}$ & $0.079\pm{0.068}$  & 52\\
Gaussian $0.15 \leq z < 0.3$ & $6.557\pm{0.038}$ & $0.190\pm{0.053}$ & $0.097\pm{0.052}$& 131\\
Gaussian $z < 0.15$ & $6.501\pm{0.028}$ & $0.137\pm{0.024}$ & $0.149\pm{0.026}$& 208\\
Gaussian $z \geq 0.3$, $\lambda \geq 47.2$ & $6.511\pm{0.163}$ & $0.267\pm{0.133}$ & $0.090\pm{0.064}$ & 52\\
Gaussian $0.15 \leq z < 0.3$, $\lambda \geq 47.2$ & $6.584\pm{0.038}$ & $0.174\pm{0.055}$ & $0.096\pm{0.057}$& 114\\
Gaussian $z < 0.15$, $\lambda \geq 47.2$ & $6.573\pm{0.039}$ & $0.066\pm{0.039}$ & $0.147\pm{0.040}$ & 73 \\
Gaussian $z \geq 0.3$, $L_X \geq 10^{44}$ [ergs/s] & $6.589\pm{0.129}$ & $0.216\pm{0.099}$ & $0.104\pm{0.062}$ & 52\\
Gaussian $0.15 \leq z < 0.3$, $L_X \geq 10^{44}$ [ergs/s] & $6.554\pm{0.056}$ & $0.213\pm{0.070}$ & $0.135\pm{0.045}$ & 97\\
Gaussian $z < 0.15$, $L_X \geq 10^{44}$ [ergs/s] & $6.559\pm{0.076}$ & $0.091\pm{0.072}$ & $0.175\pm{0.047}$ & 46 \\
\hline                  
\end{tabular}
\end{table*}

\begin{table*}
\caption{Summary of results from the \textit{linmix} routine for the SPIDERS sample.}             
\label{table:linmix results SPIDERS}      
\centering          
\begin{tabular}{ l l c l l }
\hline\hline   

Relation & Test Sample & Result & Baseline sample & Parameter values  \\
\hline

$\sigma_v$ - $\lambda$ & Gaussian & lower $\sigma$ & Full  & $0.103\pm{0.017}$ vs $0.134\pm{0.014}$ \\
$\sigma_v$ - $\lambda$ & non-Gaussian & higher $\alpha$ & Gaussian  &  $6.491\pm{0.036}$ vs $6.453\pm{0.013}$ \\
$\sigma_v$ - $\lambda$ & non-Gaussian & higher $\sigma$ & Gaussian  & $0.207\pm{0.028}$ vs $0.103\pm{0.017}$ \\
$\sigma_v$ - $\lambda$ & Gaussian $\lambda \geq 47.2$ & lower $\sigma$ & Gaussian $\lambda < 47.2$   & $0.090\pm{0.022}$ vs $0.136\pm{0.026}$ \\
$\sigma_v$ - $\lambda$ &   $0.15\leq z < 0.3$ & lower $\sigma$ &  $z < 0.15$ & $0.098\pm{0.048}$ vs $0.160\pm{0.024}$ \\
$\sigma_v$ - $\lambda$ & Gaussian $0.15\leq z < 0.3$ & lower $\sigma$ & Gaussian $z < 0.15$ & $0.089\pm{0.049}$ vs $0.122\pm{0.030}$ \\

$L_X$ - $\lambda$ & Non-Gaussian & lower $\alpha$ & Gaussian  & $100.731\pm{0.081}$ vs $100.890\pm{0.034}$  \\
$L_X$ - $\lambda$ & $0.05 < \alpha_{AD} \leq 0.15$ & lower $\alpha$ &  $\alpha_{AD} < 0.05$  & $100.741\pm{0.098}$ vs $100.890\pm{0.034}$ \\

$L_X$ - $\lambda$ & $\Delta \theta/R_{200c} \geq 0.3$ & lower $\alpha$ & $\Delta \theta/R_{200c} < 0.3$ & $100.564\pm{0.089}$ vs $100.766\pm{0.038}$ \\
$L_X$ - $\lambda$ & $z < 0.15$ & higher $\sigma$ & $0.15\leq z < 0.3$ & $0.727\pm{0.054}$ vs $0.505\pm{0.055}$ \\
$L_X$ - $\lambda$ & $z < 0.15$, $\sigma_v \geq 650$ [km/s] & higher $\sigma$ & $0.15\leq z < 0.3$, $\sigma_v \geq 650$ [km/s]  & $0.844\pm{0.117}$ vs $0.527\pm{0.075}$ \\

$\sigma_v$ - $L_X$ & $0.15\leq z < 0.3$ & lower $\sigma$ & $z < 0.15$  & $0.109\pm{0.045}$ vs $0.179\pm{0.023}$\\
$\sigma_v$ - $L_X$ & Gaussian $0.15\leq z < 0.3$ & lower $\sigma$ & Gaussian $z < 0.15$   & $0.097\pm{0.052}$ vs $0.149\pm{0.026}$ \\
$\sigma_v$ - $L_X$ & Gaussian $0.15\leq z < 0.3$, $\lambda \geq 47.2$ & lower $\sigma$ & Gaussian $z < 0.15$ $, \lambda \geq 47.2$ & $0.096\pm{0.057}$ vs $0.147\pm{0.040}$\\
$\sigma_v$ - $L_X$ & G $0.15\leq z < 0.3$, $L_X \geq 10^{44}$ [ergs/s] & lower $\sigma$ & G $z < 0.15$, $L_X \geq 10^{44}$ [ergs/s] & $0.135\pm{0.045}$ vs $0.175\pm{0.047}$ \\

\hline                  
\end{tabular}
\end{table*}
\end{appendix}

\end{document}